\documentclass[12pt,a4paper]{article}

\usepackage[british]{babel}

\usepackage[a4paper,top=2cm,bottom=2cm,left=2cm,right=2cm,marginparwidth=1.75cm]{geometry}

\usepackage{graphicx}
\usepackage[title]{appendix}
\usepackage{mathrsfs}
\usepackage{amsmath,amsfonts,amsthm,amssymb}
\usepackage{stmaryrd} 
\usepackage{bbold}
\usepackage{booktabs} 
\usepackage{caption}  
\usepackage{threeparttable} 
\usepackage{algorithm}
\usepackage{algorithmicx}
\usepackage{algpseudocode}
\usepackage{listings}
\usepackage{enumitem}
\usepackage{chngcntr}
\usepackage{booktabs}
\usepackage{lipsum}
\usepackage{subcaption}
\usepackage{authblk}
\usepackage[T1]{fontenc}    
\usepackage{csquotes}       
\usepackage{diagbox}
\usepackage{float}
\usepackage{multirow}
\usepackage{tikz}
\usetikzlibrary{shapes.geometric, arrows}

\usepackage[colorinlistoftodos,prependcaption,textsize=tiny]{todonotes}

\tikzstyle{startstop} = [rectangle, rounded corners, minimum width=3cm, minimum height=1cm, text centered, text width = 10cm, draw=black, fill=white]
\usetikzlibrary{positioning, calc}
\tikzstyle{process} = [rectangle, minimum width=3cm, minimum height=1cm, text centered, text width = 6cm, draw=black, fill=white, text width = 10cm]
\tikzstyle{arrow} = [ultra thick,->,>=stealth]

\definecolor{nottblue}{HTML}{4b8893}
\definecolor{darkred}{HTML}{8B0000}
\definecolor{lightgray}{RGB}{240, 240, 240}
\definecolor{lightorange}{RGB}{255, 245, 242}

\usepackage{csquotes}
\usepackage{xcolor}

\usepackage{natbib}
\let\cite\citealt

\usepackage[
  colorlinks=true,
  linkcolor=teal,
  citecolor=teal,
  urlcolor=teal
]{hyperref}

\usepackage{setspace}
\usepackage{titlesec}
\titleformat{\section}
  {\normalfont\Large\bfseries}{\thesection.}{1em}{}
  
\usepackage{float}   
\usepackage{caption} 


\newcommand{\R}{\mathbb{R}}

\newcommand{\Sset}{\mathcal{S}}
\newcommand{\Tset}{\mathcal{T}}
\newcommand{\Eset}{\mathcal{E}}
\newcommand{\Tlags}{\Lambda_\mathcal{T}}
\newcommand{\Slags}{\Lambda_\mathcal{S}}

\newcommand{\h}{\boldsymbol{h}}
\newcommand{\V}{\boldsymbol{V}}
\newcommand{\X}{\boldsymbol{X}}
\newcommand{\Y}{\boldsymbol{Y}}
\newcommand{\Z}{\boldsymbol{Z}}
\newcommand{\W}{\boldsymbol{W}}

\newcommand{\E}{\boldsymbol{E}}
\newcommand{\s}{\boldsymbol{s}}

\newcommand{\hnorm}{\lVert \h \rVert}
\newcommand{\norm}[1]{\lVert #1 \rVert}

\AtBeginDocument{%
}

\title{Spatio-temporal modeling of urban extreme rainfall events at high resolution}
\author[1,*]{Chloe Serre-Combe}
\author[1]{Nicolas Meyer}
\author[2]{Thomas Opitz}
\author[1]{Gwladys Toulemonde}
\affil[1]{\small IMAG, Univ. Montpellier, CNRS, Inria LEMON team, Montpellier, France}
\affil[2]{BioSP, Inrae, Avignon, France}
\affil[*]{Corresponding author: \texttt{chloe.serre-combe@umontpellier.fr}}

\date{}  

\begin{document}
\maketitle

\begin{abstract}
Modeling precipitation and its accumulation over time and space is essential for flood risk assessment.
 In this paper, we analyze rainfall data collected over several years through a micro-scale precipitation sensor network in Montpellier, France.
  A novel spatio-temporal stochastic model is proposed for high-resolution urban extreme rainfall and combines realistic marginal behaviour and flexible dependence structure.
  Marginally, rainfall intensities are described by the Extended Generalized Pareto Distribution (EGPD), capturing both moderate and extreme events without threshold selection. Based on peaks-over-threshold theory for spatial processes, dependence during extreme 
  episodes is modeled by an $r$-Pareto process with a non-separable variogram allowing for episode-specific advection, such that the displacement of rainfall cells is
  represented explicitly. 
  Based on a catalog of extreme space-time episodes extracted from observations, parameters are estimated by a new composite likelihood based on joint exceedance indicators. Empirical advection velocities are derived beforehand from a radar reanalysis dataset.
  We show that the model accurately reproduces the spatio-temporal structure of extreme rainfall observed in the Montpellier OMSEV network and enables realistic stochastic scenario generation for flood risk assessment.
\end{abstract}

\textbf{Keywords}: Spatio-temporal extremes, rainfall modeling, high spatio-temporal resolution, $r$-Pareto process, advection

\section{Introduction}\label{sec:intro}

Rainfall modeling is essential for evaluating flood risks and analyzing how it is modulated by increasing urbanization,
particularly in areas such as the Cévennes mountain range and the nearby coastal agglomeration of the city of Montpellier in the south of France.
The Cévennes massif is known for  specific climatological features, experiencing some of the most extreme rainfall amounts in France,
especially during autumn. The topography of these mountains acts as a barrier to relatively warm and humid air masses from the Mediterranean Sea,
leading to atmospheric instability and orographic uplift. These conditions lead to intense rainfall events that can sometimes exceed 
$500$~mm within $24$~hours, as evidenced by historical occurrences, such as the significant floods in Anduze in 2002 with $680$ mm of 
rain in $24$ hours (\cite{fouchier_inondations_2004}), which is more than the annual average rainfall amount at other French sites such as  Paris (\cite{rainparis2021}).
These events, also called Cévenol episodes, are characterized by very large amounts of rainfall precipitating over short durations and
are often highly localized, possibly leading to severe flooding, especially flash floods.

The city of Montpellier (see \autoref{fig:maps}) also faces such extreme rainfall events, for instance in September 2014 with a total of $260$ mm of rain 
falling in only four hours (\cite{brunet_retour_2018}), or more recently with a total of $100$ mm in seven hours recorded on December 22, 2025 by Météo France at the station 
of Montpellier airport (\url{meteo.data.gouv.fr}).
The city's proximity to the Cévennes mountains exacerbates the risk of flooding, highlighting the need for accurate rainfall modeling and simulations
for effective flood risk assessment and urban planning. The city is traversed by the Lez river, whose source is to the northeast of Montpellier towards the Cévennes, and another smaller affluent.
The highly localized nature of these weather events emphasizes the importance of modeling and understanding
rainfall behavior at a high spatio-temporal resolution.

In this context, a rain-gauge network has been deployed in Montpellier by the Observatoire Montpelliérain et au Sud de l'Eau dans la Ville (OMSEV), detailed in Section \ref{sec:datapres}. 
This network provides high-resolution 
rainfall measurements at 1-minute frequency to help in the analysis of urban flood processes and their impacts within a hydrological run-off model developed 
at Inria Montpellier (\cite{steinstraesser2022sw2d}). However, these data present several challenges, including a relatively short observation period, 
a high proportion of missing values, the sparsity of non-zero rainfall observations and relatively strong discretization of values of small positive rainfall amounts.
While these challenges complicate statistical modeling and inference, our objective is to design 
a stochastic precipitation generator. 
This tool will simulate realistic rainfall 
fields at high spatio-temporal resolution, 
while accounting for advection effects
driving the spatial displacement of rainfall events.

Physics-based climate model simulations provide valuable large-scale information with spatial resolution from $10$ to $100$ km
and temporal resolution from hours to days or months.
However, their spatial and temporal resolutions remain too coarse to capture short, intense, and localized rainfall,
despite recent improvements with high-resolution (kilometer-scale) models (\cite{lucas2021convection}).
Their high computational cost also limits the generation of large ensembles.

Stochastic weather generators offer a complementary and efficient alternative (\cite{wilks1999weather}).
Within this class of models, stochastic precipitation generators (SPGs) focus on rainfall and aim to 
reproduce its main characteristics (occurrence, intensity, duration, dependence structure).
Designing SPGs is a challenging task due to heavy-tailed rainfall distributions, 
many zeros due to intermittence of rainfall, and complex space–time dependence patterns.
Early developments and methodological overviews highlight the diversity of SPG frameworks and their ability to capture rainfall variability across
a range of spatial and temporal scales (\cite{allard_stochastic_2015}). 
Previous works highlight the importance of advection for fine-scale rainfall dynamics
(\cite{leblois_space-time_2013,schleiss_stochastic_2014,huser2014space}).
Geostatistical approaches have also shown good performance for spatial structures
(\cite{benoit2017generating}), and meta-Gaussian frameworks provide realistic high-resolution simulations considering advection effects 
(\cite{benoit_stochastic_2018}).

Rainfall intensities exhibit different regimes, ranging from dry periods to moderate and extreme precipitation. 
Rainfall distributions at relatively high frequency are typically quite heavy-tailed. Therefore, accurate modeling of tail properties is very important to appropriately model the distribution of large rainfall amounts aggregated over spatiotemporal domains of interest, which motivates the use of extreme value theory.
Classical approaches rely on block maxima models based on the Generalized Extreme Value (GEV) 
distribution and Peaks-Over-Threshold (POT) 
models based on the Generalized Pareto Distribution (GPD).
POT approaches are often preferred for estimating parameters as they retain more extreme observations and offer greater flexibility, and directly model original extreme events instead of extracting the maximum over a temporal block.
The performance of POT methods depends on threshold choice which is challenging and significantly affects inference results, as highlighted in reviews by
\cite{coles2001introduction} and \cite{scarrott2012review}.
To avoid explicit threshold selection, the entire range of rainfall intensities can be modeled.
Mixture models combining a distribution for moderate rainfall with a GPD tail have been proposed
(\cite{carreau2009hybrid}).
These approaches still require a transition threshold and induce a dependence between bulk and tail behavior (\cite{naveau_modeling_2016}).
Instead, we here adopt extended Generalized Pareto Distributions (EGPDs) to avoid threshold selection
and model both moderate and extreme rainfall in a unified model
(\cite{papastathopoulos_extended_2013,naveau_modeling_2016}).
EGPDs allow a flexible representation of rainfall and have already proven their efficiency 
in precipitation modeling applications (\cite{haruna2023modeling, carrer_distributional_2022}).

Beyond marginals, realistic simulations require modeling spatio-temporal dependence.
Extreme dependence is often described using asymptotic models such as max-stable or Pareto processes
(\cite{davison2012statistical, ferreira2014generalized}).
Max-stable models are often difficult to interpret and simulate, especially in high dimension.
 (\cite{davison2013geostatistics}).
We instead rely on the Brown–Resnick $r$-Pareto process
(\cite{de_fondeville_high-dimensional_2018, HandbookExtremesCh16}),
a limiting model for threshold exceedances of stochastic processes.
This model has dependence structure parametrized as the max-stable Brown-Resnick process 
while offering greater flexibility and improved computational efficiency.

In the present work, we propose a framework for simulating high-resolution extreme rainfall fields.
Marginals are modeled with EGPD (\cite{naveau_modeling_2016}).
In contrast to recent high-resolution stochastic precipitation models, our approach focuses on
reproducing the spatio-temporal dependence of extremes.
This is achieved by estimating variogram parameters using a composite likelihood
based on joint exceedances from identified extreme episodes.
A key contribution is the explicit inclusion of event-specific advection
in the dependence structure through a non-separable spatio-temporal variogram.
Similar ideas have been explored for max-stable processes using wind information
at coarser scales (\cite{huser2014space}).
Here, we adopt an $r$-Pareto framework and focus on threshold exceedances
in high-resolution urban rainfall data.

The framework allows the simulation of realistic, moving extreme rainfall fields
over a given spatial domain.
We apply the proposed methodology to rainfall simulation on a grid covering the OMSEV network 
in the Montpellier area. 

This paper is organized as follows. Section \ref{sec:datapres} presents the data used in this study
for the proposed stochastic precipitation generator. 
Section \ref{sec:model} introduces the complete statistical model for extreme rainfall, 
combining an EGPD for the marginal distributions with the $r$-Pareto process to capture spatio-temporal dependence. 
Estimation procedures for the model parameters are detailed in Section \ref{sec:inference}, 
including the incorporation of advection effects. 
Section \ref{sec:application} applies the proposed framework to the OMSEV rainfall data 
to estimate the model parameters. Section \ref{sec:swg_application} presents the simulations
of extreme rainfall fields based on the fitted model over the OMSEV network and
evaluates its performance in reproducing the spatio-temporal structure of extreme rainfall events.
Finally, Section \ref{sec:discussion} concludes the paper with a discussion of the results and potential future research directions.

\section{Case study and datasets}
\label{sec:datapres}

\begin{figure}[H]
    \centering
    \begin{subfigure}{0.45\textwidth}
        \includegraphics[width=\linewidth]{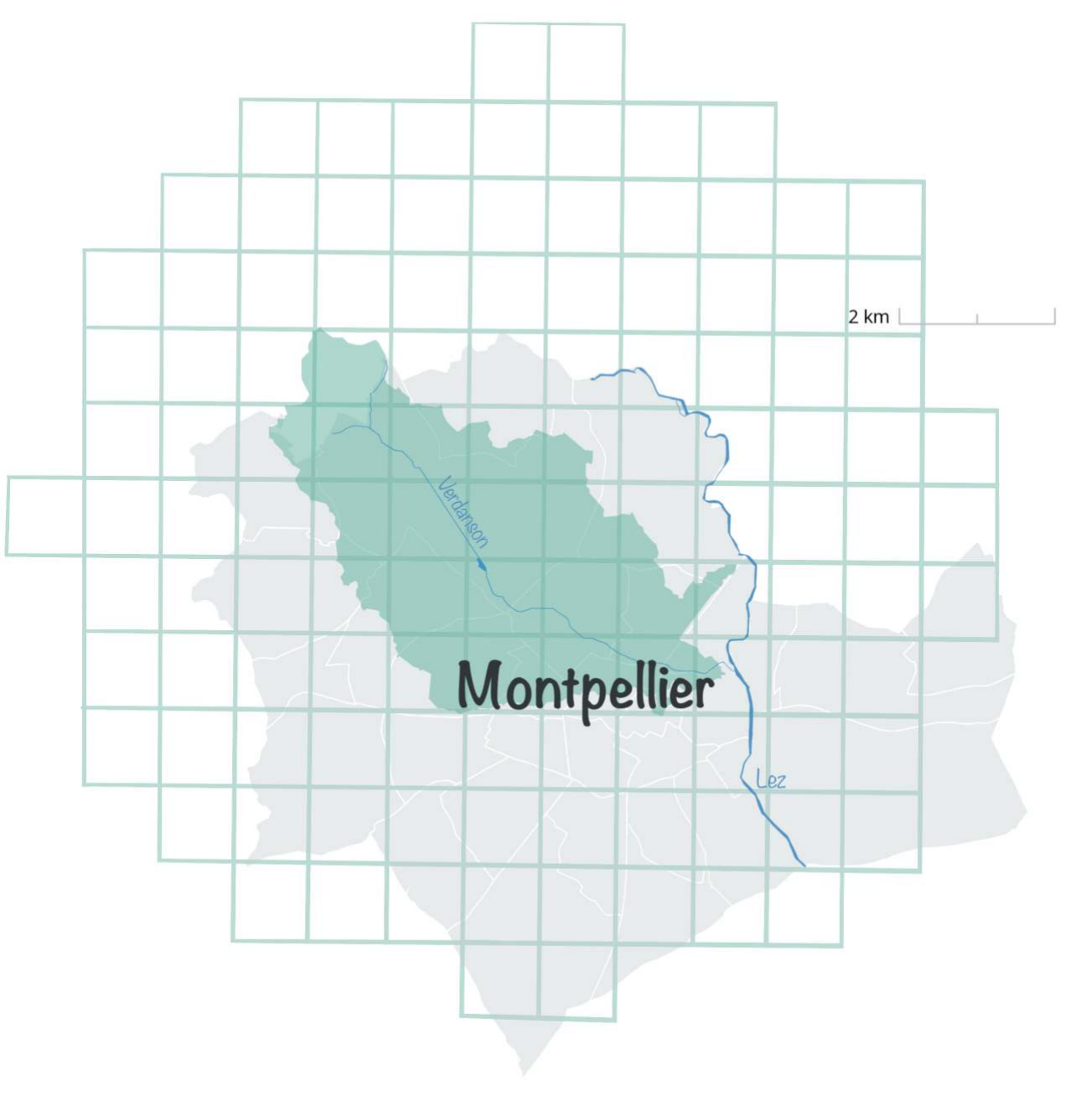}
        \caption{Verdanson river catchment}\label{fig:mtpquartier}
    \end{subfigure}
    \hfill
    \begin{subfigure}{0.48\textwidth}
        \includegraphics[width=\linewidth]{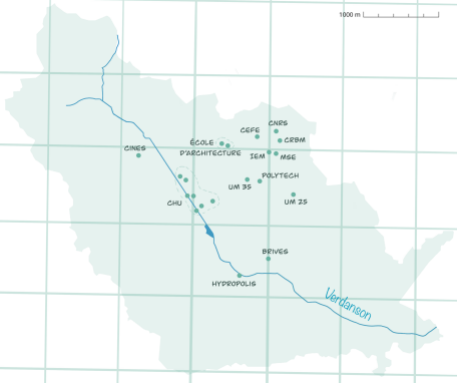}
        \caption{Rain gauges}\label{fig:stationames}
    \end{subfigure}

    \caption{Location of the 20 rain gauges of the OMSEV network over the Verdanson river catchment
    in Montpellier with COMEPHORE grid pixels of 1 km$^2$ resolution.}\label{fig:maps}
\end{figure}

Our study focuses on the OMSEV rain gauge network, located within the Verdanson river catchment, 
a tributary of the Lez river (see~\autoref{fig:maps}), flowing into the Mediterranean Sea around 10 km South of Montpellier. 
This network consists of 20 tipping-bucket rain gauges installed by the urban observatory of HydroSciences Montpellier (\cite{data2023pluvio}). 
Although tipping-bucket rain gauges provide high-resolution measurements, 
they may introduce measurement biases in recorded rainfall intensities. A proper calibration was 
therefore done by hydrologists of each rain gauge in order to obtain the most reliable estimates of rainfall intensity
(\cite{robin2021}). 
The OMSEV dataset globally spans the period from September 2019 to December 2024. We exclude three rain gauges installed only recently that do not yet provide sufficiently long data series (CINES, Brives and Hydropolis, 
see \autoref{fig:stationames}).
The dataset provides rainfall measurements at high temporal resolution, recorded every minute leading to discretized low values, even after hydrologists' calibration.
In order to reduce measurement errors, to decrease data volume and to avoid overly discretized values, the data are aggregated in $5$-minute
intervals which preserve a high temporal resolution. It allows to reduce the number of missing values that are considered as $0$ over these intervals when there is at least one non-missing value, 
which are more frequent at the original minute resolution. 
At this temporal resolution, approximately $23\%$ of data are missing, due to activation dates of each rain gauges and measurement errors and are not imputed in the present work.
With this fine resolution, a kind of discretization is still present for very low values, and the proportion of non-zero values is very low, representing only $1.2\%$ of the data.
Spatial granularity is also very fine with inter-site distance ranging from $77$ to $1531$ meters (without removed gauges).
Data from urban microscale rain gauge networks like this are very scarce (e.g. \cite{casas2010analysis} for Barcelona, Spain) but can provide important insights rainfall patterns known 
to be highly variable and complex even at very small spatio-temporal scales.

The rain gauges of the OMSEV network are irregularly distributed. A more
regular spatial grid would be desirable for rainfall field modeling. 
In addition, the relatively short temporal coverage of the OMSEV dataset may not fully capture important
features and the variability 
of extreme rainfall events. 
To address these limitations, we enrich these data with the French COMEPHORE\footnote{COmbinaison en vue de la 
Meilleure Estimation de la Précipitation HOraiRE. The dataset is available on AERIS platform  
(\href{https://radarsmf.aeris-data.fr/}{https://radarsmf.aeris-data.fr/}) } mosaic from Météo France (\cite{tabary201210}).
COMEPHORE is a reanalysis product combining radar and rain gauge observations over France, provided on a regular grid with a spatial resolution of 1 km$^2$ per pixel 
and an hourly temporal resolution.
The dataset spans from January 1997 to December 2025. We restrict the analysis to after 2008 due to a change in the reanalysis model in 2007 inducing possible
distributional changes.
We extract the data over the same area as the OMSEV network. It corresponds to 136 pixels within a $5.5$ km radius of the network centroid, 
which is sufficient to represent the local spatial variability. Only five of these pixels contain the OMSEV rain gauges or seven 
pixels when considering also removed gauges (see \autoref{fig:maps}).

\section{Spatio-temporal modeling for extreme precipitation}
\label{sec:model}

Let $\X = \{X_{\s, t},\, (\s, t) \in \Sset \times \Tset\}$ be a nonnegative
spatio-temporal random field representing the rainfall intensity
for all locations $\s$ in a spatial domain $\Sset \subset \R^2$ 
and all times $t$ in a temporal domain $\Tset \subset \R_+$.
Let $\Lambda_{\Sset} \subset \R^2$ and $\Lambda_{\Tset} \subset \R$ be the sets of spatial and 
temporal lags, respectively, with $\Lambda_{\Sset}$ gathering all possible differences $s_2-s_1$ in 
the space domain for $\s_1,\s_2\in \Sset$ and  $\Lambda_{\Tset}$ gathering all possible differences 
$t_2-t_1$ in the time domain for $t_1,t_2\in \Tset$. 
In the following, we introduce the proposed model for constructing the stochastic process $\X$ for extreme rainfall. 
It combines a flexible marginal distribution 
using the Extended Generalized Pareto Distribution (EGPD) and a spatio-temporal 
dependence structure for extreme-event episodes based on the $r$-Pareto process.

\subsection{Marginal rainfall distributions}
\label{sec:margins}

Rainfall at high temporal resolution is characterized by a particularly large proportion of zero values,
corresponding to periods without precipitation, to which we simply refer as dry periods.
To account for this structure, rainfall at each site is modeled using a mixture distribution
combining a discrete point mass at zero and a distribution for positive values. 
Depending on the application, the marginal distribution parameters are allowed to vary
in space and time, or may be assumed constant.
In this work, we adopt the latter approach and assume stationary marginal distributions
with parameters that are constant over the spatial domain and over time.
This assumption is further assessed and discussed in Section~\ref{sec:egpd_omsev}.
\subsubsection{Modeling positive rainfall intensities}
\label{sec:egpd}

Let $X_{\s}$ be the rainfall intensity at a 
fixed location $\s$ for a fixed time interval.
To model the whole marginal distribution of positive rainfall intensities, we use the EGPD.
It is constructed as an extension of the classical Generalized Pareto
Distribution (GPD), which corresponds to the only possible limit distribution of excesses above a high threshold
as the latter tends to the upper bound of the support of the distribution.
In practice, for a sufficiently large threshold $u$, the distribution of the
exceedances is approximated by a GPD with shape parameter $\xi$ and scale
parameter $\sigma_u>0$:
\begin{eqnarray*}
\mathbb{P}\left(X_{\s} - u > y \mid X_{\s} > u \right)
\approx
\overline{H}_{\xi, \sigma}\!\left(y\right)
=
\begin{cases}
\left(1 + \xi \dfrac{y}{\sigma_u}\right)_{+}^{-1/\xi},
& \text{if } \xi \neq 0,\\[0.2cm]
\exp\!\left(-\dfrac{y}{\sigma_u}\right),
& \text{if } \xi = 0,
\end{cases}
\qquad y>0,
\label{eq:gpd}
\end{eqnarray*}
where $(a)_+ = \max(a,0)$. 
Various more flexible extensions of the GPD can be considered to ensure a satisfactory fit when using relatively 
low thresholds or no threshold at all. 
The EGPD we use is defined through a polynomial transformation of the GPD distribution, namely $P(x) = x^{\kappa}$ with $\kappa > 0$. 
This choice provides a simple one-parameter extension that performs well in practice while remaining simple compared to alternative EGPD 
specifications \cite{naveau_modeling_2016}.
We define $F_{\mathrm{EGPD}}$ as the EGPD cumulative distribution function with parameters $(\xi, \sigma, \kappa)$, that is,
\[
F_{\mathrm{EGPD}}(\cdot ;\xi, \sigma, \kappa) = H_{\xi, \sigma}\!\left(\cdot\right)^{\kappa},
\]
where the scale parameter $\sigma$ no longer depends on the threshold $u$.
The EGPD retains a GPD-type upper tail. More precisely, 
its tail is asymptotically equivalent to a GPD with the same shape parameter $\xi$ 
and a modified scale parameter, ensuring that the extreme-value behavior is unchanged. 
At the same time, the parameter $\kappa$ controls the distribution 
at moderate intensities through a smooth transformation of the GPD, allowing a better fit over the 
whole range of positive rainfall values.
In our work, this distribution is used to model the marginal distribution of positive rainfall intensities, that is $X_{\s} \mid X_{\s} > 0$.

\subsubsection{Complete marginal model}
\label{sec:overall_marginal_model}

Let $p_{0}$ denote the marginal probability of zero rainfall, assumed to be spatially and temporally constant 
in our framework. 
Then, the complete marginal distribution of rainfall intensity, is given by 
a mixture of a point mass at zero and an EGPD for positive values:
\[
F(x)=
\begin{cases}
p_{0}, & x = 0,\\[0.2cm]
p_{0} + (1 - p_{0})\,F_{\mathrm{EGPD}}(x;\xi, \sigma, \kappa), & x > 0,
\end{cases}
\]
with constant parameters $p_0$, $\xi$, $\sigma$ and $\kappa$ over the spatial domain and over time.
This expression allows the dry-period frequency and the distribution of positive rainfall
intensities to be modeled separately, while keeping a coherent marginal model.

\subsection{Spatio-temporal extremal dependence}
\label{sec:extreme_dependence}

We consider the $r$-Pareto process introduced by \cite{de_fondeville_high-dimensional_2018} and \cite{palacios2020generalized}. 
To model the spatio-temporal dependence of extreme rainfall, 
this process provides a natural framework for modeling the dependence structure of exceedances above high 
thresholds.

We consider a spatio-temporal process $\X$
and work with the common assumption that it is in the maximum domain of attraction of a \emph{simple} max-stable process having unit Fréchet margins, i.e., $\X$ is a regularly varying process. We use a nonnegative and 
$1$-homogeneous risk function $r$, that is,
a function satisfying $r(a\X) = a\,r(\X)$ for any $a>0$.
As shown by \cite{HandbookExtremesCh16}, the $r$-exceedances of $\X$
converge in distribution to a $r$-Pareto process
$\Y = \{Y_{\s,t},\, (\s, t) \in \Sset \times \Tset \}$:
\[
u^{-1} \X \mid r(\X) > u
\;\xrightarrow{d}\;
\Y,
\qquad u \to \infty.
\]
In practice, we apply $r$ to subperiods of the whole study period, 
and those subperiods where the risk functional exceeds a 
high fixed threshold $u$ are considered as extreme episodes with respect to $r$. 

To model the dependence structure of $\Y$, we assume a
Brown-Resnick representation induced by an underlying zero-mean Gaussian process
$\W = \{W_{\s,t}, (\s,t)\in\Sset\times\Tset\}$ with stationary increments.
The dependence structure of the process is then characterized through the
spatio-temporal variogram $\gamma$ of the underlying Gaussian process defined 
for any $(\h,\tau) \in \Slags \times \Tlags$ as
\[
\gamma(\h,\tau)
=
\frac{1}{2}
\mathbb{V}\!\left(
W_{\s+\h,\,t+\tau} - W_{\s,t}
\right).
\]
In this paper, for inference on the model parameters we consider the risk function $r(\X) = X_{\s_0,t_0}$ where $(\s_0,t_0) \in \Sset \times \Tset$
corresponds to a fixed reference spatio-temporal location.
The corresponding $r$-Pareto process can be represented as
\[
Y_{\s,t}
=
R \exp\!\left(
W_{\s,t} - W_{\s_0,t_0} - \gamma(\s-\s_0,\,t-t_0)
\right),
\]
where $R$ follows a univariate Pareto distribution, that is,
$\mathbb{P}(R>v)=v^{-1}$ for $v\ge1$.
This approach is relevant since it links the variogram of the underlying Gaussian process
to the dependence structure of the $r$-Pareto process $\Y$. It allows capturing the main features
of the extremal dependence structure of the process $\X$ with respect 
to the conditional exceedance at $(\s_0,t_0) \in \Sset \times \Tset$ given by the risk function.

Extremal dependence can be summarized through the spatio-temporal extremogram (\cite{buhl2019semiparametric})
which corresponds to the probability that an exceedance at one location and time
occurs given an exceedance at another location and time.
Following \cite{coles1999dependence}, the extremal dependence coefficient is defined
for a spatial lag $\h\in\Lambda_{\Sset}$ and a temporal lag
$\tau\in\Lambda_{\Tset}$ as
\begin{equation}\label{eq:extremogram}
\chi(\h,\tau)
=
\lim_{q\to1}
\mathbb{P}\!\left(
X^*_{\s+\h,\,t+\tau} > q
\;\middle|\;
X^*_{\s,\,t} > q
\right),
\end{equation}
where $X^*_{\s,t}$ denotes the uniform marginal transformation of $X_{\s,t}$.
The coefficient $\chi(\h,\tau)$ ranges from $0$ (asymptotic independence) to $1$
(perfect extremal dependence). 
For the chosen $r$-Pareto process,
the extremal dependence of $\X$ can be summarized through the
$r$-extremogram
\[
\chi_r(\h,\tau)
=
\lim_{q\to1}
\mathbb{P}\!\left(
X^*_{\s_0+\h,t_0+\tau}>q \mid X^*_{\s_0,t_0}>q
\right)
=
\mathbb{P}\!\left(
Y_{\s_0+\h,t_0+\tau}>1
\right),
\]
which describes the extremal dependence structure of $\X$ with respect to the conditional exceedance.
Under the Brown-Resnick model representation, this coefficient depends
directly on the variogram and can be expressed as
\begin{equation}\label{eq:chivario}
\chi_r(\h,\tau)
=
2\left(
1-\Phi\!\left(\sqrt{\tfrac12\,\gamma(\h,\tau)}\right)
\right),
\end{equation}
where $\Phi$ denotes the standard normal distribution function (\cite{buhl2019semiparametric}) and 
$(\h,\tau)=(\s-\s_0,t-t_0)\in\Slags\times\Tlags$
represents the space-time lag
with respect to the reference location.
This formulation entails that small values of $\gamma(\h,\tau)$ correspond 
to strong extremal dependence,
whereas large variogram values imply weak dependence.
Since $\gamma(\h,\tau)$ is obtained through the extremogram,
which itself depends on a high threshold defining exceedances,
the resulting quantity is denoted as an \emph{extremal variogram},
\textit{i.e}, a variogram describing the dependence structure of the extremes.
For rainfall, it is natural to expect extremal dependence to vanish at large spatial or temporal lags. 
From \eqref{eq:chivario}, this requires the variogram $\gamma(\h,\tau)$ to diverge as $\|\h\|$ or $|\tau|$ increase. 
Consequently, the underlying Gaussian process $\W$ cannot be stationary, but must instead have stationary increments.
This is consistent with the \emph{intrinsic} (stationary-increment) structure assumed for $\W$.

\subsection{Non-separable space-time variogram with advection}
\label{sec:variogram_model}

A common approach to model spatio-temporal dependence consists in considering the variogram 
as a sum of purely spatial and purely temporal components.
In these so-called separable models, the spatio-temporal variogram can be expressed as
\begin{equation}\label{eq:gammasep}
  \gamma(\h, \tau) = \gamma_{\Sset}(\h) + \gamma_{\Tset}(\tau), \quad \h \in \Slags,\, \tau \in \Tlags,
\end{equation}
where $\gamma_{\Sset}$ and $\gamma_{\Tset}$ are purely spatial and purely temporal variograms, respectively.

Separable spatio-temporal variogram models could provide a simple and reasonable description of the overall 
dependence structure of rainfall data (\cite{buhl2019semiparametric}).
However, they typically fail to capture the complex interactions between space and time,
especially in the context of meteorological phenomena and in particular for extreme rainfall events.
Extreme precipitation is often driven by dynamical mechanisms, such as advection that induce
strong interactions between space and time.
In meteorology, it corresponds to the horizontal transport of properties (heat, moisture) by 
the movement of air masses, such as wind and clouds. 
This phenomenon therefore influences the spatio-temporal behavior of precipitation, but 
cannot be adequately captured by a separable model and requires more flexible
non-separable spatio-temporal structures.

To construct a non-separable variogram accounting for advection, 
we start from a spatio-temporal process with stationary increments, 
whose spatial and temporal variograms are, for spatial lag $\h \in \Slags$ and temporal lag $\tau \in \Tlags$,
\[
\gamma_{\Sset}(\h) = 2 \beta_1 \|\h\|^{\alpha_1},
\qquad
\gamma_{\Tset}(\tau) = 2 \beta_2 |\tau|^{\alpha_2},
\]
with $\beta_1, \beta_2 > 0$ and $\alpha_1, \alpha_2 \in (0,2]$. The corresponding Gaussian processes are known as spatial and temporal fractional Brownian motion, respectively. 

Advection is incorporated through a deterministic space-time shift defined by a velocity vector $\V \in \mathbb{R}^2$, yielding the non-separable variogram
\[
\gamma(\h, \tau; \Theta, \V)
=
2 \left(
\beta_1 \|\h - \tau \V\|^{\alpha_1}
+
\beta_2 |\tau|^{\alpha_2}
\right),
\]
where $\Theta = (\beta_1, \beta_2, \alpha_1, \alpha_2)$.
The corresponding $r$-extremogram follows from Equation~\eqref{eq:chivario} and is denoted by
$\chi_r(\h,\tau;\Theta,\V)$.

\subsection{Stochastic precipitation generator}
\label{sec:swg_model}

The stochastic precipitation generator is constructed by combining the marginal
model for rainfall amounts with the dependence model for extremes based on
the $r$-Pareto process.

The $r$-Pareto process $\Y$ describes the dependence structure of extreme
events on a standardized Pareto scale and does not carry information on
their real marginal scales and magnitude.
To reintroduce the intensity level of extreme episodes, simulations are
rescaled by a high threshold $u$,  leading to the spatio-temporal
process
\[
\Z = \{ Z_{\s,t} = u\,Y_{\s,t},\, (\s,t)\in\Sset\times\Tset \}
\]
with risk return period given by $u$.
By construction, $Z_{\s_0,t_0}$ 
has Pareto tails with \(\mathbb{P}(Z_{\s_0,t_0}>z)=\frac{u}{z}\), for $z>u$.

To combine this dependence model with the rainfall marginal distributions,
the simulated values corresponding to the standardized scale are mapped to the unit
interval using a standardization function $G$ which is equal to the distribution function of the standard Pareto distribution at moderate to large quantiles (\cite{palacios2020generalized}). 
The function $G$ is defined as
\[
G(x)=
\begin{cases}
0, & x<0,\\
p_0, & x=0,\\
p_0+\dfrac{(1-p_0)^2}{4}x, & 0<x<\dfrac{2}{1-p_0},\\
1-\dfrac{1}{x}, & x>\dfrac{2}{1-p_0}.
\end{cases}
\]
This transformation preserves
the tail behavior of the $r$-Pareto model. At smaller quantiles, it ensures a monotonic transition towards the point mass $p_0$ at $0$ corresponding to absence of precipitation. 
Applying this transformation $G$ to the process $\Z$, we obtain
the simulated values on the unit interval,
\[
U_{\s,t}=G(Z_{\s,t}), \qquad \s\in\Sset,\ t\in\Tset.
\]
The simulated rainfall process is obtained by applying the inverse
marginal distribution function of rainfall,
\[
X_{\s,t}=F^{-1}(U_{\s,t}), \qquad \s\in\Sset,\ t\in\Tset,
\]
where
\[
F^{-1}(x)=
\begin{cases}
0, & x\le p_0,\\[1.2ex]
F_{\mathrm{EGPD}}^{-1}\!\Bigl(\dfrac{x-p_0}{1-p_0};\xi,\sigma,\kappa\Bigr),
& x>p_0.
\end{cases}
\]
This approach allows the generation of spatio-temporal rainfall
fields reproducing both the intensity and the dependence structure of
extreme precipitation events.

\section{Statistical estimation of model parameters}
\label{sec:inference}

In this section, we present the estimation procedure for the parameters of the proposed model, 
including the marginal parameters, the episode selection procedure 
and the estimation of the variogram parameters with advection effects using a composite likelihood 
based on joint exceedances.

\subsection{Marginal parameter estimation}
\label{sec:marginal_estimation}

The marginal parameters $(p_0, \xi, \sigma, \kappa)$ are estimated
over all locations to obtain a single set of parameters for the whole spatial domain.
The probability of zero rainfall $p_0$ is estimated by the empirical
proportion of zero values and
the EGPD parameters $(\xi, \sigma, \kappa)$ are estimated by maximum
independence likelihood (\textit{i.e.}, not considering spatiotemporal dependence) using only positive rainfall values.

\subsection{Selection of extreme episodes}
\label{sec:episode_selection}

Let $\delta>0$ denote a fixed temporal duration for extreme episodes.
For each conditioning exceedance occurring at location $\s_0 \in \Sset$ and time $t_0 \in \Tset$,
we define an extreme episode from the process $\X$ starting at time $t_0$ with a duration $\delta$:
\[
\E = \{ X_{\s,t}\;\mid\;  X_{\s_0,t_0} > u, \, \s \in \Sset,\ t \in [t_0,\, t_0+\delta[ \}.
\]
Extreme episodes are selected following a procedure described by \cite{palacios2020generalized}, which ensures 
weak dependence between episodes, \textit{i.e.}, we perform spatiotemporal declustering where each episode represents a cluster. 
First, we identify all possible conditioning exceedances above a high threshold $u$,
\textit{i.e.} all spatio-temporal locations $(\s_0,t_0) \in \Sset \times \Tset$ such that $X_{\s_0,t_0} > u$.
Then, these conditioning exceedances are ordered by time and processed in chronological order.
An episode with conditioning location $(\s_0,t_0) \in \Sset \times \Tset$ is retained if for every 
already selected episode with conditioning location $(\s_0',t_0') \in \Sset \times \Tset$ 
we have
\[
\|\s_0'-\s_0\| \ge d_{\min}
\qquad\text{or}\qquad
|t_0'-t_0| \ge \delta,
\]
where $d_{\min}>0$ is a minimum spatial separation and the temporal separation is given by the episode duration $\delta$.
In other words, all potential episodes falling within the spatio-temporal neighborhood 
of an already selected episode are discarded from further consideration.
The procedure stops when either a fixed maximum number of episodes is reached
or no exceedance above $u$ remains. The resulting set of selected episodes is
denoted by $\Eset$.

\subsection{Composite likelihood based on exceedances}
\label{sec:composite_likelihood}

For each episode $\E \in\Eset$, spatio-temporal neighbors are defined
relatively to the conditioning point $(\s_0,t_0) \in \Sset \times \Tset$.
The corresponding neighborhood, \textit{i.e.} the set of all spatio-temporal locations
at given lags from the conditioning point, is defined as
\[
\mathcal{N}_{\E}(\h,\tau)
=
\left\{ (\s,t)\in\Sset\times\Tset,\,
\|\s-\s_0\|\in C_{\h},\ |t-t_0|\in C_\tau
\right\},
\]
where $C_{\h}$ and $C_\tau$ denote predefined spatial and temporal lag classes.
In practice, if a regular spatio-temporal grid is considered, these classes are defined as
direct spatial and temporal lag sets.
For each location $(\s,t)\in\mathcal{N}_{\E}(\h,\tau)$, we define the indicator
variable of joint exceedances as
\[
k_{\E}(\s,t)
=
\mathbb{1}_{\{X_{\s,t}>u, X_{\s_0,t_0}>u\}} = \mathbb{1}_{\{X_{\s,t}>u\}}
\]
where the conditioning exceedance $X_{\s_0,t_0}>u$ holds by construction.
With the $r$-Pareto model we propose, $k_{\E}(\s,t)$ follows a Bernoulli distribution with
success probability $\chi_r(\h,\tau ; \Theta, \mathbf{V}_{\E})$
with parameters $\Theta =(\beta_{1},\beta_{2},\alpha_{1},\alpha_{2})$ and
episode-specific velocity vectors $\mathbf{V}_{\E}$.

The sum of these exceedances provide the basis for the composite
likelihood used to estimate the variogram parameters.
Under approximate independence between episodes and for large $u$, Bernoulli
likelihood contributions are used to construct the composite
likelihood.
The resulting composite log-likelihood satisfies
\begin{equation}\label{eq:logcll}
\begin{aligned}
  \ell_C(\Theta)\propto
\sum_{\E\in\Eset}
\sum_{(\h,\tau)\in\Lambda_{\Sset}\times\Lambda_{\Tset}}
\sum_{(\s,t)\in\mathcal{N}_{\E}(\h,\tau)}
&\;k_{\E}(\s,t)\log \chi_r(\h,\tau ; \Theta, \mathbf{V}_{\E}) \\
&+\big(1-k_{\E}(\s,t)\big)\log\!\big(1-\chi_r(\h,\tau ; \Theta, \mathbf{V}_{\E})\big).
\end{aligned}
\end{equation}
The final inference step consists in maximizing the composite log-likelihood
$\ell_C(\Theta)$ with respect to the variogram parameters $\Theta$.

\subsection{Advection and parametric transformation}
\label{sec:advection_transform}

Advection velocities may be either known and directly specified in the model, 
or estimated from the data when unavailable. 
In our case, episode-specific empirical velocity vectors are inferred from rainfall barycenter displacements (see Section \ref{sec:application}), 
and their estimation uncertainty is handled with
a parametric transformation acting on their magnitude.
For each episode, an empirical velocity vector
$\mathbf{V}^{\rm emp}$ is estimated from the data.
This empirical velocity vector is not used directly in the model but is mapped to an
effective velocity vector with a parametric transformation
$\mathcal{A}(\cdot)$. 
The transformation acts only on the magnitude of the velocity vector, while preserving its direction.
For any non-zero empirical velocity vector $\mathbf{V}^{\rm emp}$, we consider
\[
\mathcal{A}\!\left(\mathbf{V}^{\rm emp}\right)
=
\eta_1 \, \norm{\mathbf{V}^{\rm emp}}^{\eta_2}
\, \frac{\mathbf{V}^{\rm emp}}{\norm{\mathbf{V}^{\rm emp}}}.
\]

The advection associated with episode $\E \in \Eset$ is given by
$\mathbf{V}_{\E}=\mathcal{A}(\mathbf{V}^{\rm emp}_{\E})$.
The apparent episode-specific advection $\mathbf{V}_{\E}$ is therefore fully
determined by the parameters $\eta_1$ and $\eta_2$. 
The new extended parameter vector becomes $\widetilde{\Theta}=(\beta_1,\beta_2,\alpha_1,
\alpha_2,\eta_1,\eta_2)$ for all episodes and is estimated by maximizing the
composite log-likelihood $\ell_C(\widetilde{\Theta})$ given in Equation~\eqref{eq:logcll} with
the empirical advection $\mathbf{V}^{\rm emp}_{\E}$.

\subsection{Validation on simulated $r$-Pareto processes}
\label{sec:simulation_validation}

The proposed inference methodology is validated using simulated $r$-Pareto processes
with known parameters. The simulation of $r$-Pareto processes is based on the spectral representation of
spatio-temporal Brown-Resnick processes.
Specifically, we adapt the simulation algorithm proposed by \cite{lebermaster2015},
which itself relies on exact simulation techniques for spatial max-stable processes
developed in \cite{dombry_exact_2016}.
The $r$-Pareto processes are simulated on a regular grid of $n$ spatial locations
over a period of $\delta$ time steps.
A threshold $u = 1$ is used to define the $r$-exceedances, and the risk function is
chosen as $r(\X) = X_{\s_0,t_0}$, where $\s_0$ is randomly selected among
the spatial locations and $t_0 = 0$ corresponds to the first time step of each
simulated episode.
Different sets of simulations are generated using known variogram parameters
$\widetilde{\Theta}$.

\begin{figure}[H]
  \centering
  \begin{subfigure}[b]{0.48\textwidth}
    \centering
    \includegraphics[width=0.7\textwidth]{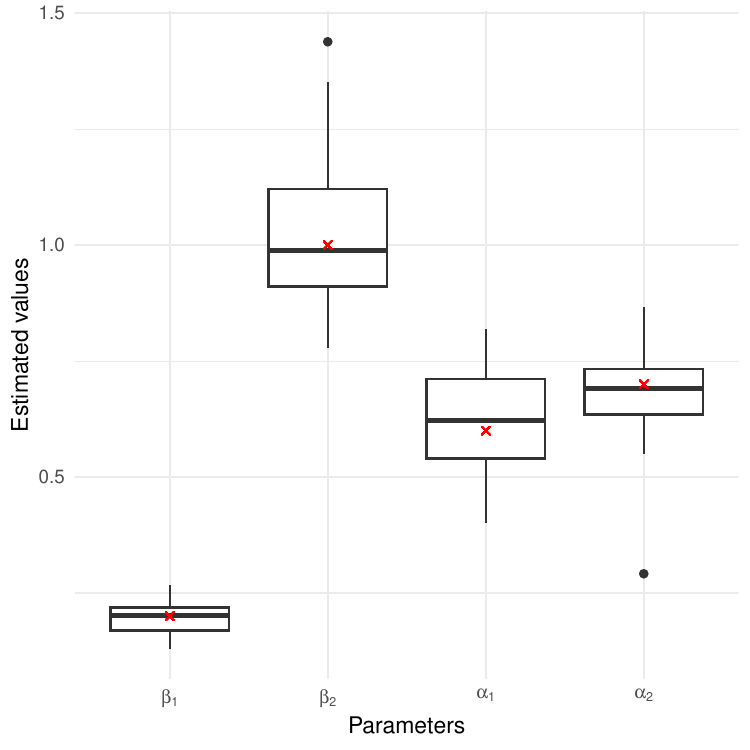}
    \caption{$\beta_1 = 0.2, \beta_2 = 1, \alpha_1 = 0.6, \alpha_2 = 0.7$}
  \end{subfigure}
  \hfill
  \begin{subfigure}[b]{0.48\textwidth}
    \centering
    \includegraphics[width=0.7\textwidth]{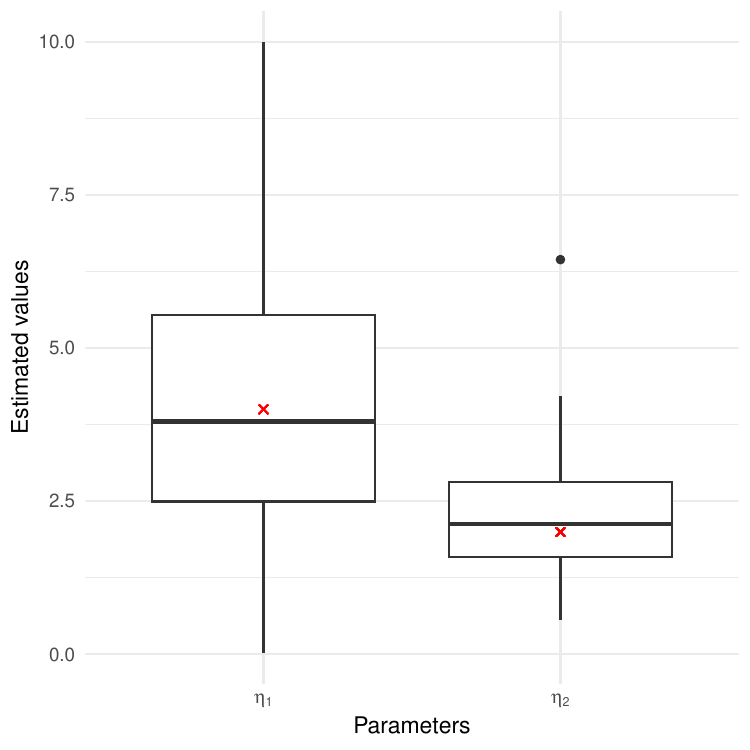}
    \caption{$\eta_1 = 4, \eta_2 = 2$}
  \end{subfigure}
  \caption{Estimations of variogram parameters with maximum likelihood optimization on $r$-Pareto simulations with $49$ sites and $24$ time observations. 
  The true parameters are indicated by red crosses. Here random advection by replicate (episode) is considered.}
  \label{fig:simu_vario_simicom}
\end{figure}

\begin{figure}[H]
  \centering
  \begin{subfigure}[b]{0.48\textwidth}
    \centering
    \includegraphics[width=0.7\textwidth]{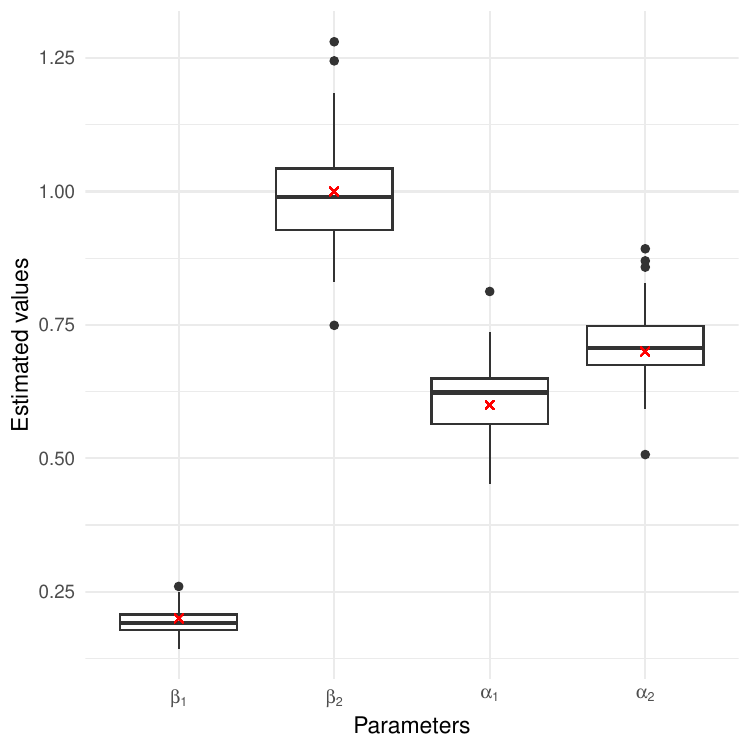}
    \caption{$\beta_1 = 0.2, \beta_2 = 1, \alpha_1 = 0.6, \alpha_2 = 0.7$}
  \end{subfigure}
  \hfill
  \begin{subfigure}[b]{0.48\textwidth}
    \centering
    \includegraphics[width=0.7\textwidth]{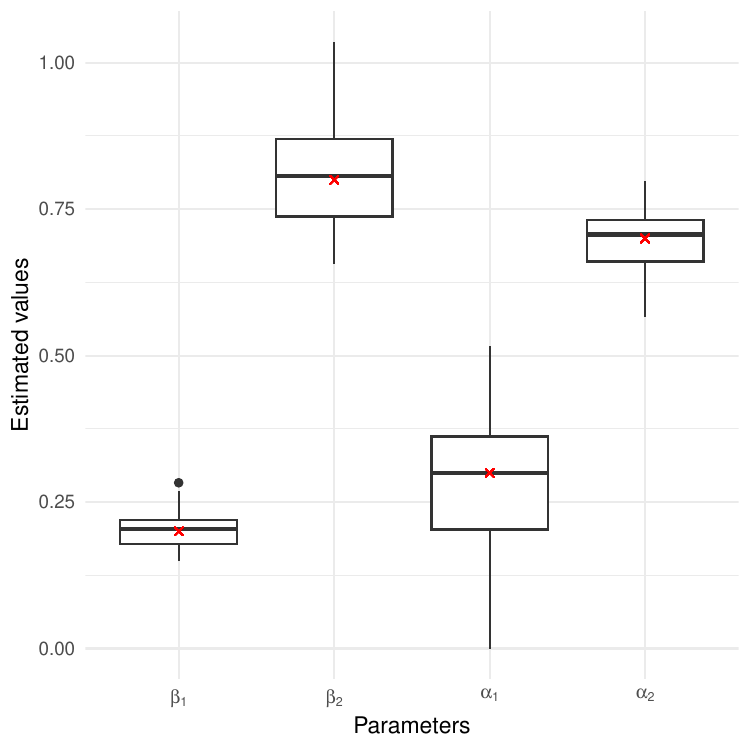}
    \caption{$\beta_1 = 0.2, \beta_2 = 0.8, \alpha_1 = 0.3, \alpha_2 = 0.7$}
  \end{subfigure}
  \caption{Estimations of variogram parameters with maximum likelihood optimization on $r$-Pareto simulations with $49$ sites with $24$ time observations for 
  panel (a) and $12$ time observations for panel (b). 
  The true parameters are indicated by red crosses. We use a random advection by replicate (episode) with fixed $\eta_1=4$ and $\eta_2=2$.}
  \label{fig:simu_vario_simiomsev}
\end{figure}

Three simulation settings are considered to assess the performance of the
proposed inference methodology under different advection configurations. 
For each setting, $50$ independent simulations of $r$-Pareto processes are generated,
each consisting of $500$ replicates on a regular grid of $n=7^2=49$ spatial locations over $\delta = 24$
or $12$ time steps.
\autoref{fig:simu_vario_simicom} illustrates the results obtained when the advection
varies across episodes and the parameters $(\eta_1,\eta_2)$ are jointly estimated
together with the variogram parameters.
Despite the increased model complexity, the variogram parameters
$(\beta_1,\beta_2,\alpha_1,\alpha_2)$ are well recovered, with estimates centered around
their true values and moderate variability.
In contrast, the advection parameters $(\eta_1,\eta_2)$ exhibit higher variability.
This is expected, as they are only indirectly identified through the transformation
of empirical velocities and can compensate each other:
different combinations of $(\eta_1,\eta_2)$ may lead to similar effective advection vectors.
This results in a weaker identifiability and larger dispersion of their estimates,
while still capturing the overall advection structure.
\autoref{fig:simu_vario_simiomsev} illustrates the estimation results obtained
when the advection parameters $(\eta_1,\eta_2)$ are fixed to their true values.
In this setting, the estimation of the variogram parameters remains accurate,
with slightly reduced variability compared to previous case.
Fixing $(\eta_1,\eta_2)$ removes some of the uncertainty associated with the advection estimation, 
leading to 
more stable estimates of the variogram parameters.
In practice, this approach can be advantageous when the advection is particularly difficult to estimate from the data.

\section{Application to rainfall data}
\label{sec:application}

In this section, the proposed spatio-temporal stochastic precipitation generator
is applied to rainfall data from the OMSEV network in Montpellier.
The marginal distributions are estimated and the use of a common distribution is justified.
The dependence structure of extremes is investigated by evaluating the adequacy 
of the non-separable variogram model, 
describing the selection of extreme episodes, and the advection estimation procedure
is explained.
This is followed by the inference of the spatio-temporal extremal dependence structure using 
both COMEPHORE and OMSEV datasets.
Finally, the performance of the stochastic precipitation generator is assessed by comparing the simulated and observed rainfall fields 
at OMSEV sites, with a particular focus on the representation of extreme rainfall events. 
A final stochastic precipitation generator is constructed on a regular grid covering the entire area of interest.

\subsection{Estimation of marginal distributions (OMSEV)}
\label{sec:egpd_omsev}

\begin{figure}[H]
    \centering
    \begin{subfigure}[b]{0.45\textwidth}
        \centering
        \includegraphics[width=0.65\textwidth]{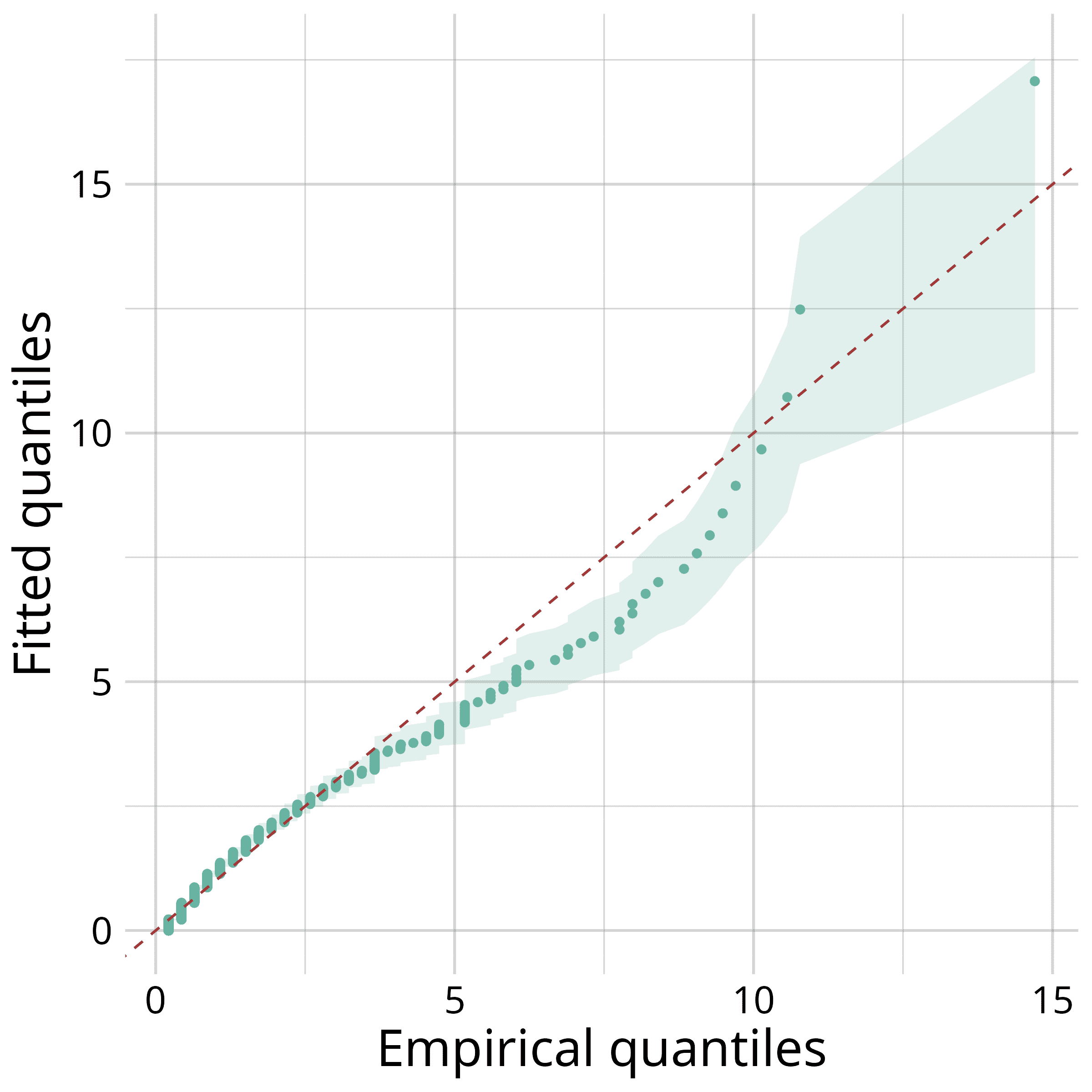}
        \caption{CRBM, left-censoring at $0.2153$ mm}
    \end{subfigure}%
    \hfill
    \begin{subfigure}[b]{0.45\textwidth}
        \centering
        \includegraphics[width=0.65\textwidth]{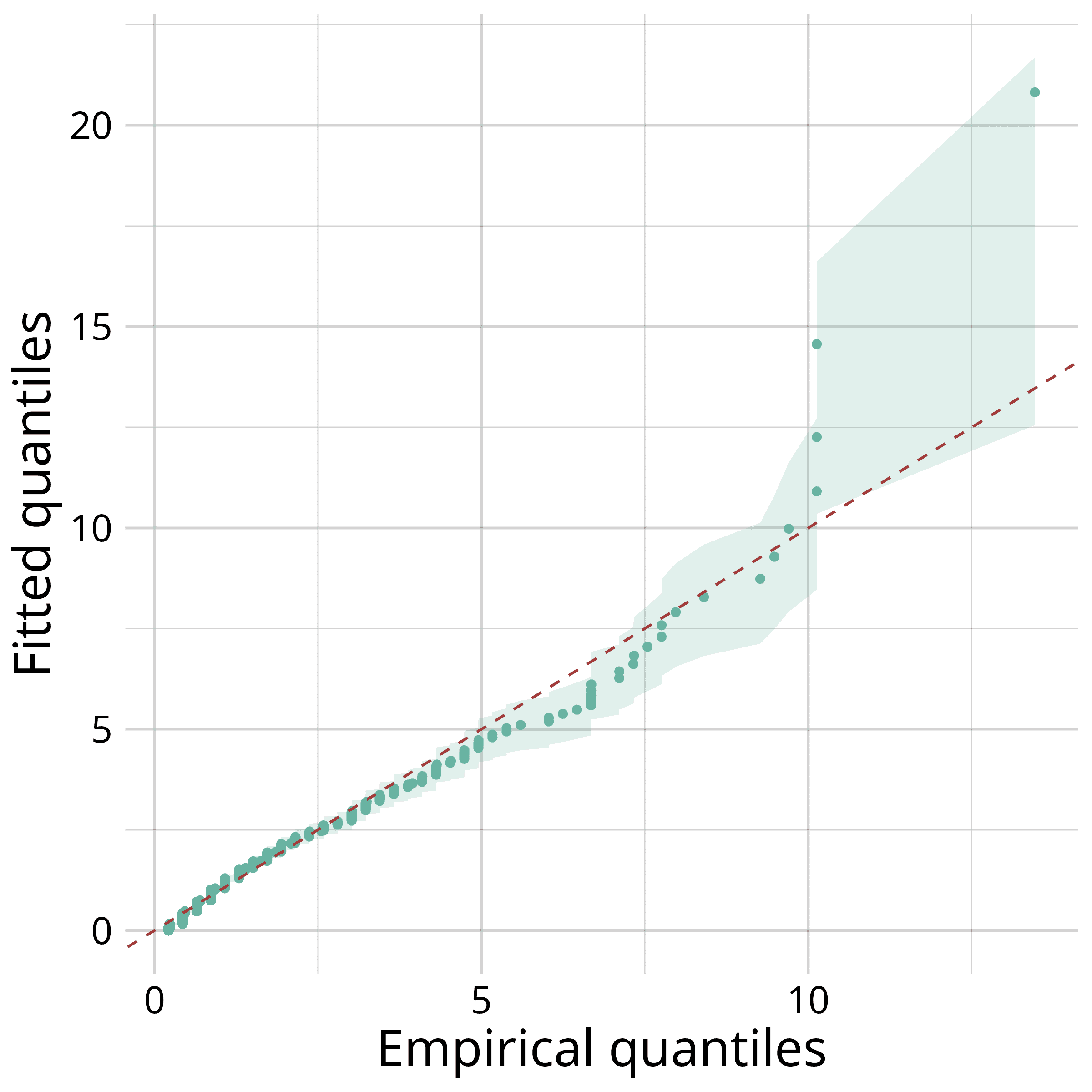}
        \caption{CEFE, left-censoring at $0.4306$ mm}
    \end{subfigure}
    \vfill
    \begin{subfigure}[b]{0.45\textwidth}
        \centering
        \includegraphics[width=0.65\textwidth]{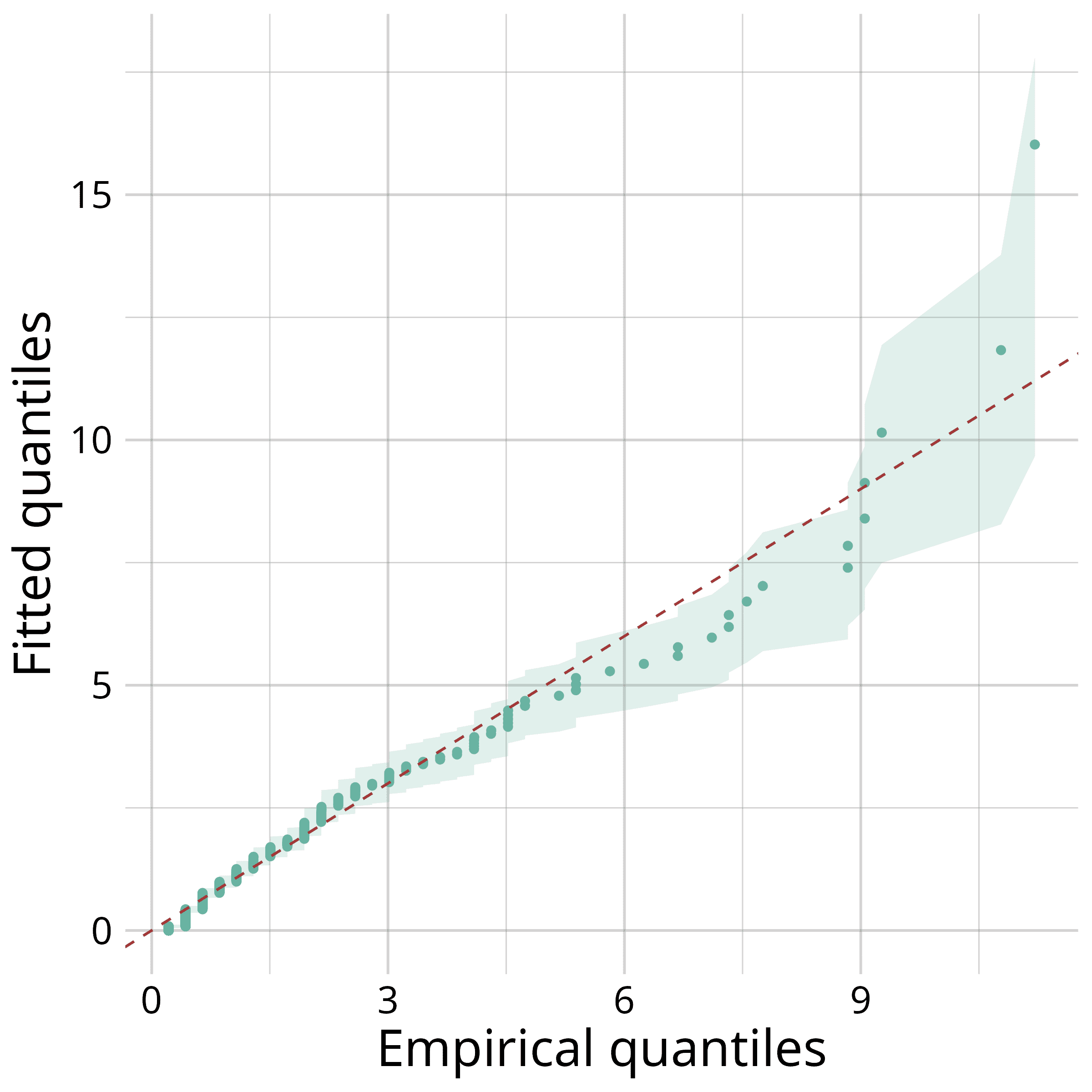}
        \caption{Archie, left-censoring at $0.4306$ mm}
    \end{subfigure}%
    \hfill
    \begin{subfigure}[b]{0.45\textwidth}
        \centering
        \includegraphics[width=0.65\textwidth]{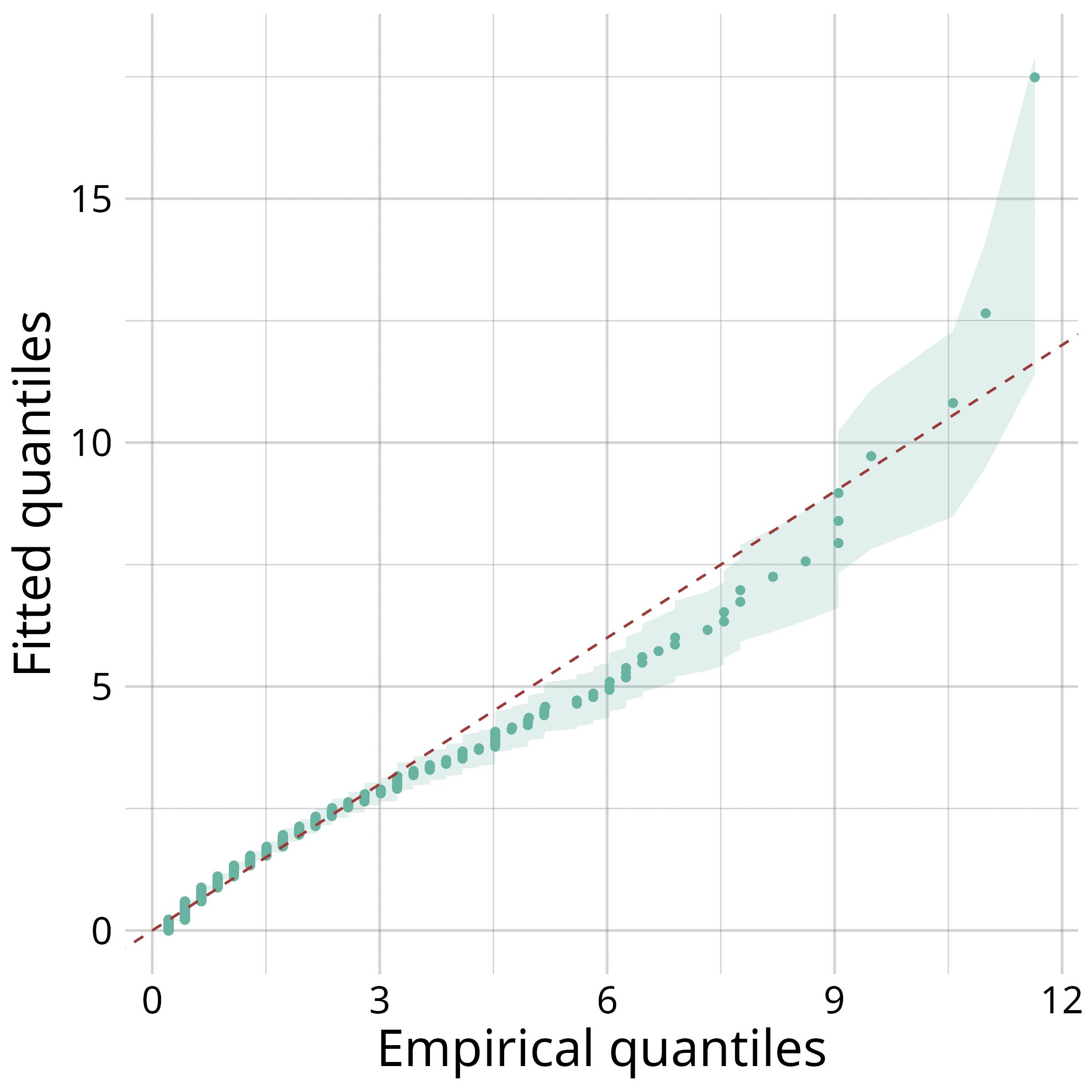}
        \caption{UM, left-censoring at $0.2153$ mm}
    \end{subfigure}
    \caption{Quantile-quantile plot of the EGPD fitting with bootstrap confidence intervals on four rain gauges of the OMSEV network
    with site-specific left-censoring.}\label{fig:egpdfit}
\end{figure}

\begin{figure}[H]
    \centering
    \includegraphics[width=0.45\textwidth]{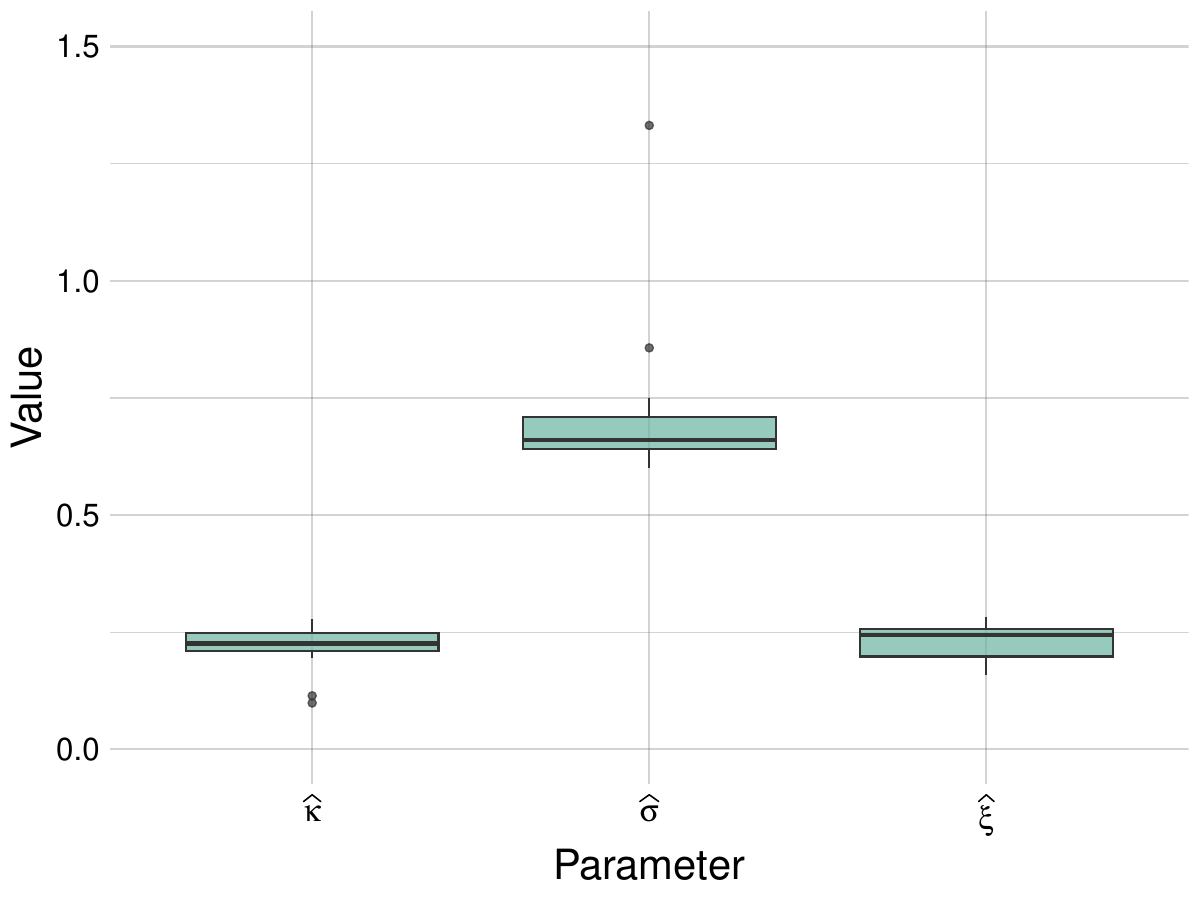}
    \caption{Boxplot of EGPD parameter estimates across the OMSEV network with site-specific 
    left-censoring.} \label{fig:egpdparam}
\end{figure}

The EGPD model is fitted to rainfall data from the OMSEV network in order to assess its ability 
to represent the marginal distribution of positive rainfall intensities at each site.
To account for the discretization induced by the finite measurement precision of the rain gauges,
a small left-censoring is introduced, chosen according to a local goodness-of-fit criterion based on 
the Root Mean Square Error (RMSE), following \cite{haruna2023modeling}.
Possible censoring thresholds are restricted to multiples of the gauge precision,
$p \approx 0.2153$~mm, \textit{i.e.} values of the form $k \times p$ with $k \in \mathbb{N}$.
The final censoring threshold is set to $p$ for most sites, and to $2p$ for three sites (CEFE, Archie, and CNRS).
The quantile-quantile plots displayed in \autoref{fig:egpdfit} illustrate an overall good agreement
between the fitted EGPD and the empirical rainfall distributions for four sites of the OMSEV network.
The same level of agreement is observed at all other sites (see Appendix \ref{sec:appendix_episode_swg_validation} for additional examples).
Relatively good fits are observed on the upper tail,
suggesting that the extreme rainfall behaviour is well captured by the EGPD model.
The distribution of the estimated EGPD parameters across the OMSEV network is summarized in
\autoref{fig:egpdparam}. 
The shape parameter $\xi$ is ranging approximately between
$0.25$ and $0.30$, indicating heavy-tailed rainfall intensity distributions.
Estimated scale parameter $\sigma$ mostly lie between $0.60$ and $0.75$, reflecting moderate
variability in rainfall intensities across all sites. It is consistent with short-duration rainfall data, which typically exhibit 
lower variability compared to longer-duration accumulations.
Finally, the lower-tail parameter $\kappa$ generally fall between $0.20$ and $0.25$, indicating heavier
mass near small rainfall intensities relatively to the central part of the distribution. This is consistent
with the short-duration nature of the data that leads to a larger proportion of small rainfall values 
compared to longer-duration accumulations.
The largest $\kappa$ outlier corresponds to the Archie site.
This site is different from the others because of an interruption of measurements in 2023,
which can explain differences in parameter estimates.
Overall, these results seem to indicate that the EGPD provides a relevant representation
of the marginal rainfall distributions at all sites in the network.

The use of common EGPD parameters across all sites is motivated 
by the absence of strong evidence of spatial heterogeneity in extreme rainfall marginals. 
Global Moran's $I$ test is applied to the estimated EGPD parameters to detect possible
spatial autocorrelation.
For all parameters, the associated $p$
-values are large, indicating that the null
hypothesis of no global spatial autocorrelation cannot be rejected for each parameter 
(see \autoref{tab:moran_results}).
The EGPD is fitted to the full range of positive rainfall values,
while primarily targeting upper-tail behaviour.
For this reason, spatial homogeneity is assessed on exceedances above a high threshold
(the $q=0.90$ quantile) using pairwise Anderson–Darling tests.
For approximately $76\%$ of the 136 site pairs, the null hypothesis of identical upper-tail
distributions cannot be rejected, indicating similar tail behaviour across sites.

Taken together, these results with the similar estimates (\autoref{fig:egpdparam}) 
suggest that the marginal distributions of extreme rainfall 
can be reasonably approximated as identical across the OMSEV network.

\begin{table}[H]
\centering
\caption{Moran's $I$ test for global spatial autocorrelation in EGPD parameters.
The test is performed using the $4$ nearest neighbors for each site.
Reported $p$-values correspond to two-sided tests assessing the null hypothesis of no global spatial autocorrelation.}
\label{tab:moran_results}
\begin{tabular}{lcc}
\toprule
Parameter & Moran's $I$ & $p$-value \\
\midrule
$\xi$    & $-0.211$ & 0.252 \\
$\sigma$ & $-0.002$ & 0.446 \\
$\kappa$ & $-0.098$ & 0.765 \\
\bottomrule
\end{tabular}
\end{table}

The marginal common parameters for the OMSEV network are estimated by fitting an EGPD across all sites together, with left-censoring at the gauge precision 
$p$, and by jointly estimating the probability of zero rainfall. The final parameter estimates are reported in \autoref{tab:egpd_params_omsev}.
These are used as the common marginal parameters in the final stochastic precipitation generator 
for the transformation of simulated values 
from the Pareto scale to the rainfall scale, described in Section \ref{sec:model}.

\begin{table}[H]
\centering
\caption{Estimated common marginal parameters for the OMSEV network.}
\label{tab:egpd_params_omsev}
\begin{tabular}{lccccc}
\toprule
Parameter & $\widehat{p}_0$ & $\widehat{\xi}$ & $\widehat{\sigma}$ & $\widehat{\kappa}$ \\
\midrule
Estimate & $0.989$ & $0.262$ & $0.591$ & $0.270$ \\
\bottomrule
\end{tabular}
\end{table}

\subsection{Empirical evidence of space-time non-separability}
\label{sec:nonseparability_empirical}

\begin{figure}[H]
    \centering
        \includegraphics[width=0.7\linewidth]{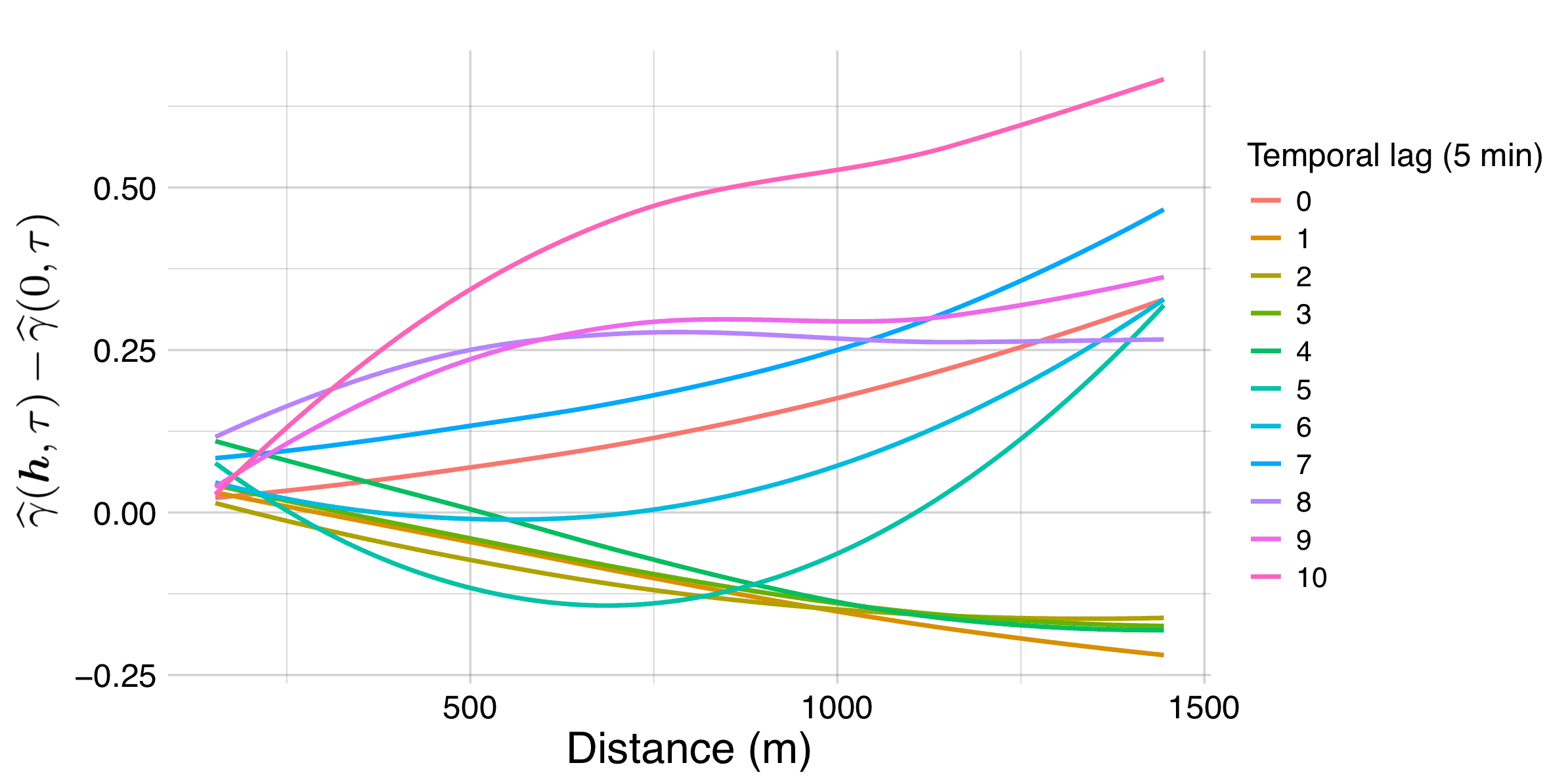}
    \caption{Empirical spatio-temporal extremal variogram differences $\widehat{\gamma}(\h,\tau) - \widehat{\gamma}(0,\tau)$ for the OMSEV data, shown as a function of spatial distance $\hnorm$ for different temporal 
    lags $\tau$, at quantile $q=0.95$.}
     \label{fig:empiricalvariospatemp}
\end{figure}

We investigate the space-time structure of extremal dependence using the empirical
spatio-temporal extremal variogram.
It is estimated from the empirical extremogram obtained by replacing 
probabilities from Equation~\eqref{eq:extremogram} by observed proportions 
of exceedances at a fixed high quantile $q$.
The transformation in Equation~\eqref{eq:chivario} is then applied to obtain the variogram.
To assess additive separability we consider the quantity 
$\gamma(\h,\tau) - \gamma(0,\tau)$.
Under an additive separable structure, described in Equation~\eqref{eq:gammasep}, this difference becomes 
$\gamma_{\Sset}(\h) - \gamma_{\Sset}(0)$, independent of the temporal lag $\tau$.
\autoref{fig:empiricalvariospatemp} shows that the empirical curves do not collapse to a single function of $\h$, 
and exhibit clear changes in shape across temporal lags.
This indicates a strong interaction between spatial and temporal components.
This behaviour provides strong empirical evidence against an additive separable
spatio-temporal variogram for extreme rainfall in the OMSEV data, supporting a non-separable variogram in our model.

\subsection{Episode configuration}
\label{sec:configuration_episodes}

Extreme episodes are identified by selecting conditioning points exceeding 
a large spatio-temporal quantile $q \in [0,1]$, a duration $\delta \in \R_{+}^*$ and a minimum spatial separation 
$d_{\min} \in \R_{+}^*$ between the episodes.
This aims to ensure weak dependence between selected episodes while retaining a sufficient number of episodes for inference.

To select an appropriate quantile threshold $q$ for defining extreme episodes in the OMSEV data,
we examine the number of joint exceedances observed at various spatial distances and temporal lags.
For this purpose, we compute the number of joint exceedances above different quantiles $q$ and 
we keep the quantile $q = 0.95$ as a compromise between the representation of extremes and sample size requirements.
It allows a minimum of $40$ spatial joint
exceedances per spatial lag and a minimum of $20$ temporal joint exceedances per temporal lag (see Section \ref{sec:appendix_episode_config}).

An episode duration of $\delta = 1$ hour (\textit{i.e.}, $12$ time steps of $5$ minutes) is selected for the OMSEV data.
After several tests, the choice of $\delta = 1$ hour 
seems to be a reasonable compromise between temporal interpretability,
approximate independence between episodes, and a sufficient sample size for inference.
This duration is consistent with the temporal resolution of the COMEPHORE 
data used for the advection estimation.
A similar approach is performed to determine the minimum spatial separation $d_{\min}$ 
between episodes. A value of $d_{\min}=1200$ meters 
is chosen to balance independence and sample size for inference (see Appendix \ref{sec:appendix_episode_config}).

Regarding the COMEPHORE data, a similar analysis is employed to determine the episode configuration parameters.
The same quantile threshold $q=0.95$ is adopted to define extreme episodes, ensuring consistency with the OMSEV data analysis.
Given the hourly time resolution of the COMEPHORE data, episodes are defined over $\delta = 24$-hour periods,
starting from a conditioning exceedance, corresponding to a daily episode duration.
The minimum spatial separation between episodes is set to $d_{\min}=5$~km preserving a sufficient number of episodes for inference
and ensuring approximate independence between episodes.

For both datasets, temporal overlaps between episodes are allowed, which retain some redundancy between selected episodes.
Allowing overlaps is motivated by the need to retain a sufficient number of episodes for inference. 
The impact of this redundancy is reduced by the use of composite likelihood and by conditioning on exceedances at sufficiently separated locations, 
so that each episode provides distinct information for inference.
A summary of the episode configuration parameters adopted for both the OMSEV and COMEPHORE datasets is provided in \autoref{tab:episode_config_summary}.
Under such configurations, a total of $333$ extreme episodes are considered in the OMSEV dataset 
and $1846$ in the COMEPHORE dataset.

\begin{table}[H]
\centering
\caption{Summary of the episode configuration parameters adopted for the OMSEV and COMEPHORE datasets and the resulting number of selected episodes.}
\label{tab:episode_config_summary}
\begin{tabular}{lcc}
\toprule
 Dataset & \textbf{OMSEV} & \textbf{COMEPHORE} \\
\midrule
Quantile $q$ & $0.95$ & $0.95$ \\
Episode duration $\delta$ & $60$ minutes & $24$ hours \\
Minimum spatial separation $d_{\min}$ & $1200$ m & $5$ km \\
Number of selected episodes $|\Eset|$ & $333$ & $1846$ \\
\bottomrule
\end{tabular}
\end{table}

\subsection{Estimation of advection}
\label{sec:advection_estimation}

For each extreme rainfall episode from the OMSEV dataset, a velocity vector
is required to characterize the propagation of rainfall storm.
Due to the irregular spatial distribution of rain gauges and the limited
spatial extent of the OMSEV network, such a vector cannot in general be reliably
estimated from OMSEV data alone.
To overcome this limitation, we rely on the COMEPHORE dataset which covers a
much larger spatial domain and provides rainfall estimates at higher spatial and
temporal resolution.
COMEPHORE therefore offers a more comprehensive view of the displacement of
rainfall systems.

\begin{figure}[H]
  \centering
  \begin{subfigure}[b]{0.45\textwidth}
    \centering
    \includegraphics[width=\textwidth]{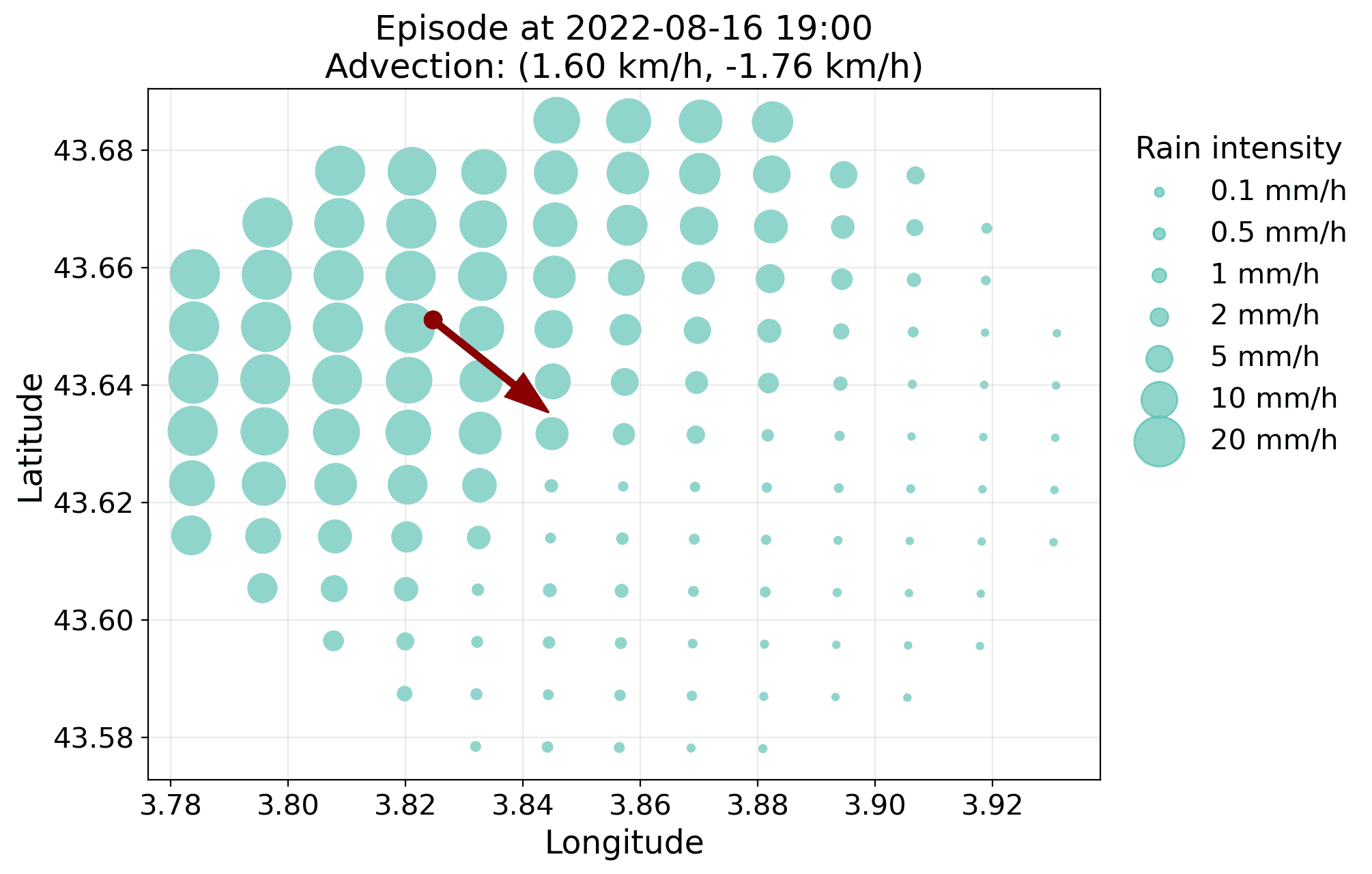}
    \caption{$t_0 - 1$}
  \end{subfigure}
  \hfill
  \begin{subfigure}[b]{0.45\textwidth}
    \centering
    \includegraphics[width=\textwidth]{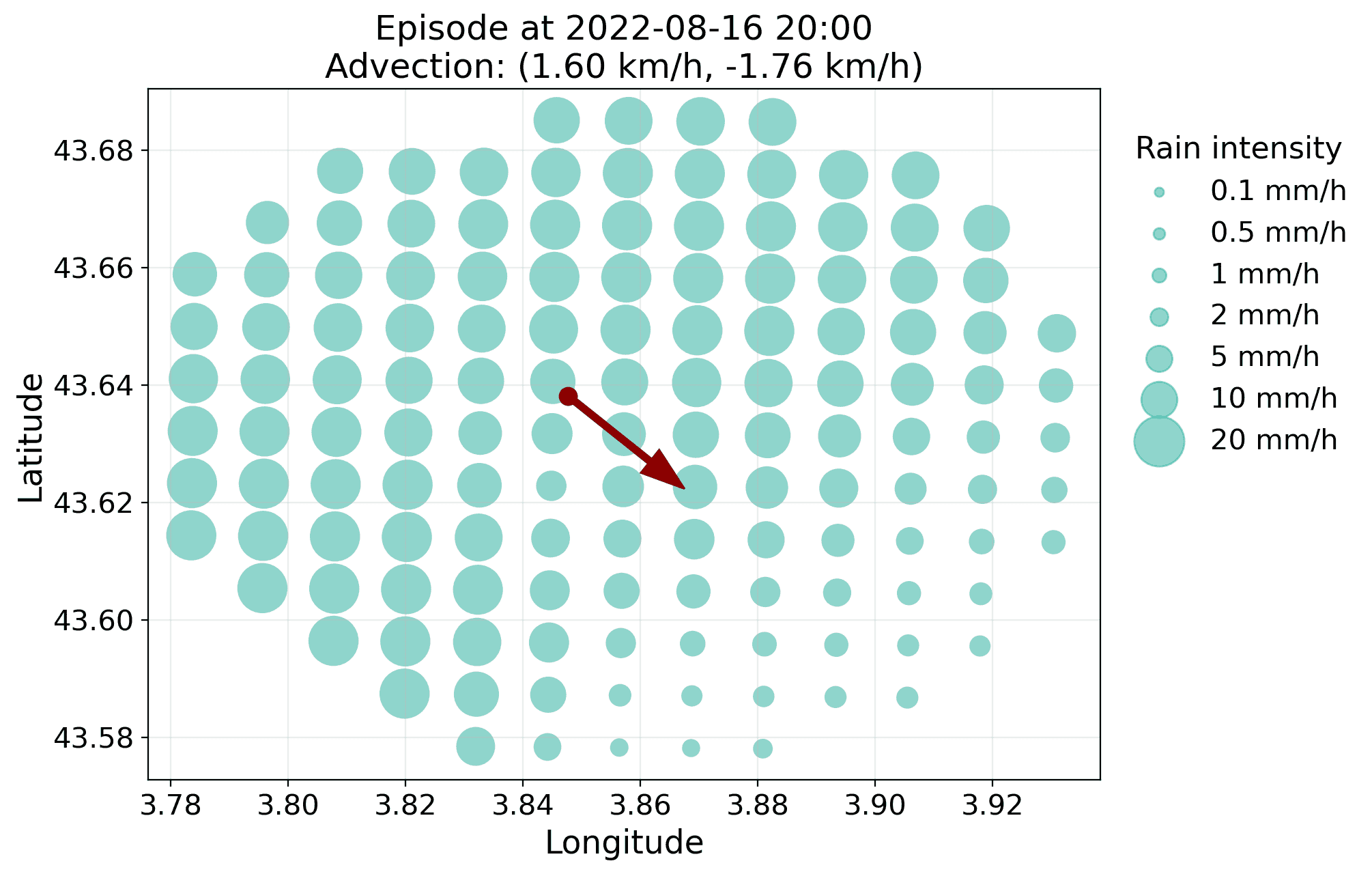}
    \caption{$t_0$}
  \end{subfigure}
  \vfill
  \begin{subfigure}[b]{0.45\textwidth}
    \centering
    \includegraphics[width=\textwidth]{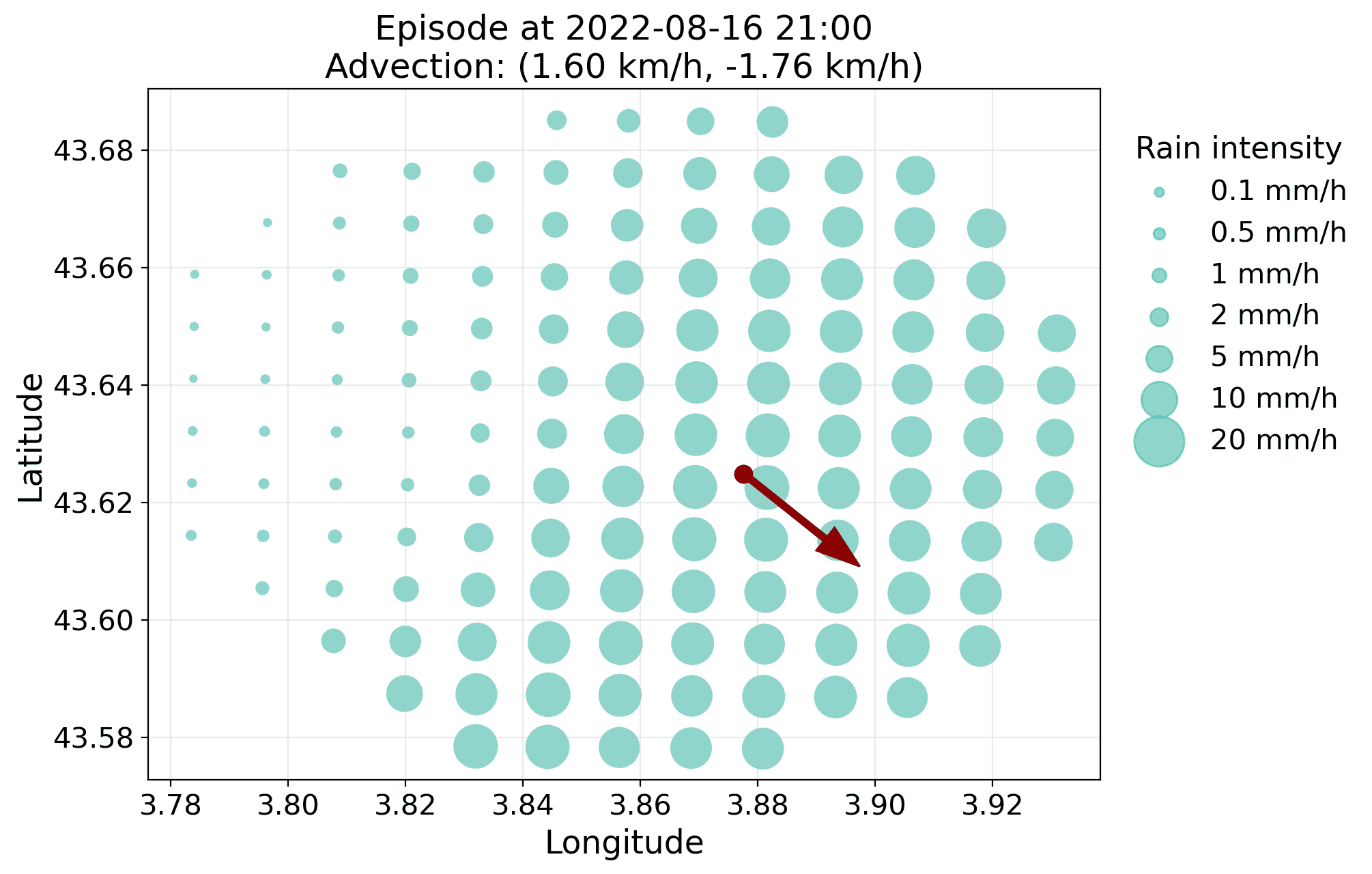}
    \caption{$t_0 + 1$}
  \end{subfigure}
  \caption{Extreme rainfall episode from COMEPHORE on August 16, 2022.
  Rainfall intensity is represented by point size.
The empirical velocity vector (red arrow), estimated over the episode, is shown
at the barycenter position at each time step.
The estimated advection is $\V^{\rm emp}_{\E} = (1.60, -1.76)$ km/h.}
  \label{fig:episode59}
\end{figure}

The duration of OMSEV extreme episodes is one hour,
which matches the time resolution of the COMEPHORE data.
For each OMSEV episode, 
the corresponding hourly COMEPHORE time is
identified and used to estimate the advection term.
The empirical velocity vector $\mathbf{V}^{\rm emp}_{\rm COM}$ is assigned based on COMEPHORE 
episodes occurring within a $\pm$2-hour time window centered on the OMSEV episode hour.
Advection is estimated by tracking the displacement of
rainfall intensity patterns over time.
At each timestamp, the rainfall field is summarized by its barycenter, defined
as the center of mass of the rainfall field, where rainfall intensities act as
weights.
The temporal evolution of these barycenters provides successive displacement
vectors. By averaging these vectors over time, we obtain an empirical advection
vector for the episode.
An illustrative example of this procedure is shown in \autoref{fig:episode59}.

Due to differences in spatial coverage and event detection, some OMSEV episodes
do not correspond to any COMEPHORE episode.
In such cases, the advection is estimated directly from the OMSEV network,
yielding an empirical vector $\mathbf{V}^{\rm emp}_{\rm OMSEV}$.
Although this estimate is much less precise because of the limited spatial
extent of OMSEV network, it still provides partial information on the local displacement
of precipitation and constitutes the only available option.
The final velocity vector $\mathbf{V}^{\rm final}$ is defined as
\begin{equation*}
\mathbf{V}^{\rm final} =
\begin{cases}
\mathcal{A}\!\left(\mathbf{V}^{\rm emp}_{\rm COM}\right) & \text{if a corresponding COMEPHORE episode exists}, \\
\mathcal{A}\!\left(\mathbf{V}^{\rm emp}_{\rm OMSEV}\right) & \text{otherwise},
\end{cases}
\end{equation*}
where the function $\mathcal{A}(\cdot)$ denotes the transformation described in Section \ref{sec:advection_transform}.

\subsection{Variogram estimation}
\label{sec:variogram_estimation}

The OMSEV variogram is estimated using a strategy based on the COMEPHORE data. 
Variogram parameters, including the advection transformation parameters $\eta_1$ and $\eta_2$, 
are estimated from COMEPHORE episodes using the composite likelihood described in Section \ref{sec:composite_likelihood}. 
Empirical velocity vectors are computed as detailed in Section \ref{sec:advection_estimation}, 
and weighted least-squares estimates are used to initialize the optimization following \cite{buhl2019semiparametric}. 
These estimates are then used both to initialize the OMSEV variogram optimization and to fix the advection parameters.
The resulting COMEPHORE parameter estimates are reported in
\autoref{tab:estcom}. 
We observe that the temporal component $\widehat{\beta}_2$ is larger than the spatial one 
$\widehat{\beta}_1$,
indicating that temporal variability dominates.
The values of
$\widehat{\alpha}_1$ and $\widehat{\alpha}_2$ are both around $0.6$, which 
suggest a high degree of irregularity in the spatial and temporal dependence structure of extremes.
This is consistent with convective extreme rainfall events, which are characterized by strong local variability and short temporal persistence.
The estimated advection transformation parameters 
$\widehat{\eta}_1$ and $\widehat{\eta}_2$ indicate a 
strong non-linear transformation of the empirical velocity vectors. 
By applying this transformation on OMSEV empirical velocity vectors, 
we still obtain a plausible range of transformed advection speeds, with a median of approximately $3$ km/h and a maximum of approximately $149$ km/h.
The COMEPHORE parameter estimates are used as initial values for the OMSEV
optimization, with the advection parameters $\eta_1$ and $\eta_2$ fixed.
The resulting OMSEV parameter estimates, with month-leave-one-out
Jackknife confidence intervals, are presented in \autoref{tab:estomsev_ci}.
As for COMEPHORE, the temporal component has more influence than the spatial one, with $\widehat{\beta}_2$ larger than $\widehat{\beta}_1$.
The value of $\widehat{\alpha}_1$ is around $0.2$, while $\widehat{\alpha}_2$ is around $0.7$, indicating a higher degree of irregularity in the temporal dependence structure compared to the spatial one.

\begin{figure}[H]
  \centering

  \begin{subfigure}[t]{0.48\linewidth}
    \centering
    \includegraphics[width=0.7\linewidth]{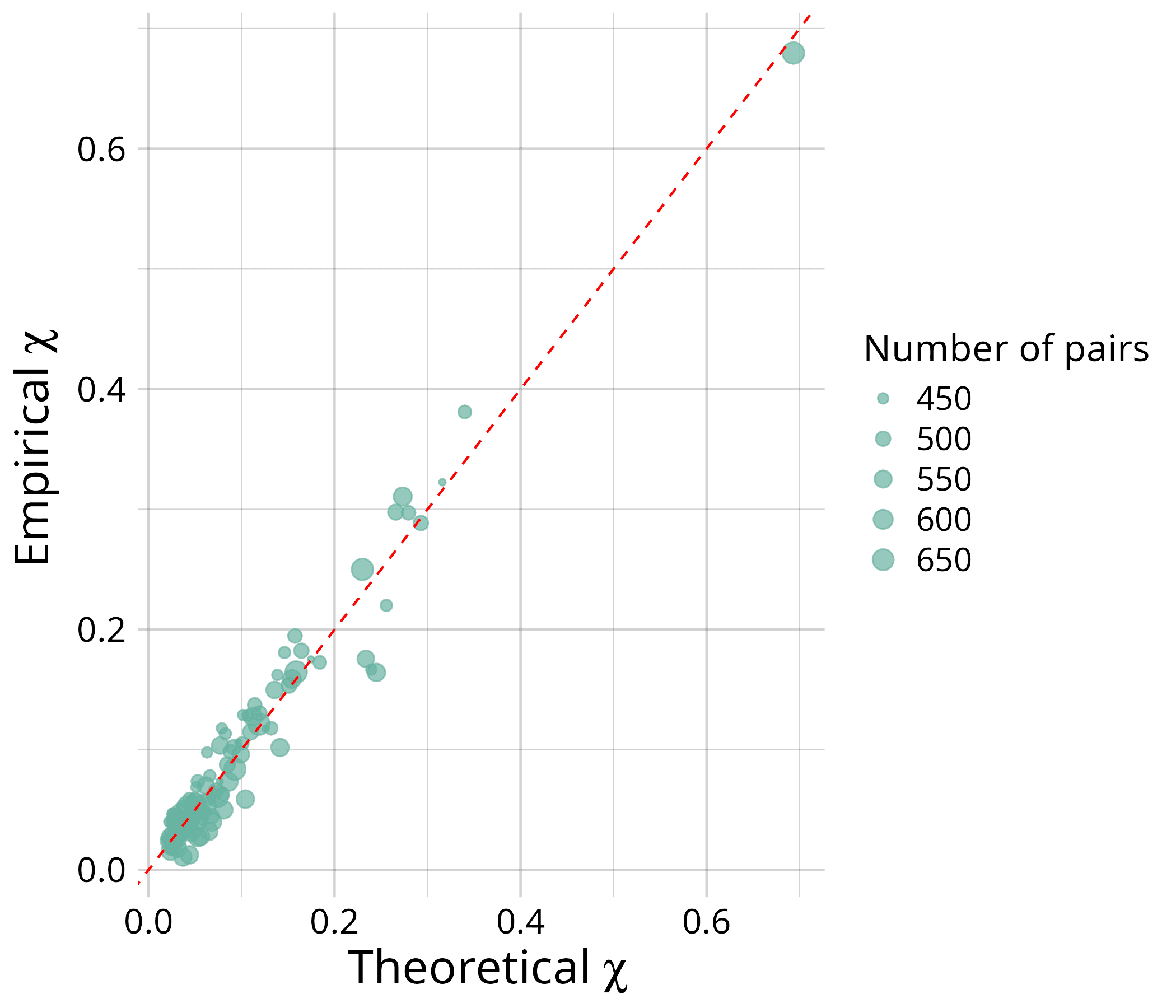}
    \caption{OMSEV data}
    \label{fig:chithempomsev}
  \end{subfigure}
  \hfill
  \begin{subfigure}[t]{0.48\linewidth}
    \centering
    \includegraphics[width=0.7\linewidth]{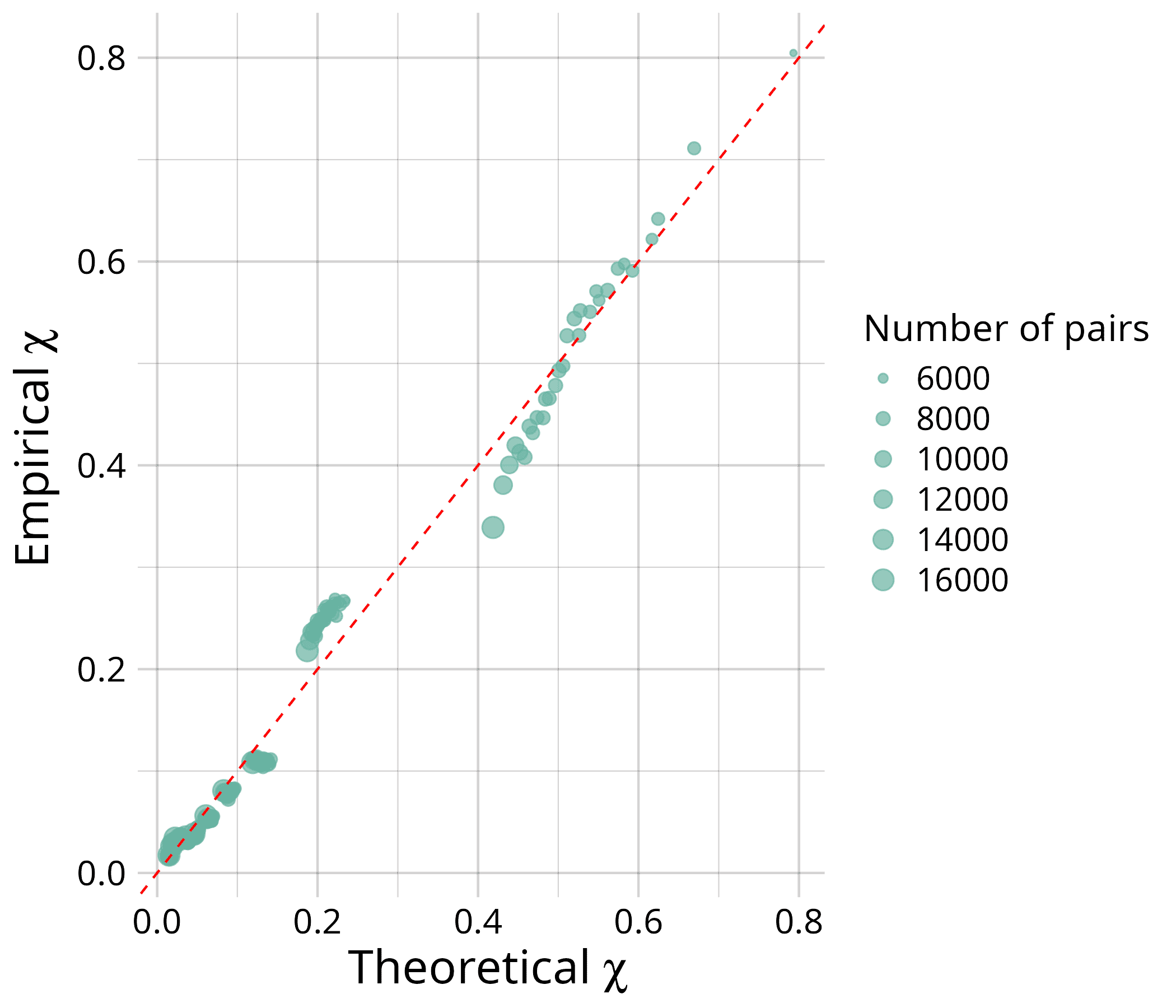}
    \caption{COMEPHORE data}
    \label{fig:chithempcom}
  \end{subfigure}

  \caption{Empirical against theoretical $r$-extremogram. 
  Spatial lags are grouped into distance classes to increase the number of pairs.}
  \label{fig:chithemp_combined}
\end{figure}

The adequacy of the fitted model is assessed by comparing empirical and fitted
$r$-extremograms for both datasets in
\autoref{fig:chithemp_combined}.
In both cases, the fitted $r$-extremograms closely match the empirical
ones across a range of spatial and temporal lags, indicating that
the proposed model effectively captures the extremal dependence structure of
both COMEPHORE and OMSEV data. 

\begin{table}[H]
\centering
\caption{Variogram parameter estimates for COMEPHORE data. Number of episodes: $|\Eset| = 1846$.}
\label{tab:estcom}
\begin{tabular}{c c c c c c c c c c c c}
\toprule
$\widehat{\beta}_1$ & $\widehat{\beta}_2$ &
$\widehat{\alpha}_1$ & $\widehat{\alpha}_2$ &
$\widehat{\eta}_1$ & $\widehat{\eta}_2$\\
\midrule
 0.158 & 0.992 & 0.580 & 0.663 & 3.896 & 2.221 \\
\bottomrule
\end{tabular}
\end{table}

\begin{table}[H]
\centering
\caption{Variogram parameter estimates for OMSEV data with month-leave-one-out
Jackknife confidence intervals.
Parameters $\eta_1$ and $\eta_2$ are fixed to the COMEPHORE estimates.
Number of episodes: $|\Eset| = 333$.}
\label{tab:estomsev_ci}
\begin{tabular}{lcc}
\toprule
 & km/h & m/5\,min\\
\midrule
$\widehat{\beta}_1$ & $1.296 ~[0.945, 1.646]$ & $0.231 ~[0.047, 1.060]$ \\
$\widehat{\beta}_2$ & $4.222 ~[3.141, 5.303]$ & $0.807 ~[0.368, 1.655]$ \\
$\widehat{\alpha}_1$ & $0.250 ~[0.064, 0.436]$ & $0.250 ~[0.064, 0.436]$ \\
$\widehat{\alpha}_2$ & $0.666 ~[0.469, 0.863]$ & $0.666 ~[0.469, 0.863]$ \\
\bottomrule
\end{tabular}
\end{table}

\section{Stochastic rainfall generation}
\label{sec:swg_application}

This section presents the full stochastic extreme space-time rainfall generator obtained by combining
the marginal and dependence models developed in the previous sections. We 
evaluate its performance in reproducing the marginal and dependence structure of extreme rainfall 
episodes observed in the OMSEV dataset.

\subsection{Validation of simulated rainfall episodes at OMSEV sites}

New rainfall simulations are obtained by generating episodes of the $r$-Pareto process fitted to OMSEV data,
conditioned on the occurrence of an extreme event at a given site in the network.
We simulate new episodes for each of the originally observed episodes. 
We condition on the same site, using the same threshold and the same empirical advection velocity vector.
This vector is transformed using the estimated parameters $\widehat{\eta}_1$ and $\widehat{\eta}_2$.
The simulated $r$-Pareto processes are then marginally transformed back to the original rainfall scale,
as described in Section \ref{sec:swg_model}.
For validation purposes, marginal distributions are fitted independently at each site,
allowing for a site-specific assessment of the model fit.

\begin{figure}[H]
  \centering

  \begin{subfigure}[b]{0.49\textwidth}
    \centering
    \includegraphics[width=0.8\textwidth]{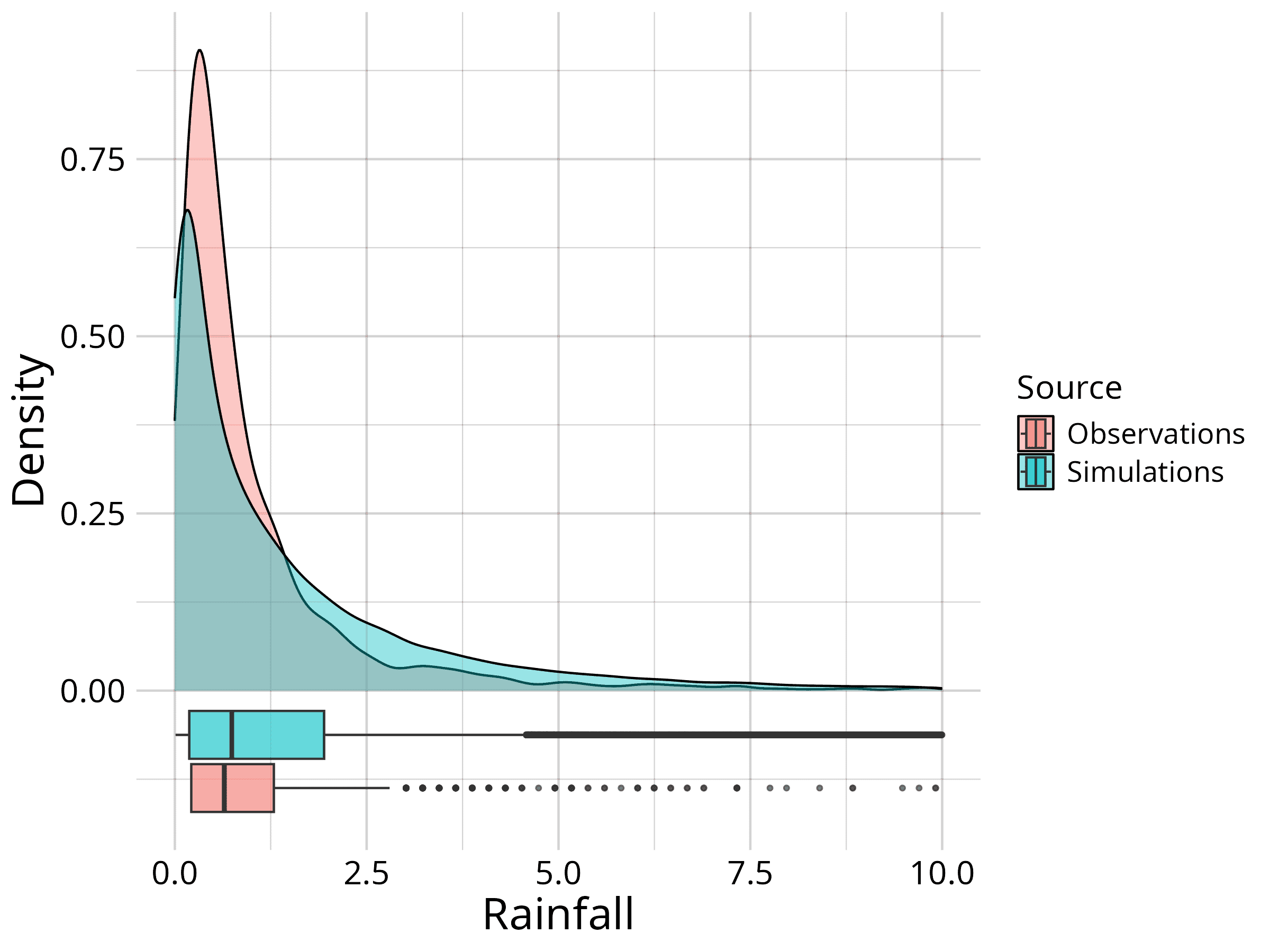}
    \caption{IEM (without correction)}
    \label{fig:iemabove0_nodisc}
  \end{subfigure}
  \hfill
  \begin{subfigure}[b]{0.49\textwidth}
    \centering
    \includegraphics[width=0.8\textwidth]{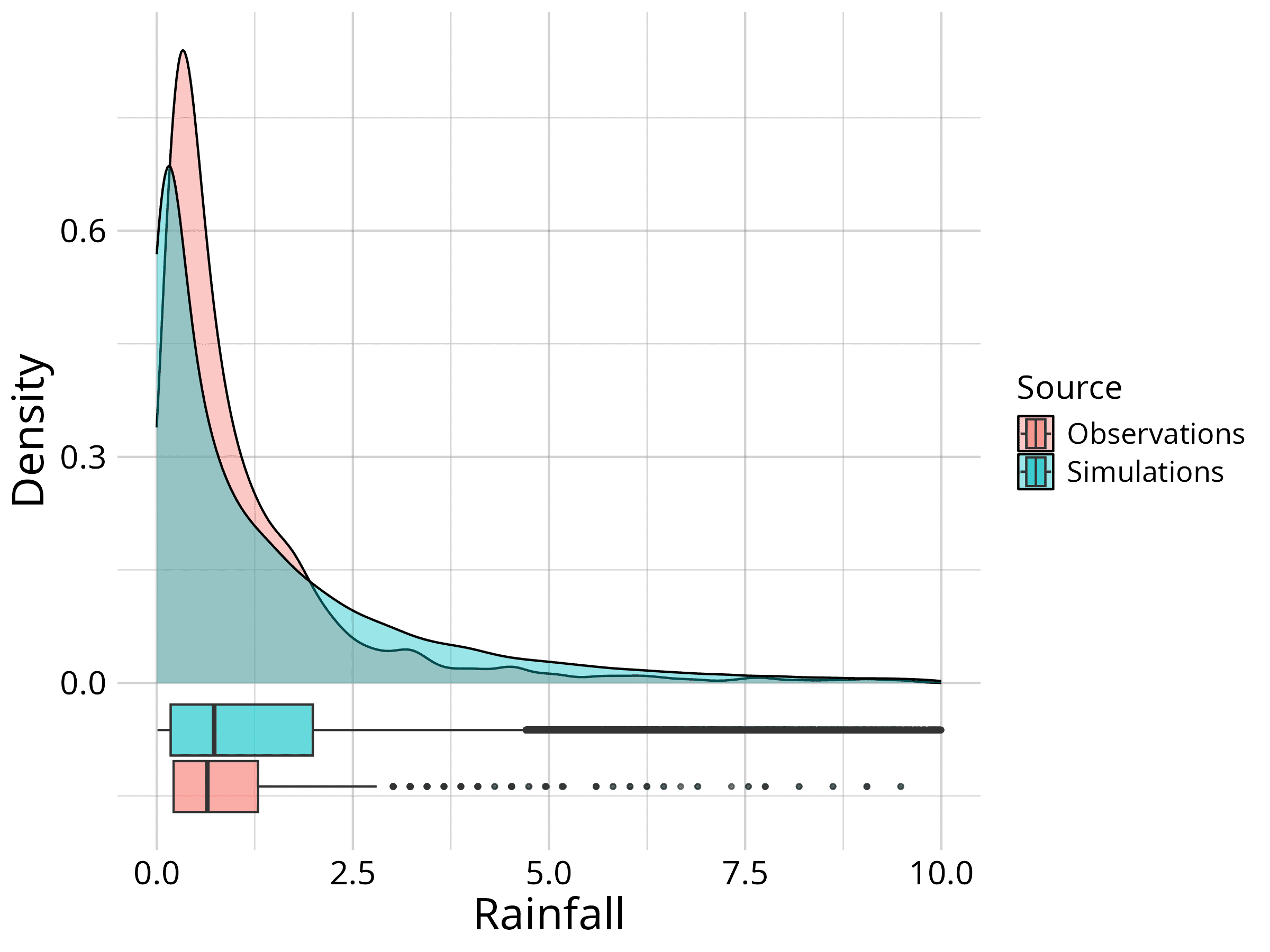}
    \caption{UM (without correction)}
    \label{fig:umabove0_nodisc}
  \end{subfigure}

  \vspace{0.6em}

  \begin{subfigure}[b]{0.49\textwidth}
    \centering
    \includegraphics[width=0.8\textwidth]{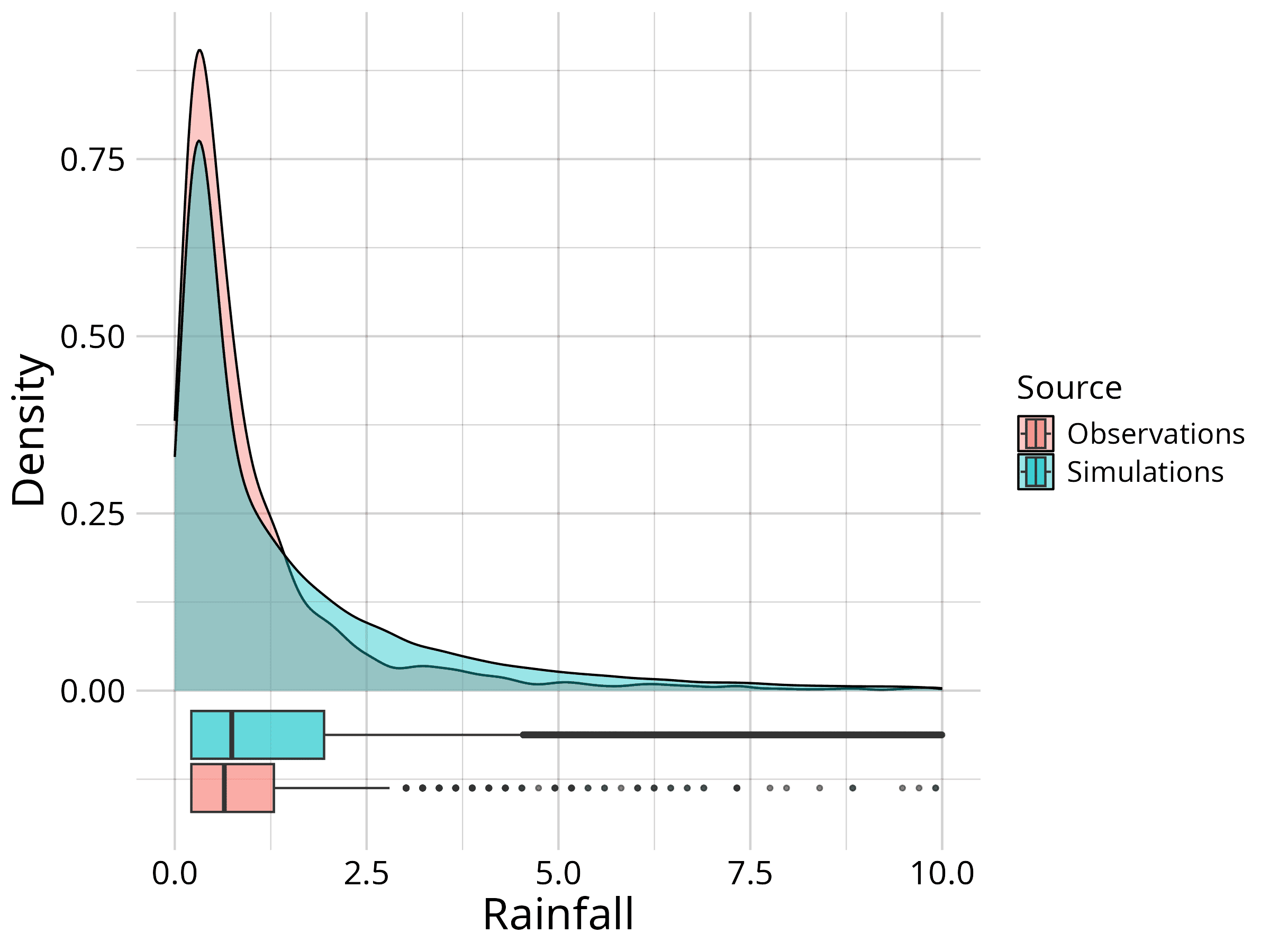}
    \caption{IEM (with correction)}
    \label{fig:iemabove0_disc}
  \end{subfigure}
  \hfill
  \begin{subfigure}[b]{0.49\textwidth}
    \centering
    \includegraphics[width=0.8\textwidth]{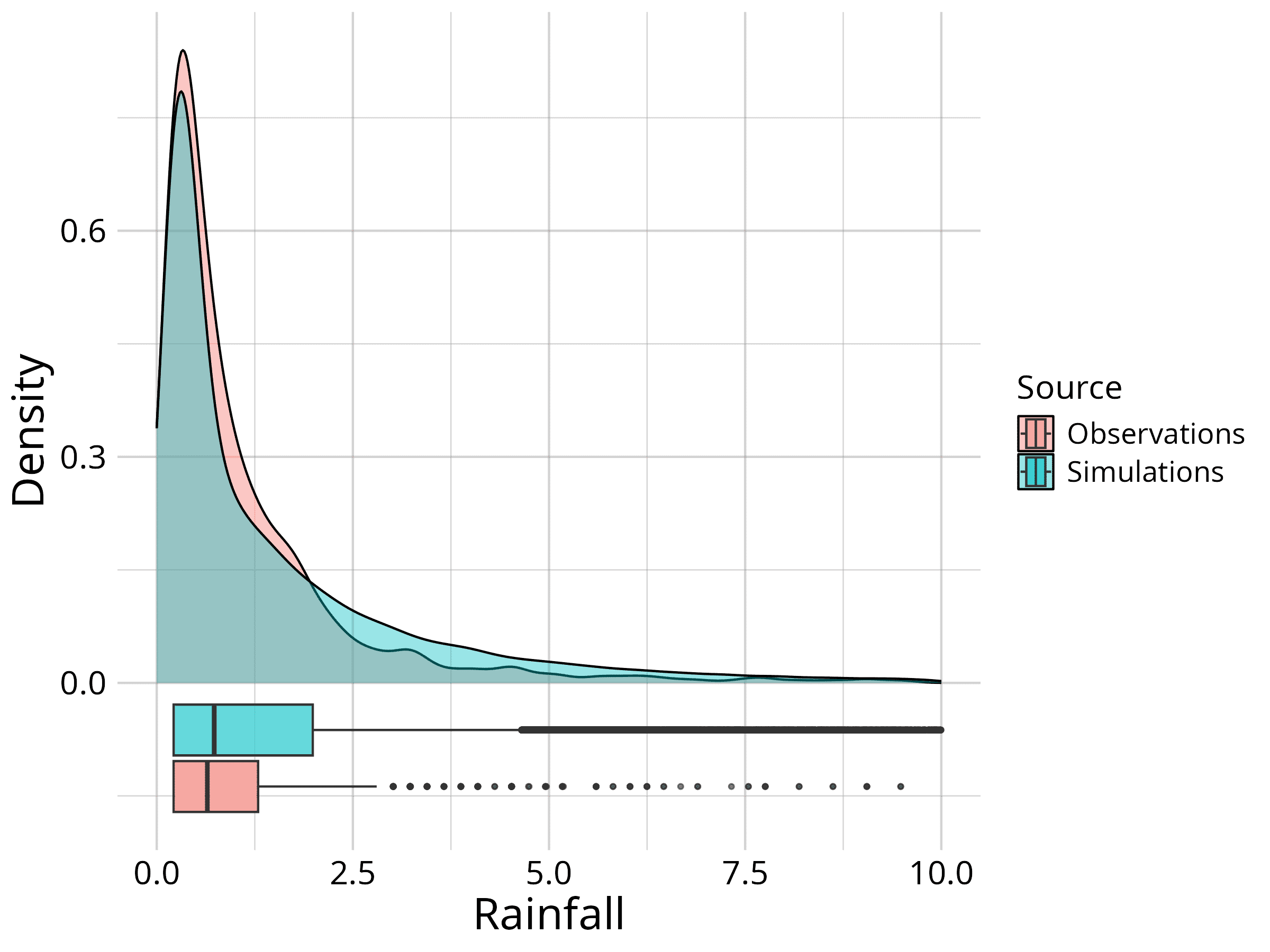}
    \caption{UM (with correction)}
    \label{fig:umabove0_disc}
  \end{subfigure}

  \caption{Density of strictly positive rainfall values for observed against simulated episodes for two selected sites of the OMSEV network.
  Top row: without correction for the discretization effect. Bottom row: with a correction on low simulated values to account for discretization in observations.
  For better visibility, only values below $10$~mm are shown.}
  \label{fig:densityabove0_all}
\end{figure}

To assess both marginal and dependence behavior, we use the $333$ extreme episodes
identified in the OMSEV dataset (see Section \ref{sec:configuration_episodes}).
For each episode, we generate $100$ conditional simulations.
First, we compare the observed and simulated distributions of strictly positive rainfall values at each site.
\autoref{fig:densityabove0_all} shows the marginal densities for two sites.
Original simulations (top row) agree well at high intensities.
However, they underestimate moderate and low values (below the $95$\% quantile).
This difference is due to the discretization of observed rainfall from tipping-bucket gauges.
Simulated values are continuous, making comparisons at low intensities unreliable.
To account for this effect, we apply a discretization correction to simulations for comparison only.
We use the gauge precision $p \approx 0.2153$ as a reference.
Very small values are set to zero.
Values in $(0, p)$ are mapped to $p$, and values in $(p, 2p)$ to $2p$.
This reproduces the discretization induced by the gauges.
As shown in the bottom row of \autoref{fig:densityabove0_all}, this correction improves the agreement at low intensities
while preserving the behavior at high values.
With this correction, the simulated marginal distributions agree well with 
observations across the network (see Appendix \ref{sec:appendix_episode_swg_validation} for 4 additional sites).
Overall, simulations show slightly more high-intensity values.
This is likely due to the absence of an upper bound of the 
estimated distribution and the presence of occasional extreme outliers.

\begin{figure}[H]
    \centering
    \includegraphics[width=0.5\linewidth]{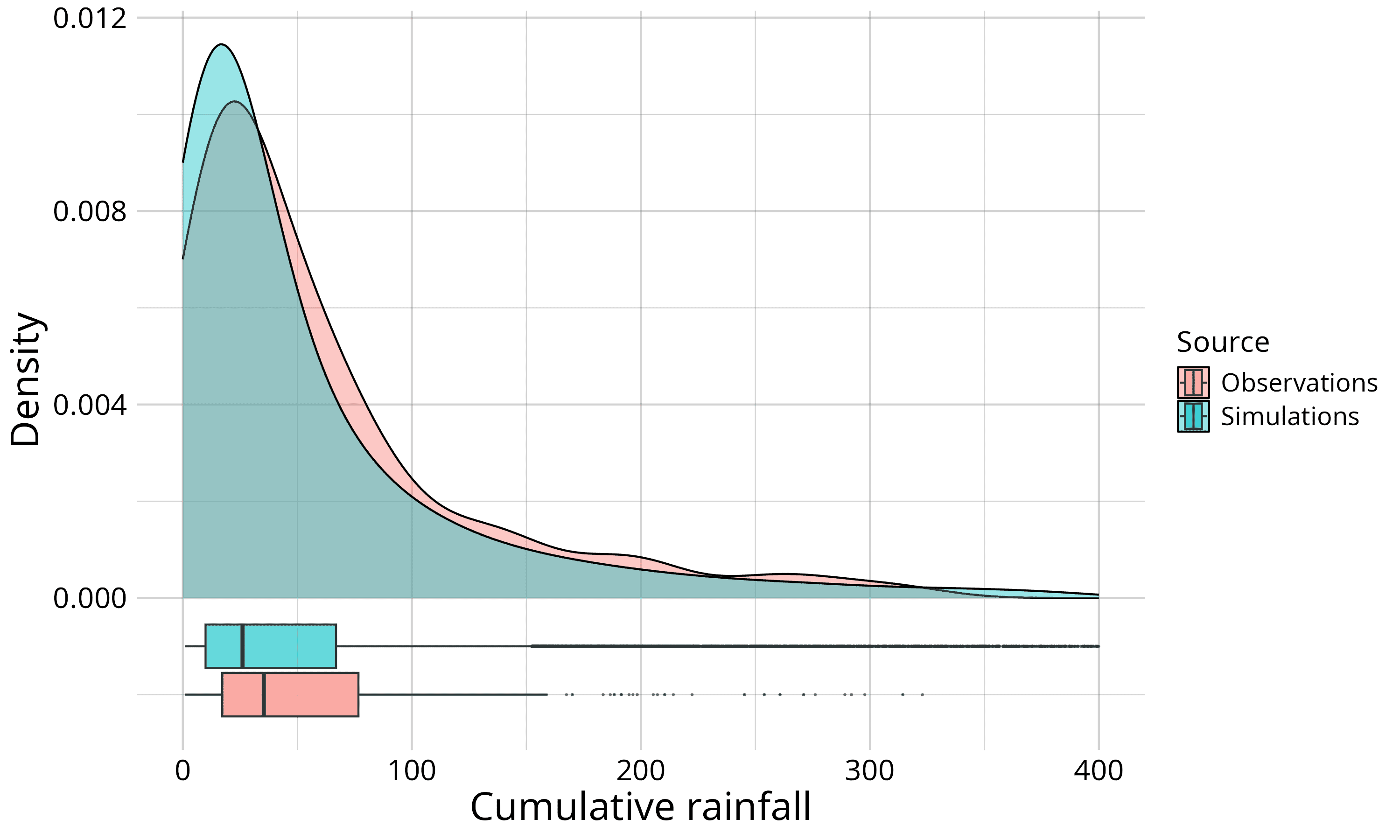}
    \caption{Density of cumulative rainfall over all sites within episodes for observed and simulated data.
    For better visibility, only values below $400$ mm are shown.
    Missing values in the observations are replicated in the simulations for better comparison.}
   \label{fig:denscumulativerain}
\end{figure}

As a complementary summary indicator,
we consider the distribution of cumulative rainfall within episodes.
This quantity indirectly reflects both the number of exceedances occurring across the network during an episode
and their intensities.
Cumulative rainfall is defined as the sum of rainfall intensities over all sites
and time steps within an episode.
It provides a global measure of event intensity.
Observed and simulated distributions are compared in \autoref{fig:denscumulativerain}.
The bulk of the distribution and the upper tail are generally well captured. Some discrepancies arise for low values of which simulations exhibit a larger proportion, leading to a lower median. This is not surprising since the dependence behavior of small values is not directly controlled by the inference method. 
Simulations are conditionally on an exceedance at one site, and some simulated episodes contain only one exceedance while rainfall at other sites remains close to zero. In contrast, observed episodes often include several exceedances across the network.
This leads to larger cumulative rainfall for moderate events relative to the threshold $u$.
As a result, simulations tend to underestimate cumulative rainfall for overall smaller events.

To further assess the dependence structure, we consider conditional exceedance probabilities.
Bivariate probabilities were already used for model estimation and are therefore generally well modeled. 
For validation, we rather focus on trivariate conditional probabilities, defined as
\[
\mathbb{P}(X_{{\s}_1}>u, X_{{\s}_2}>u | X_{\s}>u)
\]
for any pairs of sites $({\s}_1, {\s}_2) \in \mathcal{S}^2$, 
given an exceedance at a conditioning site $\s \in \mathcal{S}$.
The threshold $u$ is set to the empirical $95\%$ quantile.
These probabilities are estimated empirically from both observations and simulations.
We count joint exceedances at the three sites $({\s}_1,{\s}_2,\s)$ and 
divide by the number of exceedances at the conditioning site $\s$.
Results are shown in \autoref{fig:condprobs} for two conditioning sites (additional results are available in Appendix \ref{sec:appendix_episode_swg_validation}).
The main dependence patterns are well reproduced with large conditional probabilities well captured.

Overall, these results show satisfactory model performance in light of the complex structure of space-time precipitation extremes.
The model reproduces both the marginal distribution of rainfall intensities during extreme episodes and the main features of the extremal dependence structure.
The agreement is better for larger probabilities and higher rainfall accumulations.
This is consistent with the main objective of the model, which is to be able to 
stochastically simulate plausible space-time rainfall episodes with possibly high impacts in terms of flood risk.

\begin{figure}[H]
  \begin{subfigure}[t]{1\linewidth}
    \centering
  \includegraphics[width=0.7\linewidth]{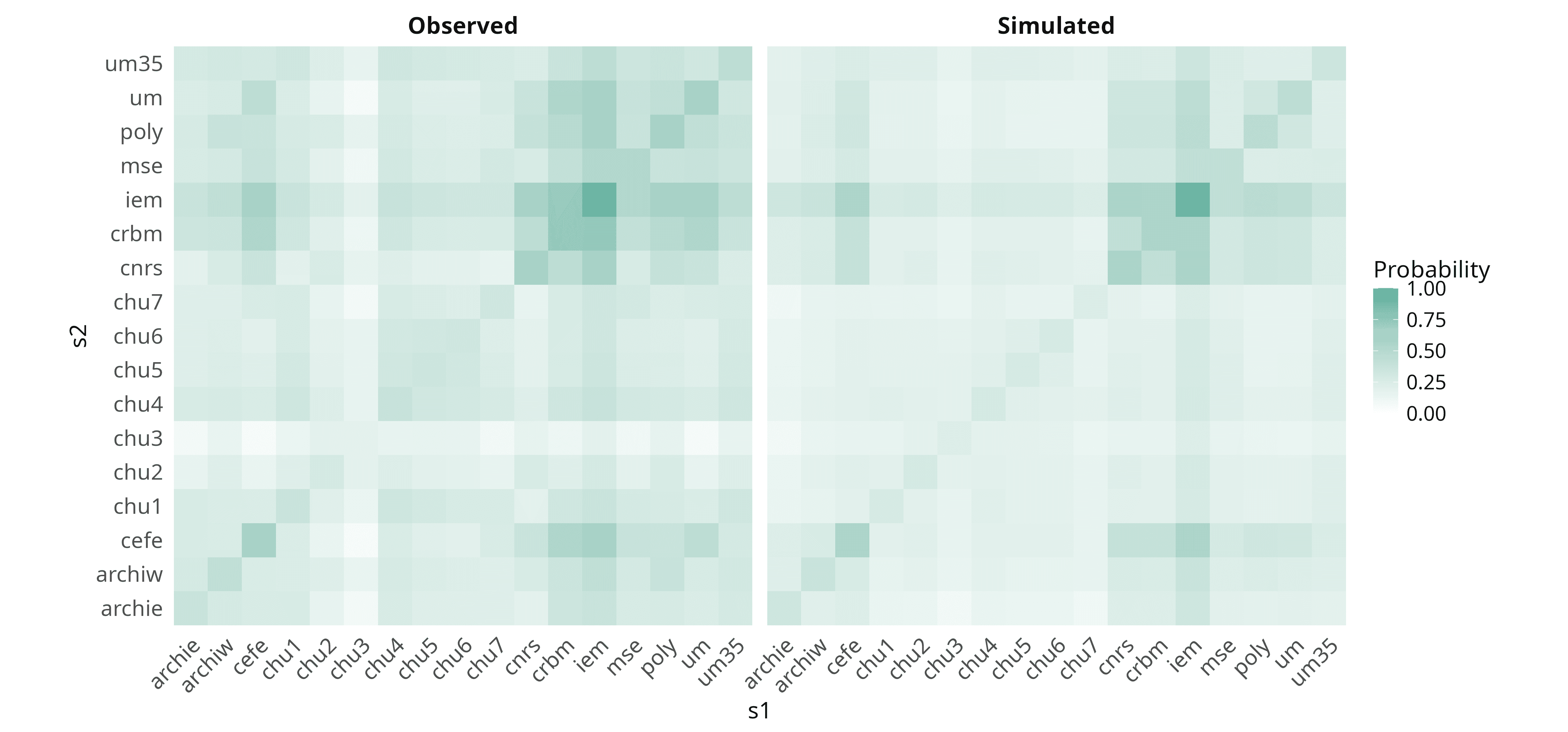}
      \caption{$\mathbb{P}(X_{s_1}>u, X_{s_2}>u | X_{\text{IEM}}>u)$}
    \label{fig:iemcondprobs}
  \end{subfigure}
  \vfill
  \begin{subfigure}[t]{1\linewidth}
    \centering
    \includegraphics[width=0.7\linewidth]{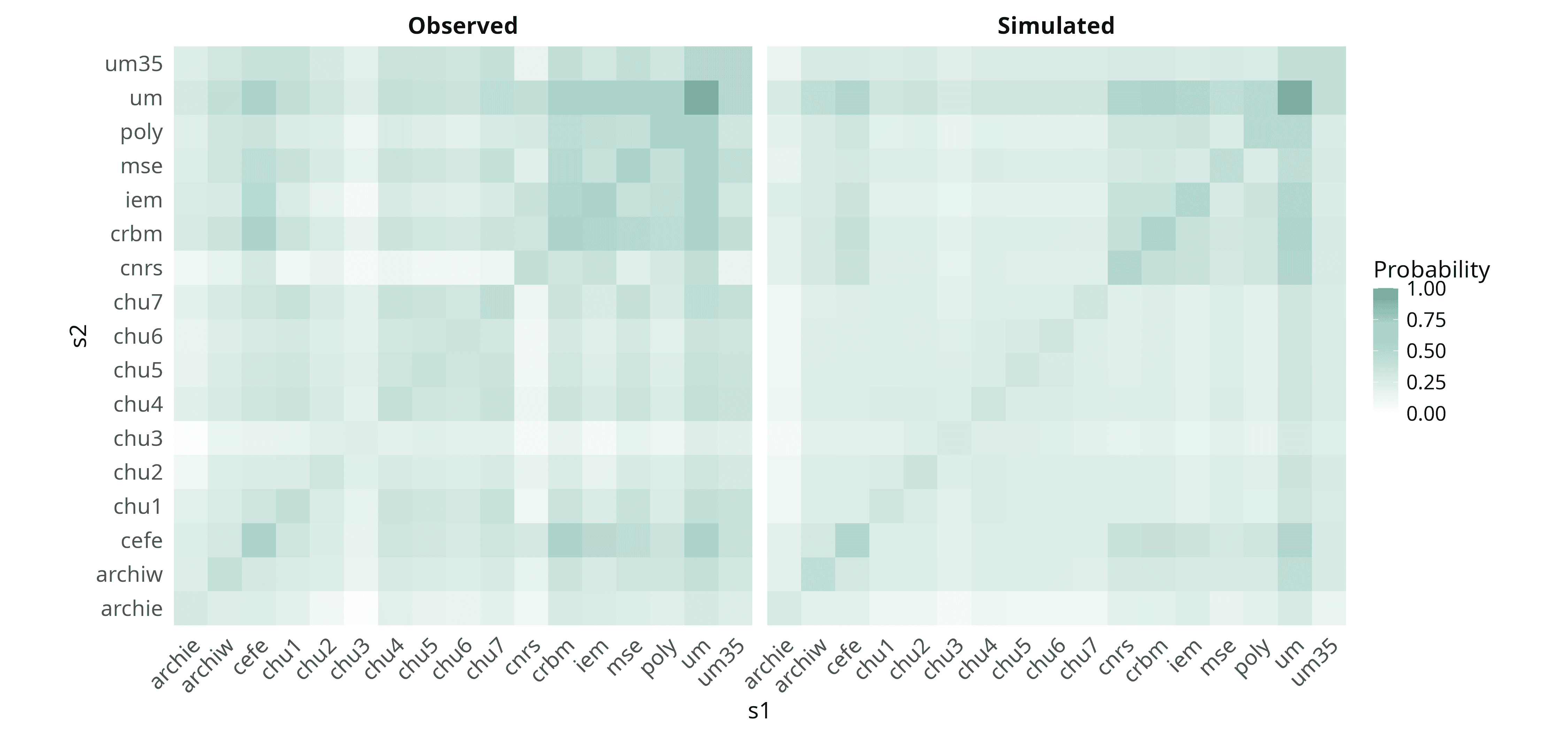}
    \caption{$\mathbb{P}(X_{s_1}>u, X_{s_2}>u | X_{\text{UM}}>u)$}
    \label{fig:umcondprobs}
  \end{subfigure}
   \caption{Conditional trivariate exceedance probabilities according to an excess at one site from observations versus simulations.
        Missing values in the observations are replicated in the simulations for better comparison.}
        \label{fig:condprobs}
\end{figure}

\subsection{New simulations on a regular space-time grid}

\begin{figure}[H]
    \begin{subfigure}[t]{0.48\linewidth}
    \centering
    \includegraphics[width=\linewidth]{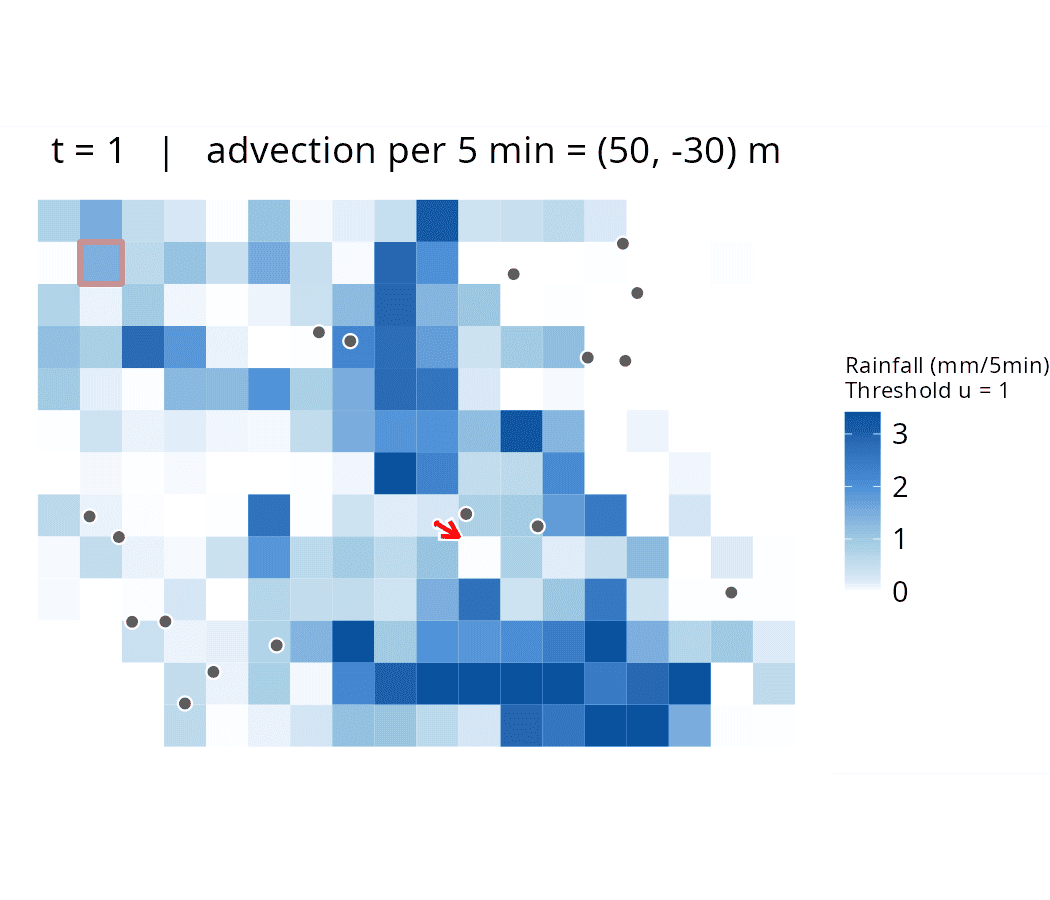}
  \end{subfigure}
  \hfill
  \begin{subfigure}[t]{0.48\linewidth}
    \centering
    \includegraphics[width=\linewidth]{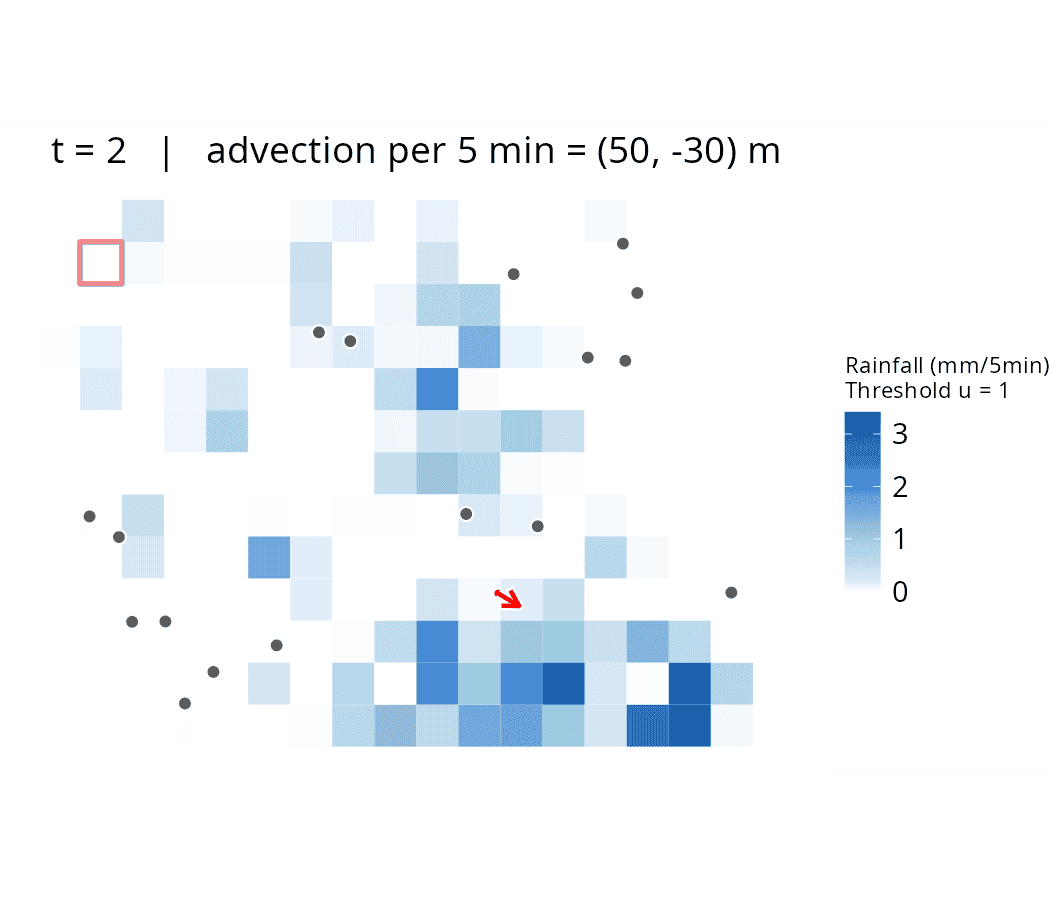}
  \end{subfigure}
  \vfill
  \begin{subfigure}[t]{0.48\linewidth}
    \centering
    \includegraphics[width=\linewidth]{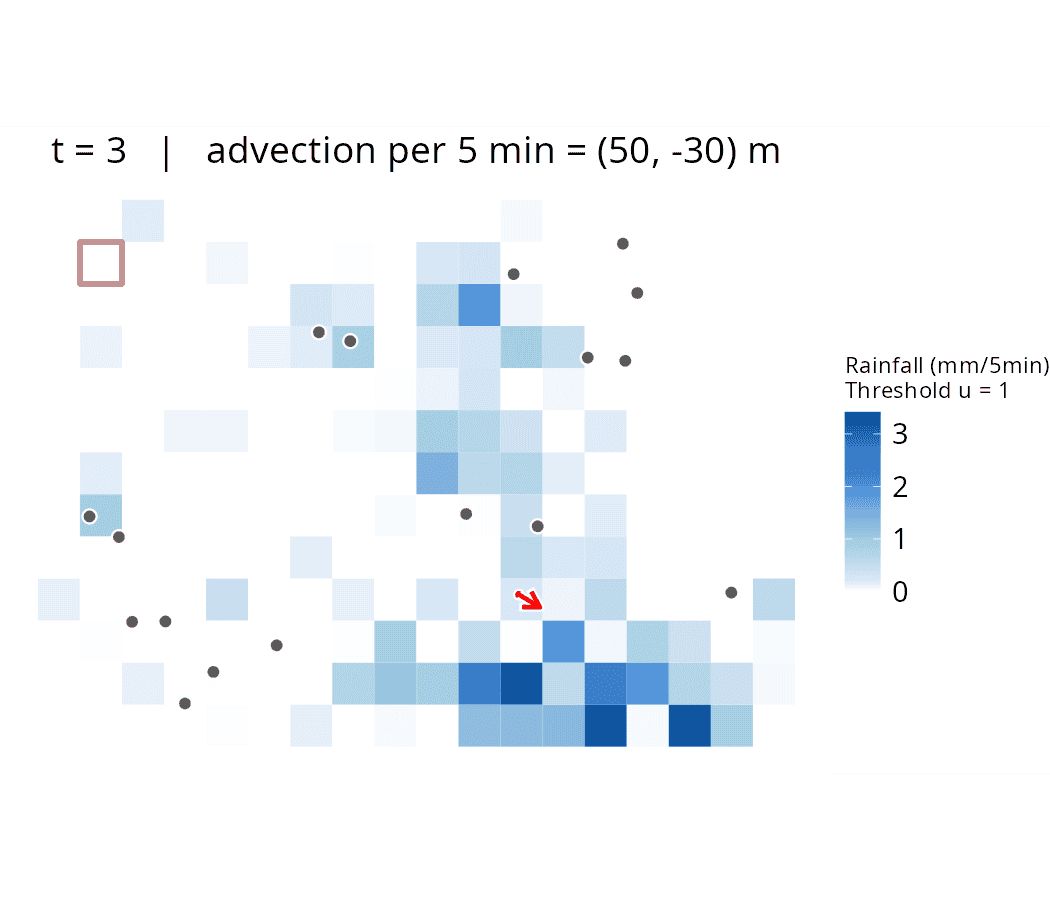}
  \end{subfigure}
  \hfill
  \begin{subfigure}[t]{0.48\linewidth}
    \centering
    \includegraphics[width=\linewidth]{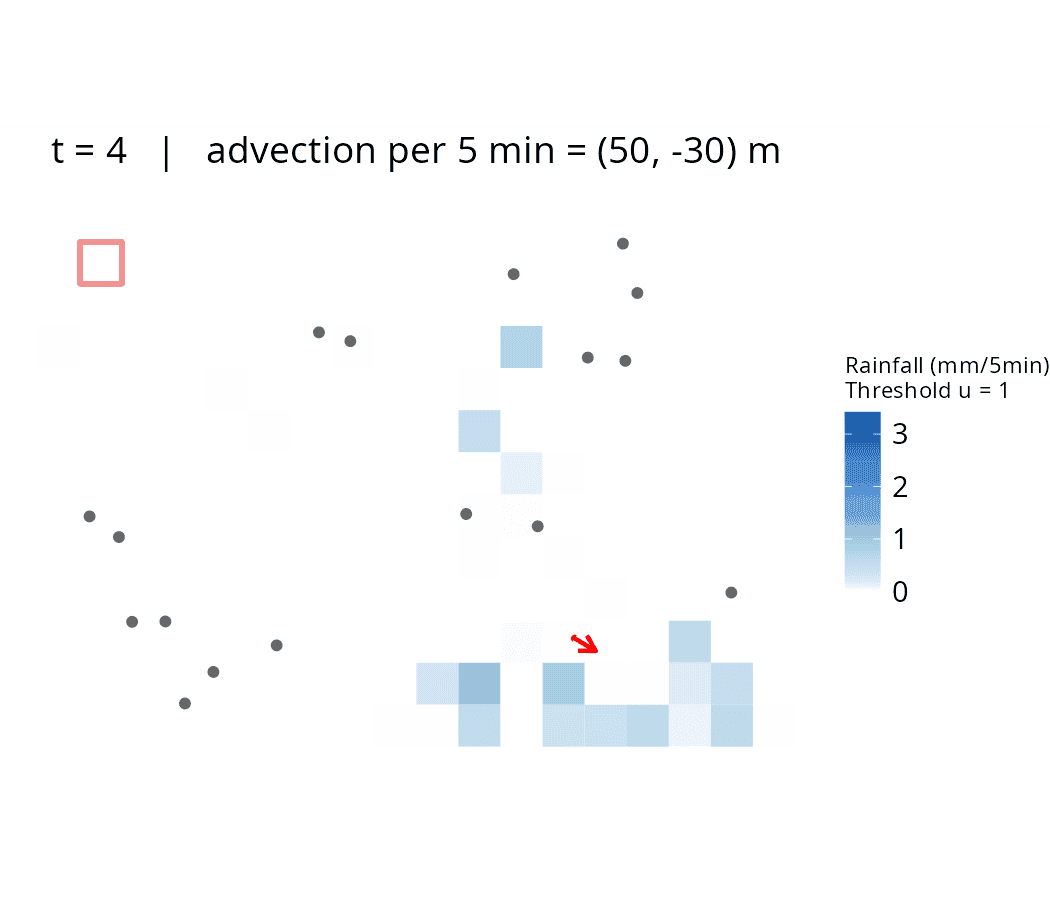}
  \end{subfigure}
 \caption{Simulated rainfall episode on a regular grid over the OMSEV 
  area at the first four time steps. 
  Rain gauge locations are shown as grey dots. The selected conditioning grid pixel is highlighted by a red square 
  and the episode advection is $\V_{\E} = (50, -30)$ meters per 5 minutes (red vector).
  A gif version is available on the github repository at \href{https://github.com/chloesrcb/extreme-rainfall-generator/blob/main/images/simulations/simulated_episode.gif}{chloesrcb/extreme-rainfall-generator/images/simulations/simulated\_episode.gif}.}
  \label{fig:simugridomsev}
\end{figure}

We now use the fitted model to simulate rainfall fields on a regular grid covering the OMSEV domain
at fine spatial resolution.
Simulations are performed on a regular grid with 100 m pixels,
assuming spatially stationary marginal distributions with parameters
given in \autoref{tab:egpd_params_omsev}.
The fitted dependence model described in \autoref{tab:estomsev_ci} is used.
\autoref{fig:simugridomsev} illustrates a simulated extreme rainfall event defined by a threshold $u=1$ mm, 
corresponding to the $95$\% quantile of the observed data.
In this example, the conditional exceedance is imposed at a randomly selected grid pixel,
and a randomly sampled velocity vector is applied.
The resulting rainfall field exhibits a coherent displacement over time,
consistent with the given advection direction.
These simulations demonstrate the ability of the framework to generate continuous
extreme rainfall episode scenarios on a regular grid at fine resolution.

\section{Discussion and perspectives}\label{sec:discussion}

In this work, we developed a spatio-temporal stochastic framework to simulate extreme rainfall episodes
at fine spatial and temporal scales with low computational cost.  
The model combines flexible marginal modeling based on the EGPD
with an advection-informed spatio-temporal dependence structure. For extreme rainfall, dependence usually arises at relatively local scale in space and time whereas one approaches independence for larger spatial or temporal lags. Purely Gaussian processes naturally provide such behaviour  when their correlation function tends to zero with increasing lags, but Gaussian processes are not appropriate for extreme precipitation. Instead, the $r$-Pareto processes we use for peaks-over-threshold modeling require more sophisticated model structures to achieve such behaviour: the Gaussian processes used within the $r$-Pareto construction must have variance that increases to infinity with increasing lags, which can only be achieved by using intrinsic Gaussian processes with unbounded variogram function. 

The proposed model reproduces well both marginal distributions and extremal dependence patterns.
The approach allows simulations on a regular grid at high resolution, which makes it suitable for applications such as risk assessment and hydrological studies.
The framework was developed using the OMSEV dataset but it could be extended to other high-resolution rainfall networks.

A key strength of this work is the joint use of two datasets.
OMSEV provides very fine-scale observations, whereas the COMEPHORE data on  a regular space-time grid available over a longer period bring additional spatial structure and support advection estimation.
Their combination enriches the modeling despite their differences in resolution and coverage.
Some limitations remain at such fine scales; for example, some temporal mismatches between OMSEV and COMEPHORE episodes can arise from postprocessing 
of radar data in the COMEPHORE ones and can degrade the signal in data for estimating advection.

A possible extension would be to incorporate real wind data or advection information into the model.
Such data should be available at high spatial and temporal resolution to match the rainfall data. Since wind patterns can significantly change with altitude in the atmosphere, one would have to identify the relevant wind components from data available in three spatial dimensions (e.g., ERA5 reanalysis data).
The advection vector is episode-specific in our approach, which to our knowledge is a novel feature for extreme precipitation generators. It is assumed constant within each episode in the current model, but we could gain further flexiblity by letting it vary within episodes with appropriate data and model extensions.
Other risk functions could also be used to define $r$-Pareto episodes.
For example, one could consider cumulative rainfall over an area, or the maximum intensity observed during a space-time window.
However, these choices would be more difficult to integrate into the current framework,
especially with the extremogram-based estimation procedure.

Overall, this framework provides a coherent and interpretable approach
for simulating spatio-temporal extreme rainfall fields.

\section*{Code availability}

The code used in this paper is available on the GitHub repository
\href{https://github.com/chloesrcb/extreme-rainfall-generator}{chloesrcb/extreme-rainfall-generator}.


\bibliography{reference}

\begin{thebibliography}{33}
\providecommand{\natexlab}[1]{#1}
\providecommand{\url}[1]{\texttt{#1}}
\expandafter\ifx\csname urlstyle\endcsname\relax
  \providecommand{\doi}[1]{doi: #1}\else
  \providecommand{\doi}{doi: \begingroup \urlstyle{rm}\Url}\fi

\bibitem[Allard et~al.(2015)Allard, Ailliot, Monbet, and Naveau]{allard_stochastic_2015}
D.~Allard, P.~Ailliot, V.~Monbet, and P.~Naveau.
\newblock \emph{{S}tochastic weather generators: {A}n overview of weather type models}, volume 156.
\newblock 2015.

\bibitem[Benoit and Mariethoz(2017)]{benoit2017generating}
L.~Benoit and G.~Mariethoz.
\newblock {G}enerating synthetic rainfall with geostatistical simulations.
\newblock \emph{Wiley Interdisciplinary Reviews: Water}, 4\penalty0 (2):\penalty0 e1199, 2017.

\bibitem[Benoit et~al.(2018)Benoit, Allard, and Mariethoz]{benoit_stochastic_2018}
L.~Benoit, D.~Allard, and G.~Mariethoz.
\newblock {S}tochastic {R}ainfall {M}odeling at {S}ub-kilometer {S}cale.
\newblock \emph{Water Resources Research}, 54\penalty0 (6):\penalty0 4108--4130, 2018.
\newblock ISSN 1944-7973.

\bibitem[Brunet et~al.(2018)Brunet, Bouvier, and Neppel]{brunet_retour_2018}
P.~Brunet, C.~Bouvier, and L.~Neppel.
\newblock {R}etour d'expérience sur les crues des 6 et 7 octobre 2014 à {M}ontpellier-{G}rabels ({H}érault, {F}rance) : caractéristiques hydro-météorologiques et contexte historique de l'épisode.
\newblock \emph{Physio-Géo. Géographie physique et environnement}, 12:\penalty0 43--59, 2018.
\newblock ISSN 1958-573X.

\bibitem[Buhl et~al.(2019)Buhl, Davis, Kl{\"u}ppelberg, and Steinkohl]{buhl2019semiparametric}
S.~Buhl, R.~A. Davis, C.~Kl{\"u}ppelberg, and C.~Steinkohl.
\newblock {S}emiparametric estimation for isotropic max-stable space-time processes.
\newblock \emph{Bernoulli}, 25:\penalty0 2508--2537, 2019.

\bibitem[Carreau and Bengio(2009)]{carreau2009hybrid}
J.~Carreau and Y.~Bengio.
\newblock {A} hybrid {P}areto model for asymmetric fat-tailed data: the univariate case.
\newblock \emph{Extremes}, 12\penalty0 (1), 2009.

\bibitem[Carrer and Gaetan(2022)]{carrer_distributional_2022}
N.~L. Carrer and C.~Gaetan.
\newblock Distributional regression models for {E}xtended {G}eneralized {P}areto distributions.
\newblock \emph{arXiv preprint arXiv:2209.04660}, 2022.

\bibitem[Casas et~al.(2010)Casas, Rodr{\'\i}guez, and Reda{\~n}o]{casas2010analysis}
M.~C. Casas, R.~Rodr{\'\i}guez, and {\'A}.~Reda{\~n}o.
\newblock {A}nalysis of extreme rainfall in {B}arcelona using a microscale rain gauge network.
\newblock \emph{Meteorological Applications: A journal of forecasting, practical applications, training techniques and modelling}, 17\penalty0 (1):\penalty0 117--123, 2010.

\bibitem[Coles et~al.(1999)Coles, Heffernan, and Tawn]{coles1999dependence}
S.~Coles, J.~Heffernan, and J.~Tawn.
\newblock {D}ependence measures for extreme value analyses.
\newblock \emph{Extremes}, 2:\penalty0 339--365, 1999.

\bibitem[Coles et~al.(2001)Coles, Bawa, Trenner, and Dorazio]{coles2001introduction}
S.~Coles, J.~Bawa, L.~Trenner, and P.~Dorazio.
\newblock \emph{{A}n introduction to statistical modeling of extreme values}, volume 208.
\newblock Springer, 2001.

\bibitem[Davison et~al.(2012)Davison, Padoan, and Ribatet]{davison2012statistical}
A.~C. Davison, S.~A. Padoan, and M.~Ribatet.
\newblock {S}tatistical modeling of spatial extremes.
\newblock \emph{Statistical Science}, 27\penalty0 (2):\penalty0 161--186, 2012.

\bibitem[Davison et~al.(2013)Davison, Huser, and Thibaud]{davison2013geostatistics}
A.~C. Davison, R.~Huser, and E.~Thibaud.
\newblock {G}eostatistics of dependent and asymptotically independent extremes.
\newblock \emph{Mathematical Geosciences}, 45\penalty0 (5):\penalty0 511--529, 2013.

\bibitem[De~Fondeville and Davison(2018)]{de_fondeville_high-dimensional_2018}
R.~De~Fondeville and A.~C. Davison.
\newblock {H}igh-dimensional peaks-over-threshold inference.
\newblock \emph{Biometrika}, 105\penalty0 (3):\penalty0 575--592, 2018.
\newblock ISSN 0006-3444.

\bibitem[Dombry et~al.(2016)Dombry, Engelke, and Oesting]{dombry_exact_2016}
C.~Dombry, S.~Engelke, and M.~Oesting.
\newblock {E}xact simulation of max-stable processes.
\newblock \emph{Biometrika}, 103\penalty0 (2):\penalty0 303--317, 2016.
\newblock ISSN 0006-3444, 1464-3510.

\bibitem[Dombry et~al.(2026)Dombry, Legrand, and Opitz]{HandbookExtremesCh16}
C.~Dombry, J.~Legrand, and T.~Opitz.
\newblock {P}areto processes for threshold exceedances in spatial extremes.
\newblock In \emph{Handbook of Statistics of Extremes}, chapter~16, pages 349--376. Chapman \& Hall/CRC, 2026.

\bibitem[Ferreira and De~Haan(2014)]{ferreira2014generalized}
A.~Ferreira and L.~De~Haan.
\newblock {T}he generalized {P}areto process; with a view towards application and simulation.
\newblock \emph{Bernoulli}, 20\penalty0 (4):\penalty0 1717--1737, 2014.

\bibitem[Finaud-Guyot et~al.(2023)Finaud-Guyot, Guinot, Marchand, Neppel, Salles, and Toulemonde]{data2023pluvio}
P.~Finaud-Guyot, V.~Guinot, P.~Marchand, L.~Neppel, C.~Salles, and G.~Toulemonde.
\newblock {R}ainfall data collected by the {HSM} urban observatory ({OMSEV}), 2023.

\bibitem[Fouchier et~al.(2004)Fouchier, Lavabre, Royet, and Felix]{fouchier_inondations_2004}
C.~Fouchier, J.~Lavabre, P.~Royet, and H.~Felix.
\newblock {I}nondations de septembre 2002 dans le {S}ud de la {F}rance - {A}nalyse hydrologique et hydraulique au niveau des barrages écrêteurs du {V}idourle.
\newblock \emph{Sciences Eaux \& Territoires}, \penalty0 (37 Ingénieries-EAT):\penalty0 23–35, 2004.

\bibitem[Haruna et~al.(2023)Haruna, Blanchet, and Favre]{haruna2023modeling}
A.~Haruna, J.~Blanchet, and A.-C. Favre.
\newblock {M}odeling intensity-duration-frequency curves for the whole range of non-zero precipitation: a comparison of models.
\newblock \emph{Water Resources Research}, 59\penalty0 (6):\penalty0 e2022WR033362, 2023.

\bibitem[Hempelmann et~al.(2021)Hempelmann, Adeline, Pagé, and Rodriguez-Dartois]{rainparis2021}
N.~Hempelmann, C.~Adeline, C.~Pagé, and Y.~Rodriguez-Dartois.
\newblock {V}ille de {P}aris : {A}ctualisation du diagnostic de vulnérabilités et de robustesses de {P}aris face aux changements climatiques et à la raréfaction des ressources - {C}ahier 2 : {L}es évolutions climatiques à {P}aris.
\newblock \emph{Ramboll}, 4, 2021.

\bibitem[Huser and Davison(2014)]{huser2014space}
R.~Huser and A.~C. Davison.
\newblock {S}pace--time modelling of extreme events.
\newblock \emph{Journal of the Royal Statistical Society Series B: Statistical Methodology}, 76\penalty0 (2):\penalty0 439--461, 2014.

\bibitem[Leber(2015)]{lebermaster2015}
D.~Leber.
\newblock Comparison of simulation methods of {B}rown--{R}esnick processes.
\newblock Master's thesis, Technische Universität München, Munich, Germany, 2015.

\bibitem[Leblois and Creutin(2013)]{leblois_space-time_2013}
E.~Leblois and J.-D. Creutin.
\newblock {S}pace-time simulation of intermittent rainfall with prescribed advection field: {A}daptation of the turning band method.
\newblock \emph{Water Resources Research}, 49\penalty0 (6):\penalty0 3375--3387, 2013.
\newblock ISSN 1944-7973.

\bibitem[Lucas-Picher et~al.(2021)Lucas-Picher, Arg{\"u}eso, Brisson, Tramblay, Berg, Lemonsu, Kotlarski, and Caillaud]{lucas2021convection}
P.~Lucas-Picher, D.~Arg{\"u}eso, E.~Brisson, Y.~Tramblay, P.~Berg, A.~Lemonsu, S.~Kotlarski, and C.~Caillaud.
\newblock {C}onvection-permitting modeling with regional climate models: {L}atest developments and next steps.
\newblock \emph{Wiley Interdisciplinary Reviews: Climate Change}, 12\penalty0 (6):\penalty0 e731, 2021.

\bibitem[Naveau et~al.(2016)Naveau, Huser, Ribereau, and Hannart]{naveau_modeling_2016}
P.~Naveau, R.~Huser, P.~Ribereau, and A.~Hannart.
\newblock {M}odeling jointly low, moderate, and heavy rainfall intensities without a threshold selection.
\newblock \emph{Water Resources Research}, 2016.

\bibitem[Palacios-Rodr{\'\i}guez et~al.(2020)Palacios-Rodr{\'\i}guez, Toulemonde, Carreau, and Opitz]{palacios2020generalized}
F.~Palacios-Rodr{\'\i}guez, G.~Toulemonde, J.~Carreau, and T.~Opitz.
\newblock {G}eneralized {P}areto processes for simulating space-time extreme events: an application to precipitation reanalyses.
\newblock \emph{Stochastic Environmental Research and Risk Assessment}, 34:\penalty0 2033--2052, 2020.

\bibitem[Papastathopoulos and Tawn(2013)]{papastathopoulos_extended_2013}
I.~Papastathopoulos and J.~A. Tawn.
\newblock {E}xtended generalised {P}areto models for tail estimation.
\newblock \emph{Journal of Statistical Planning and Inference}, 143\penalty0 (1):\penalty0 131--143, 2013.
\newblock ISSN 0378-3758.

\bibitem[Robin(2021)]{robin2021}
R.~Robin.
\newblock Calibration du réseau de pluviographes de l'observatoire hsm ``eau dans la ville'', 2021.
\newblock Projet de fin d'études, Polytech Montpellier, département STE, 63 p.

\bibitem[Scarrott and MacDonald(2012)]{scarrott2012review}
C.~Scarrott and A.~MacDonald.
\newblock {A} review of extreme value threshold estimation and uncertainty quantification.
\newblock \emph{REVSTAT-Statistical journal}, 10\penalty0 (1):\penalty0 33--60, 2012.

\bibitem[Schleiss et~al.(2014)Schleiss, Chamoun, and Berne]{schleiss_stochastic_2014}
M.~Schleiss, S.~Chamoun, and A.~Berne.
\newblock {S}tochastic simulation of intermittent rainfall using the concept of “dry drift”.
\newblock \emph{Water Resources Research}, 50\penalty0 (3):\penalty0 2329--2349, 2014.
\newblock ISSN 1944-7973.

\bibitem[Steinstraesser et~al.(2022)Steinstraesser, Delenne, Finaud-Guyot, Guinot, Casapia, and Rousseau]{steinstraesser2022sw2d}
J.~G.~C. Steinstraesser, C.~Delenne, P.~Finaud-Guyot, V.~Guinot, J.~L.~K. Casapia, and A.~Rousseau.
\newblock Sw2d-lemon: a new software for upscaled shallow water modeling.
\newblock In \emph{Advances in Hydroinformatics: Models for Complex and Global Water Issues—Practices and Expectations}, pages 23--40. Springer, 2022.

\bibitem[Tabary et~al.(2012)Tabary, Dupuy, L'Henaff, Gueguen, Moulin, and Laurantin]{tabary201210}
P.~Tabary, P.~Dupuy, G.~L'Henaff, C.~Gueguen, L.~Moulin, and O.~Laurantin.
\newblock {A} 10-year (1997―2006) reanalysis of {Q}uantitative {P}recipitation {E}stimation over {F}rance: methodology and first results.
\newblock \emph{IAHS-AISH publication}, \penalty0 (351):\penalty0 255--260, 2012.

\bibitem[Wilks and Wilby(1999)]{wilks1999weather}
D.~S. Wilks and R.~L. Wilby.
\newblock {T}he weather generation game: a review of stochastic weather models.
\newblock \emph{Progress in physical geography}, 23\penalty0 (3):\penalty0 329--357, 1999.

\end{thebibliography}


\appendix

\section{Additional figures for episode configuration selection}
\label{sec:appendix_episode_config}

\begin{figure}[H]
  \centering
  \begin{subfigure}[b]{0.48\textwidth}
    \centering
    \includegraphics[width=\linewidth]{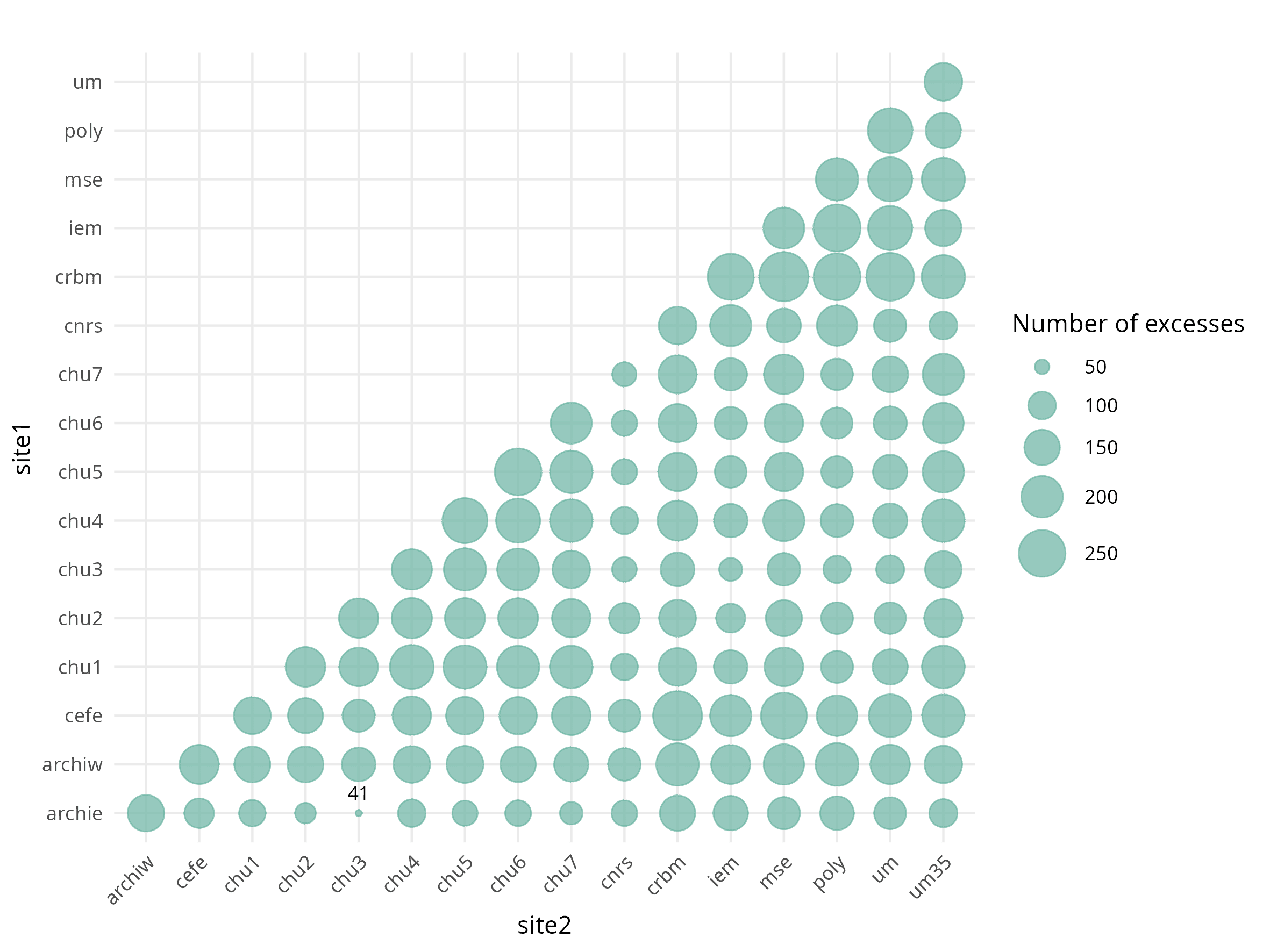}
    \caption{Spatial joint exceedances}
    \label{fig:excess_counts_spatial_0.95}
  \end{subfigure}
  \hfill
  \begin{subfigure}[b]{0.48\textwidth}
    \centering
    \includegraphics[width=\linewidth]{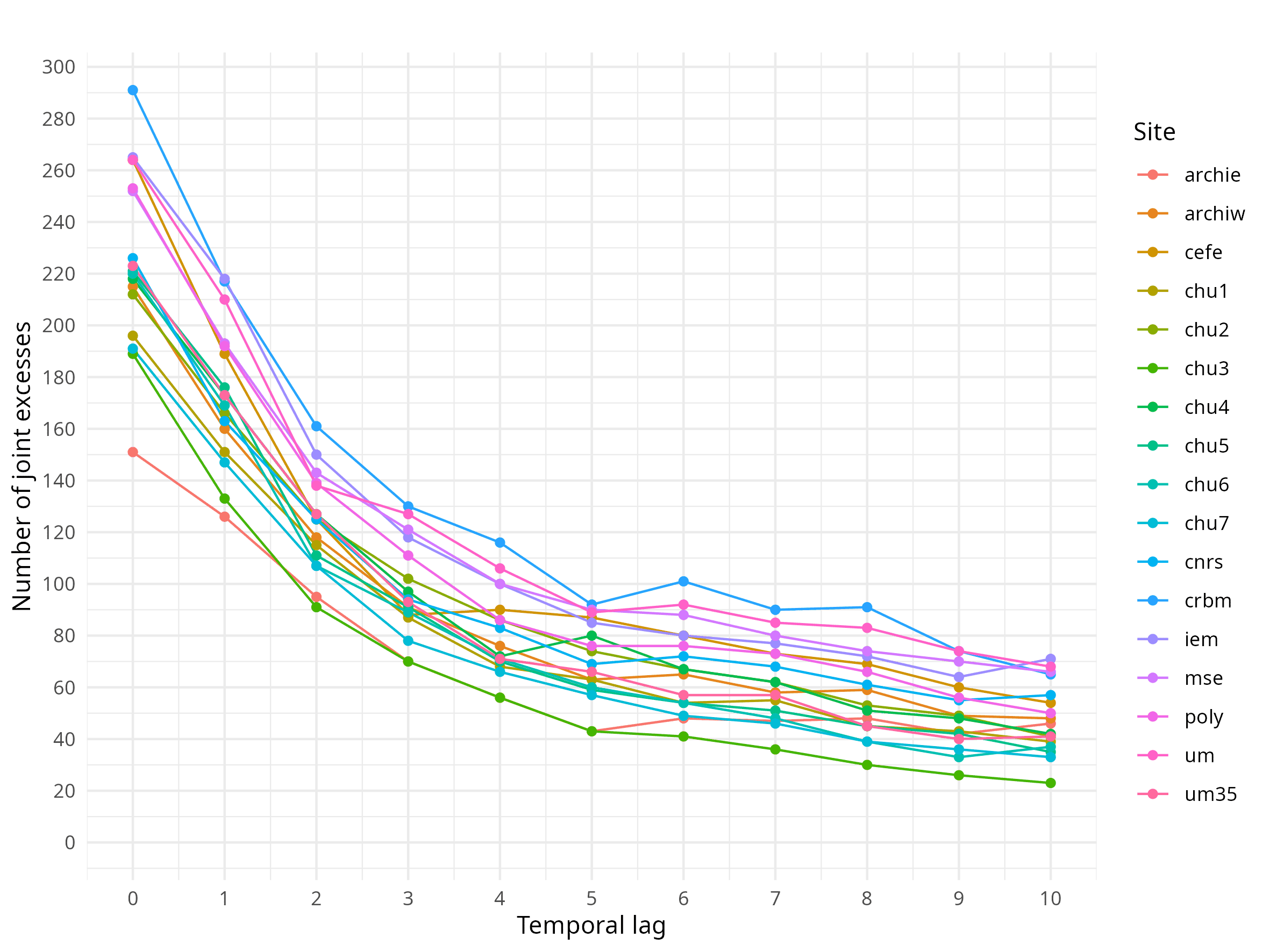}
    \caption{Temporal joint exceedances}
    \label{fig:excess_counts_temporal_0.95}
  \end{subfigure}
\caption{Number of joint exceedances at the $0.95$ quantile on the OMSEV dataset.
(a) By pair of sites (spatial distance).
(b) By temporal lag for a site with itself.}
\label{fig:excess_counts_0.95}
\end{figure}

\begin{figure}[H]
  \centering
  \begin{subfigure}[b]{0.48\linewidth}
    \centering
    \includegraphics[width=\linewidth]{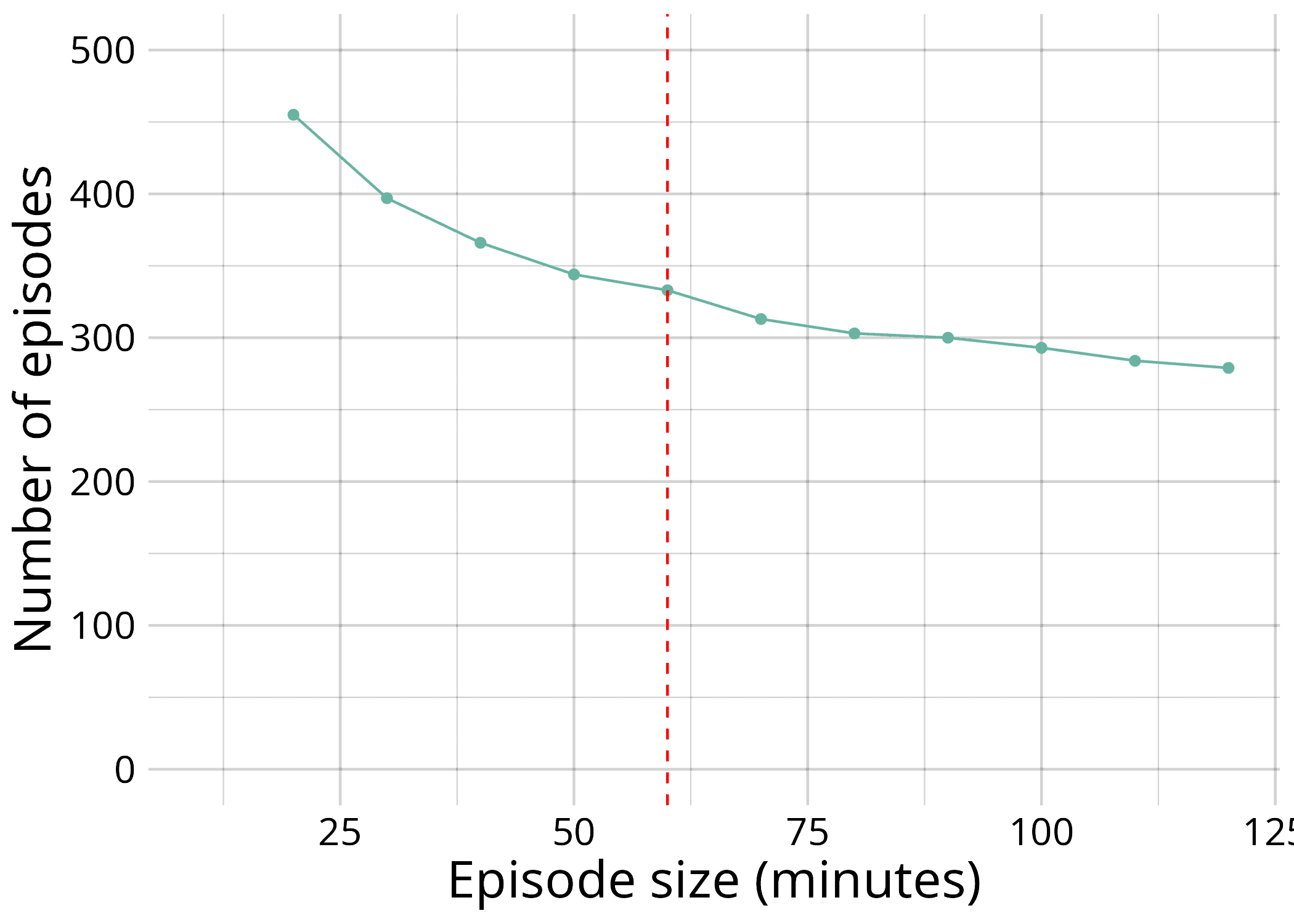}
    \caption{By episode duration $\delta$.}
    \label{fig:tradeoff_delta_omsev}
  \end{subfigure}
  \hfill
  \begin{subfigure}[b]{0.48\linewidth}
    \centering
    \includegraphics[width=\linewidth]{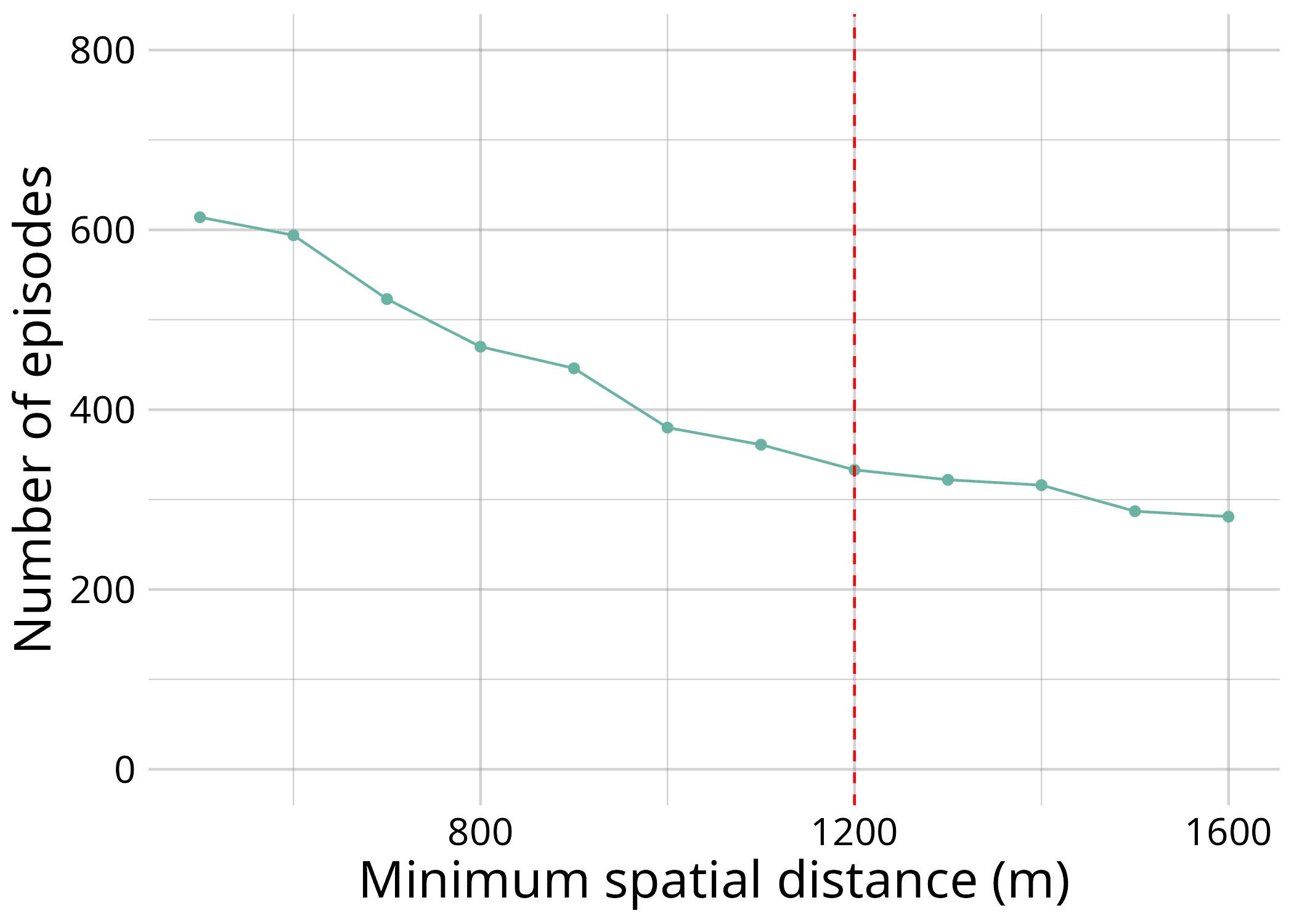}
    \caption{By minimum inter-episode distance $d_{\min}$.}
    \label{fig:tradeoff_dmin_omsev}
  \end{subfigure}
\caption{Number of selected episodes on the OMSEV dataset at the $0.95$ quantile.
(a) Effect of the minimum spatial separation $d_{\min}$ for $\delta=60$ min.
(b) Effect of the episode duration $\delta$ for $d_{\min}=1200$ m.
The selected configuration is shown by the red dashed line.}
  \label{fig:tradeoff_config_omsev}
\end{figure}

\section{Additional figures for stochastic rainfall generation validation}
\label{sec:appendix_episode_swg_validation}

\begin{figure}[H]
  \centering

  \begin{subfigure}[b]{0.49\textwidth}
    \centering
    \includegraphics[width=0.8\textwidth]{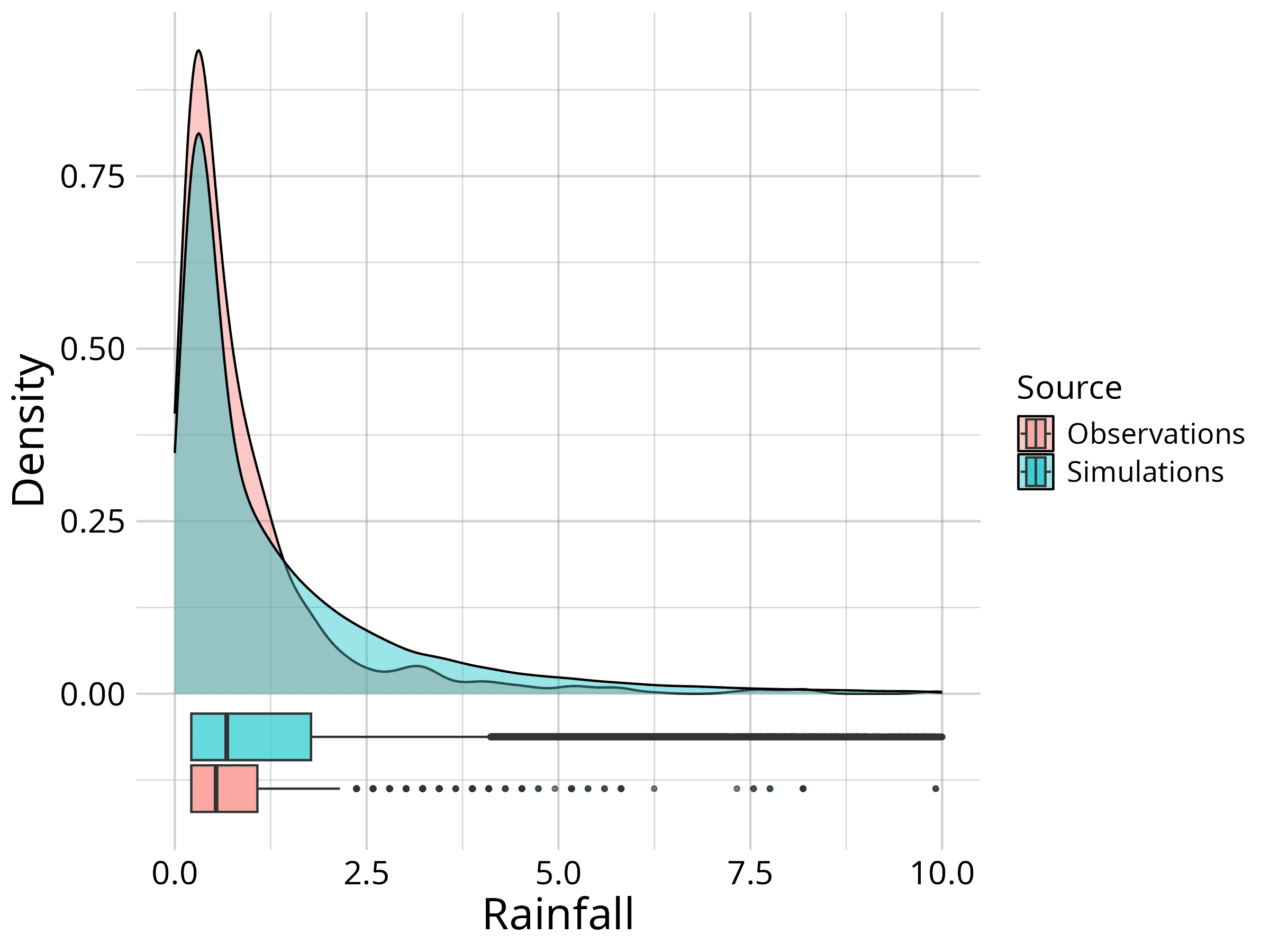}
    \caption{CHU 6}
    \label{fig:chu6above0_nodisc}
  \end{subfigure}
  \hfill
  \begin{subfigure}[b]{0.49\textwidth}
    \centering
    \includegraphics[width=0.8\textwidth]{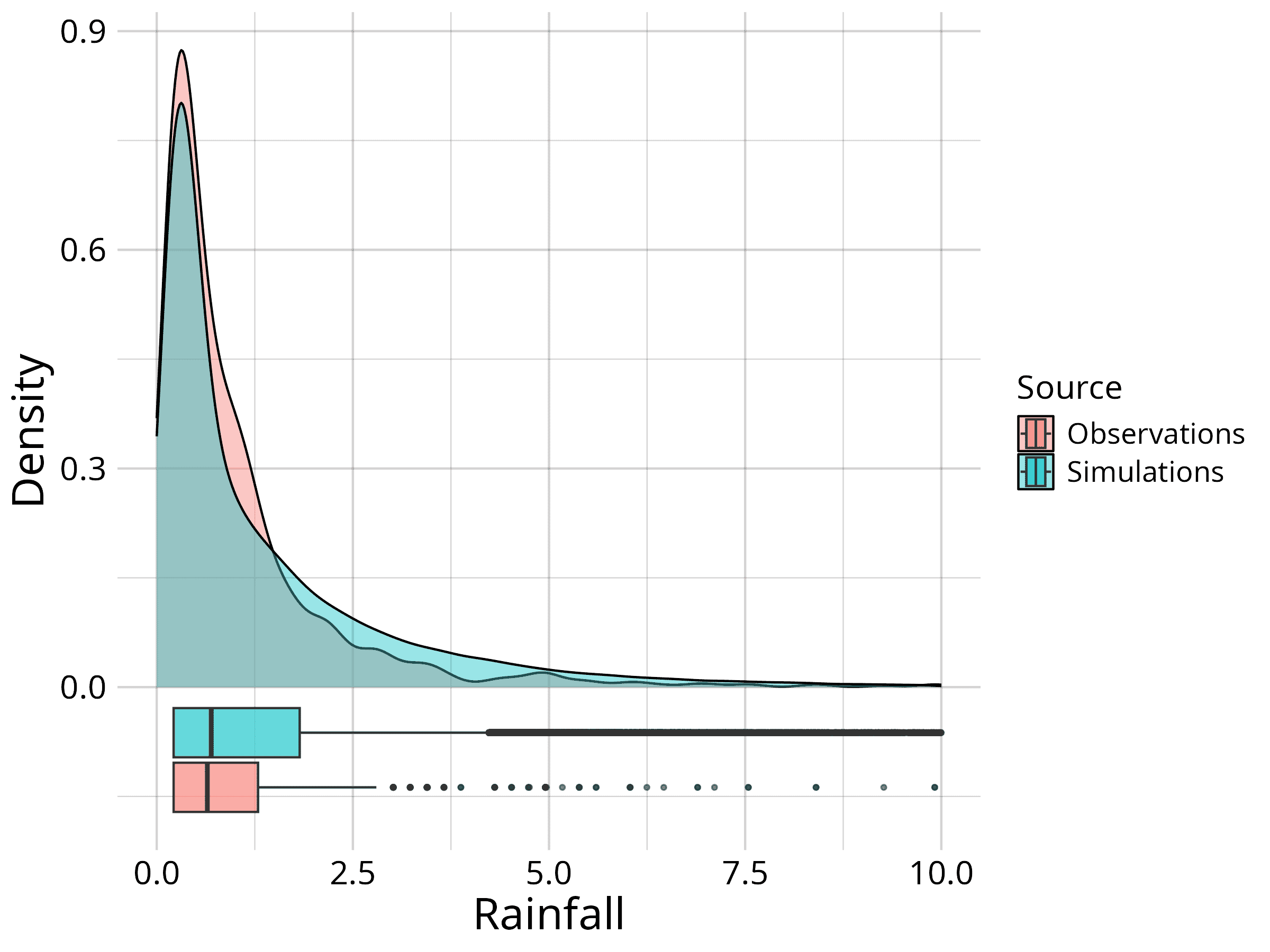}
    \caption{CHU 7}
    \label{fig:chu7above0_nodisc}
  \end{subfigure}

  \vspace{0.6em}

  \begin{subfigure}[b]{0.49\textwidth}
    \centering
    \includegraphics[width=0.8\textwidth]{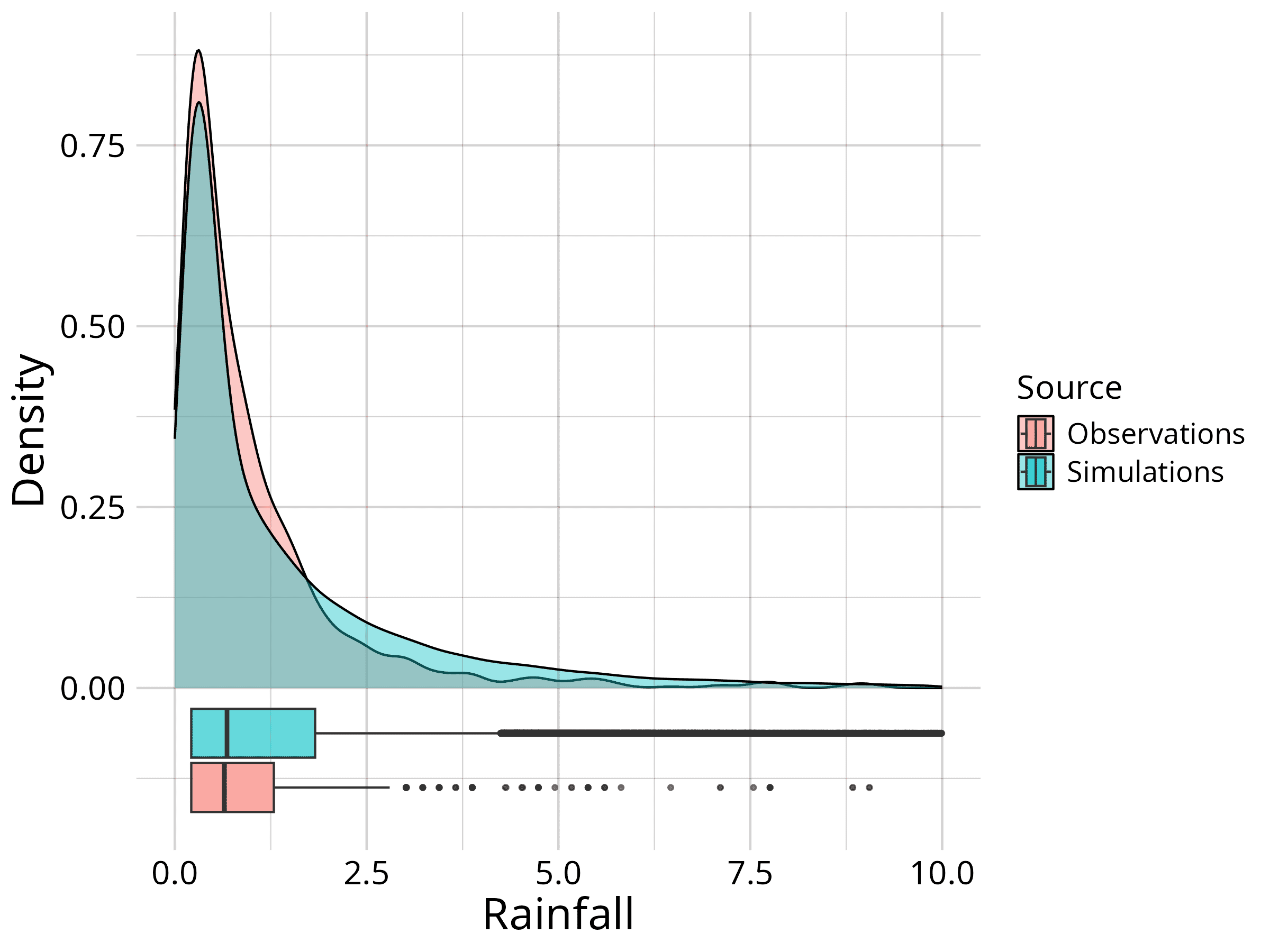}
    \caption{Polytech}
    \label{fig:polyabove0_disc}
  \end{subfigure}
  \hfill
  \begin{subfigure}[b]{0.49\textwidth}
    \centering
    \includegraphics[width=0.8\textwidth]{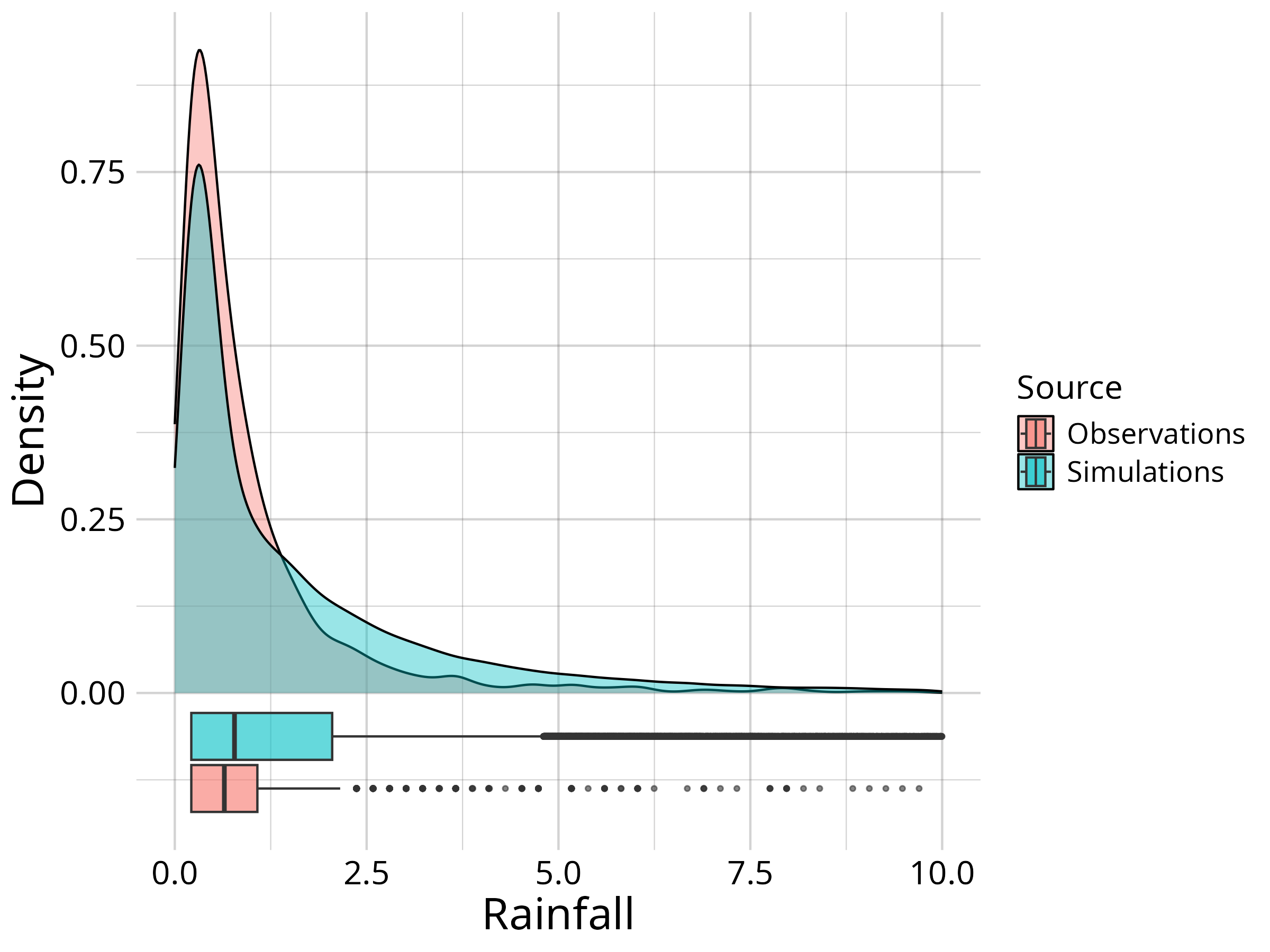}
    \caption{CRBM}
    \label{fig:crbmabove0_disc}
  \end{subfigure}

  \caption{Density of strictly positive rainfall values for observed against simulated episodes for two selected sites of the OMSEV network,
  with correction for the discretization effect. 
  For better visibility, only values below $10$~mm are shown.}
  \label{fig:densityswg_correction}
\end{figure}

\begin{figure}[H]
  \begin{subfigure}[t]{1\linewidth}
    \centering
  \includegraphics[width=0.7\linewidth]{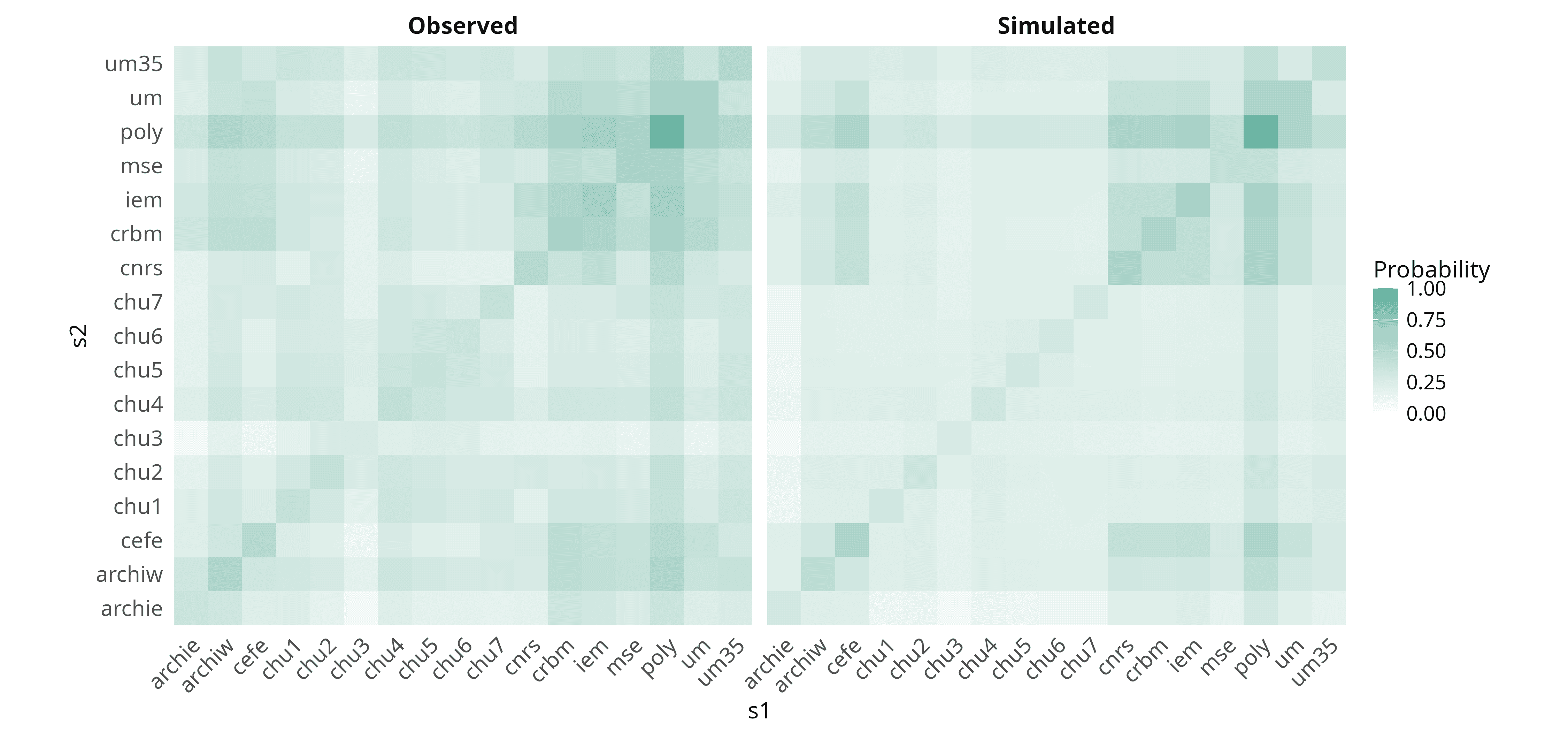}
      \caption{$\mathbb{P}(X_{s_1}>u, X_{s_2}>u | X_{\text{Polytech}}>u)$}
    \label{fig:polycondprobs}
  \end{subfigure}
  \vfill
  \begin{subfigure}[t]{1\linewidth}
    \centering
    \includegraphics[width=0.7\linewidth]{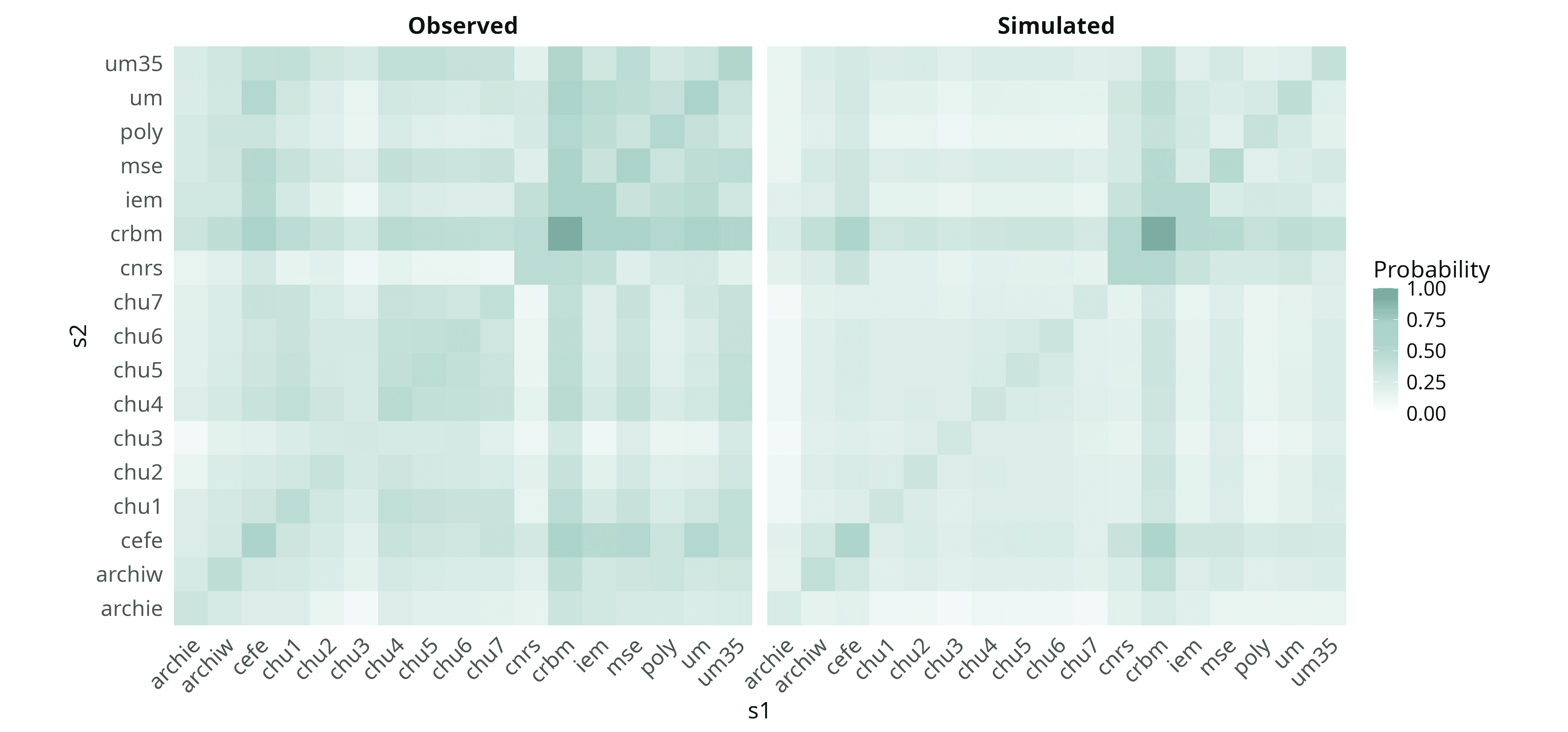}
    \caption{$\mathbb{P}(X_{s_1}>u, X_{s_2}>u | X_{\text{CRBM}}>u)$}
    \label{fig:crbmcondprobs}
  \end{subfigure}
   \caption{Conditional trivariate exceedance probabilities according to an excess at one site from observations versus simulations.
        Missing values in the observations are replicated in the simulations for better comparison.}
        \label{fig:condprobs}
\end{figure}

\end{document}